\documentclass[%
 preprint,
superscriptaddress,
 amsmath,amssymb,
 aps,
 prx,
 nofootinbib,
]{revtex4-2}

\usepackage{graphicx}
\usepackage{bm}
\usepackage{subcaption}

\begin{document}

\title{Mesoscopic Bayesian Inference by Solvable Models}

\author{Shun Katakami}
\affiliation{Graduate School of Frontier Sciences, The University of Tokyo, Kashiwa, Chiba 277-8561, Japan}
\author{Shuhei Kashiwamura}
\affiliation{Graduate School of Science, The University of Tokyo, Bunkyo, Tokyo 113-0033, Japan}
\author{Kenji Nagata}
\affiliation{Center for Basic Research on Materials, National Institute for Materials Science, Ibaraki, 305-0044, Japan}
\author{Masaichiro Mizumaki}
\affiliation{Faculty of Science, Course for Physical Sciences, Kumamoto University, Kumamoto, Kumamoto, 860-8555, Japan}
\author{Masato Okada}
\affiliation{Graduate School of Frontier Sciences, The University of Tokyo, Kashiwa, Chiba 277-8561, Japan}

\date{\today}

\begin{abstract}
The rapid advancement of data science and artificial intelligence has affected physics in numerous ways, including the application of Bayesian inference, setting the stage for a revolution in research methodology. Our group has proposed Bayesian measurement, a framework that applies Bayesian inference to measurement science with broad applicability across various natural sciences. This framework enables the determination of posterior probability distributions of system parameters, model selection, and the integration of multiple measurement datasets. However, applying Bayesian measurement to real data analysis requires a more sophisticated approach than traditional statistical methods like Akaike information criterion (AIC) and Bayesian information criterion (BIC), which are designed for an infinite number of measurements \(N\). Therefore, in this paper, we propose an analytical theory that explicitly addresses the case where \(N\) is finite in the linear regression model. We introduce\( O(1) \) mesoscopic variables for \(N\) observation noises. Using this mesoscopic theory, we analyze the three core principles of Bayesian measurement: parameter estimation, model selection, and measurement integration. Furthermore, by introducing these mesoscopic variables, we demonstrate that the difference in free energies, critical for both model selection and measurement integration, can be analytically reduced by two mesoscopic variables of \(N\) observation noises. This provides a deeper qualitative understanding of model selection and measurement integration and further provides deeper insights into actual measurements for nonlinear models. Our framework presents a novel approach to understanding Bayesian measurement results.
\end{abstract}

\maketitle

\section{Introduction}
\label{sec:intro}
The rapid development of data science and artificial intelligence has led to numerous studies in physics that actively incorporate these fields \cite{RevModPhys.91.045002, wang2023scientific}, aiming for new developments in physics. Among them, Bayesian inference shows high compatibility with traditional physics. Our group has proposed Bayesian measurement as a framework for applying Bayesian inference from statistics to measurement science \cite{nagata2012bayesian, nagata2019bayesian, tokuda2017simultaneous, katakami2022bayesian, ueda2023bayesian, nishimura2024bayesian, yokoyama2021bayesian, moriguchi2022bayesian, hayashi2023bayesian, yokoyama2023bayesian, yokoyama2021orbital, yamasaki2021bayesian, iwamitsu2020spectral, kumazoe2023quantifying, kashiwamura2022bayesian, tokuda2022intrinsic}. Bayesian measurement can be applied to almost all natural sciences, including physics, chemistry, life sciences, and earth and planetary sciences. In this framework, one can determine the posterior probability distribution of parameters for a mathematical model constituting a system. Additionally, if there are multiple mathematical models explaining the same phenomenon, one can perform model selection to determine the most appropriate model solely on the basis of measurement data. Furthermore, Bayesian integration, i.e., measurement integration, enables the integration of multiple data obtained from multiple measurements on the same system and determines how to integrate this data solely on the basis of the data itself. Bayesian measurement consists of three core principles: estimation of the posterior probability distribution of parameters, model selection, and Bayesian integration.

When performing Bayesian measurement, the results of model selection and Bayesian integration vary depending on the fluctuation of measurement data when the number of data \(N\) is finite. While Bayesian inference was first proposed by Thomas Bayes in the 18th century, its theoretical framework was traditionally developed under the assumption of an infinite number of measurement data \(N\), as represented by Bayesian information criterion (BIC) \cite{schwarz1978estimating}. Consequently, conventional BIC theory proves ineffective for model selection using Bayes Free Energy when dealing with a finite number of data N. The construction of theories that explicitly address finite N has become a crucial test of the practicality of Bayesian measurement. Our goal is to go beyond existing theories for an infinite number of data \(N\).

The purpose of this paper is to propose a novel theoretical framework for the three core principles of Bayesian measurement: estimation of the posterior probability distribution of parameters, model selection, and Bayesian integration, when the number of measurement data \(N\) is finite within the linear regression model. The proposed theory for \(N\) finiteness aims to analytically address the results of model selection and Bayesian integration. In the conventional framework that assumes the infinite limit of measurement data \(N\), which is typically seen in many theoretical frameworks of Bayesian inference, it is impossible to consider the fluctuations as random variables arising from the finiteness of \(N\). The proposed theory is an innovative framework that is fundamentally different from conventional theories. In this paper, we develop a solvable theory for the linear regression model \(y = ax + b\) with Gaussian noise as the measurement noise based on \(N\) quantity measurement data. This model, while seemingly simple, is not merely for theoretical analysis. It is widely used in real measurement settings, such as with linear system responses. Furthermore, insights gained from this model can be extended to general nonlinear models.

Let us assume that the observation noise in \(N\) observation data follows a Gaussian distribution. We define \(O(1)\) mesoscopic variables consisting of the \(N\) Gaussian noises within the linear regression model. Specifically, we define two Gaussian distributions and a chi-square distribution defined by the sum of \(N\) Gaussian noises. Using these mesoscopic variables, we propose a mesoscopic theory of the three core principles of Bayesian measurement: estimation of the posterior probability distribution of parameters, model selection, and Bayesian integration.

This paper is structured as follows. In Section \ref{sec:bayesianinference}, we develop a theory using mesoscopic variables to express the estimation of the posterior probability distribution of parameters in Bayesian measurement using the linear regression model \(y = ax + b\). In Section \ref{sec:modelselection}, we build on the mesoscopic theory in Section \ref{sec:bayesianinference} to propose a mesoscopic theory for model selection. This theory shows that the Bayesian free energy difference \(\Delta F\) that determines model selection fluctuates greatly when the number of data \(N\) is small. Furthermore, we show that by introducing mesoscopic variables, the free energy difference necessary in model selection can be analytically expressed with one mesoscopic variable of observation noise. In Section \ref{sec:bayesianintegration}, we propose a mesoscopic theory for Bayesian integration building on the mesoscopic theory in Section \ref{sec:bayesianinference}, and show that the Bayesian free energy difference \(\Delta F\) that determines the Bayesian integration fluctuates significantly when the number of data \(N\) is small. Furthermore, by introducing mesoscopic variables, we show that the free energy difference necessary in the Bayesian integration can be analytically expressed with several mesoscopic variables of \(N\) observation noises. In Sections \ref{sec:modelselection} and \ref{sec:bayesianintegration}, we provide the results of numerical calculations of model selection and Bayesian integration, respectively.

\section{Bayesian Inference with Linear Models}
\label{sec:bayesianinference}
In this section, we will demonstrate how the probability distribution of Bayesian free energy for finite data size in linear models can be described using a small number of variables within the basic framework of Bayesian inference. To advance the logic of Bayesian inference in linear models, we will first explain the mean squared error (MSE) associated with these models. Subsequently, we will derive the Bayesian posterior probability, enabling model parameter estimation, and the Bayesian free energy, which facilitates model selection.

\subsection{Mean Squared Error of Linear Models}
Here, to prepare for the discussion on Bayesian inference, we present the conventional MSE for linear models. Consider regressing data \(D = \{(x_i,y_i)\}_{i=1}^N\) with \(N\) samples using a two-variable linear model as follows:
\begin{equation}
 y = ax+b.
\end{equation}
In this context, the MSE is given by
\begin{eqnarray}
 E(a,b) &=& \frac{1}{2N}\sum_{i=1}^N \left\{y_i - (ax_i+b)\right\}^2,\\
 &=& \frac{1}{2}\left(\bar{y^2}-2a\bar{xy}-2b\bar{y}+a^2\bar{x^2}+2ab\bar{x}+b^2\right).
\end{eqnarray}
Here, we introduce the empirical means of the variables
\begin{eqnarray}
 \bar{x}&=&\frac{1}{N}\sum_{i=1}^N x_i.\\
 \bar{y}&=&\frac{1}{N}\sum_{i=1}^N y_i.\\
 \bar{x^2}&=& \frac{1}{N}\sum_{i=1}^N x_i^2.\\
 \bar{y^2}&=& \frac{1}{N}\sum_{i=1}^N y_i^2.\\
 \bar{xy}&=&\frac{1}{N}\sum_{i=1}^N x_iy_i.
\end{eqnarray}
For simplicity, let us assume the input mean of the data, \(\bar{x}=0\). Under this assumption, the MSE \(E(a,b)\) can be reformulated as:
\begin{eqnarray}
 E(a,b) = \mathcal{E}_a(a) + \mathcal{E}_b(b) + E(\hat{a},\hat{b}) \geq E(\hat{a},\hat{b}),
\end{eqnarray}
where \(\mathcal{E}_a(a)=\frac{1}{2}\bar{x^2}\left(a-\frac{\bar{xy}}{\bar{x^2}}\right)^2\), \(\mathcal{E}_b(b)=\frac{1}{2}(b-\bar{y})^2\), \(\hat{a}=\frac{\bar{xy}}{\bar{x^2}}\) and \(\hat{b}=\bar{y}\). The minimum value of the MSE \(E(\hat{a},\hat{b})\) is referred to as the residual error.

\subsection{Representation Through Microscopic Variables}
\subsubsection{Microscopic Notation of Mean Squared Error}
From this section, we introduce a noise model to facilitate the discussion of Bayesian inference. At this point, we have not addressed the noise model added to the data. Here, we assume the true parameters of \(a\) and \(b\) to be \(a_0\) and \(b_0\), respectively, and that the noise added to the data \(D\), denoted as \(\{n_i\}_{i=1}^N\), follows a normal distribution with mean zero and variance \(\sigma^2_0\). The process of generating the data is assumed to adhere to the following relation:

\begin{eqnarray}
 y_i = a_0x_i + b_0 + n_i,
\end{eqnarray}

where the probability distribution for the noise \(n_i\) is given by:
\begin{equation}
 p(n_i) = \frac{1}{\sqrt{2\pi \sigma^2_0}} \exp\left(-\frac{n_i^2}{2\sigma^2_0}\right). \label{noise_distribution}
\end{equation}

In this section, we delve deeper into understanding linear models by examining the dependency of the MSE on the stochastic variables \(\{n_i\}_{i=1}^N\). Given that \(\bar{x}=0\), the empirical means of inputs and outputs can be described as follows:

\begin{eqnarray}
 \bar{x}&=&\frac{1}{N}\sum_{i=1}^N x_i = 0.\\
 \bar{xy}&=&\frac{1}{N}\sum_{i=1}^N x_iy_i,\\
 &=&a_0\bar{x^2}+\bar{xn}.\\
 \bar{y}&=&\frac{1}{N}\sum_{i=1}^N y_i,\\
 &=&b_0+\bar{n}.\\
 \bar{y^2}&=&\frac{1}{N}\sum_{i=1}^N y_i^2,\\
 &=&a_0^2\bar{x^2}+b_0^2+\bar{n^2}+2b_0\bar{n}+2a_0\bar{xn}.
\end{eqnarray}
This can be described by introducing:
\begin{eqnarray}
 \bar{n} &=& \frac{1}{N}\sum_{i=1}^N n_i.\\
 \bar{n^2} &=& \frac{1}{N}\sum_{i=1}^N n_i^2.
\end{eqnarray}
Therefore, the MSE \(E(a,b)\) can be expressed as:
\begin{eqnarray}
 \hspace{-2cm} E(a,b) &=& \frac{1}{2}\bar{x^2}\left(a-a_0 -\frac{\bar{xn}}{\bar{x^2}}\right)^2
 + \frac{1}{2}(b-b_0-\bar{n})^2
 + \frac{1}{2}\left(-\frac{\bar{xn}^2}{\bar{x^2}}-\bar{n}^2+\bar{n^2}\right).
 \label{microE}
\end{eqnarray}

\subsubsection{Bayesian Inference for Linear Models}
From Equation (\ref{noise_distribution}), the conditional probability of observing the output \(y_i\) given the input variables and model parameters is described by
\begin{equation}
 p(y_i | a, b) = \frac{1}{\sqrt{2\pi\sigma^2_0}} \exp\left[-\frac{(y_i - a x_i - b)^2}{2\sigma^2_0}\right].
\end{equation}
Consequently, the joint conditional probability of all observed outputs \(Y = \{y_i\}_{i=1}^N\) can be expressed as
\begin{eqnarray}
 p(Y | a, b) &=& \prod_{i=1}^N p(y_i | a, b), \\
 &=& \left(\frac{1}{\sqrt{2\pi\sigma^2_0}}\right)^N \exp\left(-\frac{N}{\sigma^2_0}E(a, b)\right).
\end{eqnarray}
Utilizing the prior distributions of the linear model parameters \(a\) and \(b\), denoted as \(p(a)\) and \(p(b)\), respectively, the posterior distribution of the model parameters \(a, b\) according to Bayes' theorem can be formulated as:
\begin{eqnarray}
 p(a, b | Y) = \frac{p(Y | a, b) p(a) p(b)}{p(Y)}.
\end{eqnarray}
When the prior distributions of the model parameters \(a\) and \(b\) are independently assumed to be uniform within the ranges \([- \xi_a, \xi_a]\) and \([- \xi_b, \xi_b]\), respectively, the prior distributions for each parameter can be expressed as follows:
\begin{eqnarray}
 p(a) &=& \frac{1}{2\xi_a} \left\{ \Theta(a + \xi_a) - \Theta(a - \xi_a) \right\}, \\
 p(b) &=& \frac{1}{2\xi_b} \left\{ \Theta(b + \xi_b) - \Theta(b - \xi_b) \right\}.
\end{eqnarray}
The term \(p(Y)\), known as the marginal likelihood, is given by
\begin{eqnarray}
 p(Y) = \int \mathrm{d}a\mathrm{d}b \ p(Y|a,b)p(a)p(b). \label{posteriori1}
\end{eqnarray}
Given that the prior distributions are uniform, the posterior distribution can be expressed as
\begin{eqnarray}
 p(a,b|Y) &=& \left(\frac{1}{\sqrt{2\pi\sigma^2_0}}\right)^N \exp\left(-\frac{N}{\sigma^2_0}E(a,b)\right) \nonumber \\
 &\times& \frac{1}{2\xi_a}\left\{\Theta(a+\xi_a)-\Theta(a-\xi_a)\right\} \nonumber\\
 &\times& \frac{1}{2\xi_b}\left\{\Theta(b+\xi_b)-\Theta(b-\xi_b)\right\} \nonumber\\
 &\times& \frac{1}{p(Y)},\\
 &=& \frac{2N\sqrt{\bar{x^2}}}{\sigma^2_0\pi}\exp\left\{-\frac{N}{\sigma^2_0}\left[\mathcal{E}_a(a) + \mathcal{E}_b(b)\right]\right\} \nonumber\\ 
 &\times& \left\{\Theta(a+\xi_a)-\Theta(a-\xi_a)\right\} \left\{\Theta(b+\xi_b)-\Theta(b-\xi_b)\right\} \nonumber\\
 &\times& \left[\mathrm{erfc} \left(\sqrt{\frac{N\bar{x^2}}{2\sigma^2_0}}\left(-\xi_a-\frac{\bar{xy}}{\bar{x^2}}\right)\right)-\mathrm{erfc} \left(\sqrt{\frac{N\bar{x^2}}{2\sigma^2_0}}\left(\xi_a-\frac{\bar{xy}}{\bar{x^2}}\right)\right) \right]^{-1} \nonumber \\
 &\times& \left[\mathrm{erfc} \left(\sqrt{\frac{N}{2\sigma^2_0}}\left(-\xi_b-\bar{y}\right)\right)-\mathrm{erfc} \left(\sqrt{\frac{N}{2\sigma^2_0}}\left(\xi_b-\bar{y}\right)\right) \right]^{-1}. \label{posteriorstrict}
\end{eqnarray}
This expression enables us to compute the conditional probability of the model parameters given the data, as the posterior distribution.

Here, we derive the Bayesian free energy, which serves as an indicator for model selection and is defined as the negative logarithm of the marginal likelihood.
\begin{eqnarray}
 F(Y) &=& - \ln P(Y)\\
 &=&\frac{N}{2}\ln(2\pi\sigma^2_0) - \ln\left(\frac{\sigma^2_0\pi}{2N}\right) + \frac{1}{2}\ln\left(\bar{x^2}\right) + \ln(2\xi_a) + \ln(2\xi_b)+\frac{N}{\sigma^2_0}E(\hat{a},\hat{b})\nonumber\\
 \hspace{-4cm}&&-\ln \left[\mathrm{erfc} \left(\sqrt{\frac{N\bar{x^2}}{2\sigma^2_0}}\left(-\xi_a-\hat{a}\right)\right)-\mathrm{erfc} \left(\sqrt{\frac{N\bar{x^2}}{2\sigma^2_0}}\left(\xi_a-\hat{a}\right)\right) \right]\nonumber\\
 \hspace{-4cm}&&-\ln \left[\mathrm{erfc} \left(\sqrt{\frac{N}{2\sigma^2_0}}\left(-\xi_b-\hat{b}\right)\right)-\mathrm{erfc} \left(\sqrt{\frac{N}{2\sigma^2_0}}\left(\xi_b-\hat{b}\right)\right) \right].\label{FreeEstrict}
\end{eqnarray}

\subsection{Representation Through Mesoscopic Variables}
Up to this point, each statistical quantity has been treated empirically as an average. This section introduces the concept of mesoscopic variables, which enables a theoretical treatment of these quantities.

\subsubsection{Residual Error Through Mesoscopic Variables}
In the previous sections, the residual error was obtained as a probabilistic variable dependent on the stochastic variables \(\{n_i\}_{i=1}^N\). Here, we discuss the probability distribution of the value \(E(\hat{a},\hat{b})\times \frac{2N}{\sigma^2_0}\) and demonstrate that it follows a chi-squared distribution. The residual error was given by
\begin{eqnarray}
 E(\hat{a},\hat{b}) = \frac{1}{2}\left(-\frac{\bar{xn}^2}{\bar{x^2}}-\bar{n}^2+\bar{n^2}\right) \label{residualError}.
\end{eqnarray}
The first and second terms on the right side of Equation (\ref{residualError}) are independently distributed. Therefore, \(E(\hat{a},\hat{b}) \times \frac{2N}{\sigma^2_0}\) follows a chi-squared distribution with \(N-2\) degrees of freedom (proof \ref{append1}). Introducing a probability variable \(\upsilon\) that follows a chi-squared distribution with \(N-2\) degrees of freedom, we can write
\begin{eqnarray}
 p(\upsilon) = \frac{1}{2^{\frac{N-2}{2}}\Gamma(\frac{N-2}{2})}\upsilon^{\frac{N-4}{2}}\exp\left(-\frac{\upsilon}{2}\right).
\end{eqnarray}
Hence, the left side of Equation (\ref{residualError}), which is the residual error, can be expressed as
\begin{eqnarray}
 E(\hat{a},\hat{b}) = \frac{\sigma^2_0}{2N}\upsilon. \label{residual_meso}
\end{eqnarray}
Furthermore, the first and second terms on the right side of Equation (\ref{residualError}) can be expressed using independent stochastic variables \(\tau_1, \tau_2\), each following a normal distribution \(\mathcal{N}(0,1)\), as
\begin{eqnarray}
 \frac{\bar{xn}^2}{\bar{x^2}} &=& \frac{\sigma^2_0}{N}\tau_1^2,\\
 \bar{n}^2 &=& \frac{\sigma^2_0}{N} \tau_2^2.
\end{eqnarray}
This approach enables us to theoretically analyze the residual error, understand its distribution and behavior within the framework of Bayesian inference, and provide a more nuanced understanding of the error's properties. The respective representations of the derived micro variables and meso variables are summarized in Table \ref{table:variables}.

\begin{table}[h]
 \centering
 \caption{Summary of the respective representations of the micro and meso variables, and their respective relationships.}
 \begin{tabular}{cc}
 \begin{minipage}[t]{0.45\linewidth}
 \centering
 \vspace{1mm}
 \begin{tabular}{|c|c|}
 \hline
 \textbf{Meso $\tau_1, \tau_2, \upsilon$} & \textbf{Micro $\{n_i\}_{i=1}^N$} \\
 \hline
 $\tau_1$ & $\sqrt{\frac{N}{\sigma^2_0\bar{x^2}}}\bar{xn}$ \\
 \hline
 $\tau_2$ & $\sqrt{\frac{N}{\sigma^2_0}}\bar{n}$ \\
 \hline
 $\upsilon$ & $\frac{N}{\sigma^2_0}\left(-\frac{\bar{xn}^2}{\bar{x^2}}-\bar{n}^2+\bar{n^2}\right)$ \\
 \hline
 \end{tabular}
 \end{minipage} 
 \begin{minipage}[t]{0.5\linewidth}
 \centering
 \vspace{1mm}
 \begin{tabular}{|c|c|c|}
 \hline
 & \textbf{Micro $\{n_i\}_{i=1}^N$} & \textbf{Meso $\tau_1, \tau_2, \upsilon$} \\
 \hline
 $\hat{a}$ & $a_0 -\frac{\bar{xn}}{\bar{x^2}}$ & $a_0 + \sqrt{\frac{\sigma^2_0}{N\bar{x^2}}}\tau_1$ \\
 \hline
 $\hat{b}$ & $b_0-\bar{n}$ & $b_0 + \sqrt{\frac{\sigma^2_0}{N}}\tau_2$\\
 \hline
 $E(\hat{a},\hat{b})$ & $\frac{1}{2}\left(-\frac{\bar{xn}^2}{\bar{x^2}}-\bar{n}^2+\bar{n^2}\right)$ & $\frac{\sigma^2_0}{2N}\upsilon$\\
 \hline
 \end{tabular}
 \end{minipage} \\
 \end{tabular}
 \label{table:variables}
\end{table}

\subsubsection{Posterior Distribution Through Mesoscopic Variables}
Using the mesoscopic variables introduced in the previous section, we can reformulate the posterior distribution. From Equation (\ref{posteriorstrict}), the posterior distribution \(p(a,b|Y)\) can be rewritten as:
\begin{eqnarray}
 \hspace{-1cm}p(a,b|Y) &=& \frac{2N\sqrt{\bar{x^2}}}{\sigma^2_0\pi}\exp\left\{-\frac{N}{2\sigma^2_0}\left[\bar{x^2}\left(a-\hat{a}(\tau_1)\right)^2 + \left(b-\hat{b}(\tau_2)\right)^2\right]\right\} \nonumber\\ 
 &\times& \left\{\Theta(a+\xi_a)-\Theta(a-\xi_a)\right\} \left\{\Theta(b+\xi_b)-\Theta(b-\xi_b)\right\} \nonumber\\
 &\times& \left[\mathrm{erfc} \left(\sqrt{\frac{N\bar{x^2}}{2\sigma^2_0}}\left(-\xi_a-\hat{a}(\tau_1)\right)\right)-\mathrm{erfc} \left(\sqrt{\frac{N\bar{x^2}}{2\sigma^2_0}}\left(\xi_a-\hat{a}(\tau_1)\right)\right) \right]^{-1}\nonumber\\
 &\times& \left[\mathrm{erfc} \left(\sqrt{\frac{N}{2\sigma^2_0}}\left(-\xi_b-\hat{b}(\tau_2)\right)\right)-\mathrm{erfc} \left(\sqrt{\frac{N}{2\sigma^2_0}}\left(\xi_b-\hat{b}(\tau_2)\right)\right) \right]^{-1}.\label{posteriormeso}
\end{eqnarray}
Here, \(\hat{a}(\tau_1) = a_0 + \sqrt{\frac{\sigma^2_0}{N\bar{x^2}}}\tau_1\) and \(\hat{b}(\tau_2) = b_0 + \sqrt{\frac{\sigma^2_0}{N}}\tau_2\). Hence, the posterior distribution is determined solely by the two stochastic variables \(\tau_1\) and \(\tau_2\). Moreover, since Equation (\ref{posteriormeso}) enables independent calculations for \(a\) and \(b\), the distribution of model parameters \(a,b\) given the model, denoted as \(p_\mathrm{m}(a),p_\mathrm{m}(b)\), can be expressed as
\begin{eqnarray}
 p_\mathrm{m}(a) &=& \int \mathrm{d} \tau_1 \delta(a-\hat{a}(\tau_1))p(\tau_1),\\
 &=& \sqrt{\frac{N\bar{x^2}}{2\pi\sigma^2_0}}\exp\left(-\frac{N\bar{x^2}}{2\sigma^2_0}(a-a_0)^2\right),\\
 p_\mathrm{m}(b) &=& \int \mathrm{d} \tau_2 \delta(b-\hat{b}(\tau_2))p(\tau_2),\\
 &=& \sqrt{\frac{N}{2\pi\sigma^2_0}}\exp\left(-\frac{N}{2\sigma^2_0}(b-b_0)^2\right).
\end{eqnarray}
This shows that the posterior distribution can be represented in terms of mesoscopic variables, providing a theoretical framework to understand the distribution of model parameters \(a\) and \(b\) on the basis of observed data and assumed noise characteristics.

Here, we reformulate the Bayesian free energy using mesoscopic variables. From Equation (\ref{FreeEstrict}), the Bayesian free energy can be rewritten as
\begin{eqnarray}
 \hspace{-1cm} F(Y) &=& \frac{N}{2}\ln(2\pi\sigma^2_0) - \ln\left(\frac{\sigma^2_0\pi}{2N}\right) + \frac{1}{2}\ln\left(\bar{x^2}\right) + \ln(2\xi_a) + \ln(2\xi_b)+\frac{\upsilon}{2}\nonumber\\
 &-&\ln \left[\mathrm{erfc} \left(\sqrt{\frac{N\bar{x^2}}{2\sigma^2_0}}\left(-\xi_a-\hat{a}(\tau_1)\right)\right)-\mathrm{erfc} \left(\sqrt{\frac{N\bar{x^2}}{2\sigma^2_0}}\left(\xi_a-\hat{a}(\tau_1)\right)\right) \right]\nonumber\\
 &-&\ln \left[\mathrm{erfc} \left(\sqrt{\frac{N}{2\sigma^2_0}}\left(-\xi_b-\hat{b}(\tau_2)\right)\right)-\mathrm{erfc} \left(\sqrt{\frac{N}{2\sigma^2_0}}\left(\xi_b-\hat{b}(\tau_2)\right)\right) \right]. \label{Freemeso}
\end{eqnarray}
Note that in the limit of large \(N\), the negative logarithmic terms in the second and third lines of Equation (\ref{Freemeso}) converge to \(-\ln 2\). Therefore, the effect of stochastic fluctuations is effectively captured solely by the term \(\upsilon\).

Thus, the Bayesian free energy is determined by three stochastic variables \(\upsilon, \tau_1, \tau_2\), and can be expressed as \(F(Y) = F(\upsilon, \tau_1, \tau_2)\). The probability distribution of the Bayesian free energy is
\begin{eqnarray}
 p(F) = \int \mathrm{d}\upsilon\mathrm{d}\tau_1\mathrm{d}\tau_2 \delta(F - F(\upsilon,\tau_1,\tau_2))p(\upsilon)p(\tau_1)p(\tau_2). \label{dist_F}
\end{eqnarray}

In this section, we derive the representation of the probability distribution using mesoscopic variables. Although it is possible to describe the probability distribution without introducing mesoscopic variables, using microscopic variables leads to a computational complexity that scales proportionally with the number of data points \(N\). In contrast, the mesoscopic variable representation enables us to compute the probability distribution independently of the number of data points \(N\).

Specifically, the impact of the ln erfc term in Equation (\ref{Freemeso}) is negligible, so it can be considered a constant, resulting in the free energy distribution depending only on $\upsilon$. Since $\upsilon$ follows a chi-squared distribution, Equation (\ref{dist_F}) can be approximately analytically calculated.

\subsection{Numerical Experiments: Bayesian Inference}
Here, we numerically verify that the results of Bayesian estimation using the microscopic and mesoscopic expressions coincide. First, Figure 1 presents the probability distributions of residual errors calculated from the microscopic expression (\ref{residualError}) and mesoscopic expression (\ref{residual_meso}) for stochastically generated data. Panels (a)--(c) of Figure 1 show the probability distribution of normalized residual errors calculated using the microscopic expression (\ref{residualError}) for 100,000 artificially generated data patterns with model parameters \(a_0 = 1.0, b_0 = 0.0, \sigma_0^2 = 1.0\). On the other hand, panels (d)--(f) display the probability distribution obtained from 100,000 samplings of the probability distribution of residual errors under the mesoscopic expression (\ref{residual_meso}). Comparing the top and bottom rows of Figure 1, we can confirm that the distributions of residual errors from both microscopic and mesoscopic expressions match. As seen in Equation (\ref{residual_meso}), the residual error can be described as a chi-squared distribution, and Figure 1 demonstrates that as the number of data points increases, the chi-squared distribution asymptotically approaches a Gaussian distribution.

\begin{figure}[ht]
 \centering
 \begin{subfigure}{0.32\textwidth}
 \includegraphics[width=\linewidth]{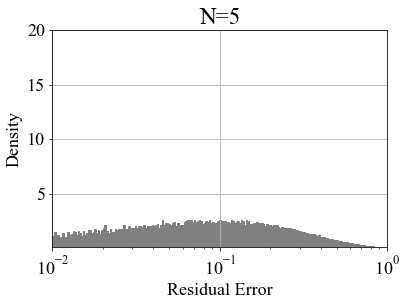}
 \caption{}
 \label{fig:image1}
 \end{subfigure}
 \hfill
 \begin{subfigure}{0.32\textwidth}
 \includegraphics[width=\linewidth]{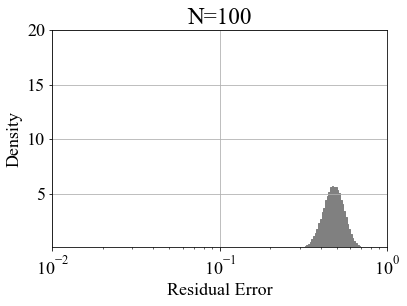}
 \caption{}
 \label{fig:image2}
 \end{subfigure}
 \hfill
 \begin{subfigure}{0.32\textwidth}
 \includegraphics[width=\linewidth]{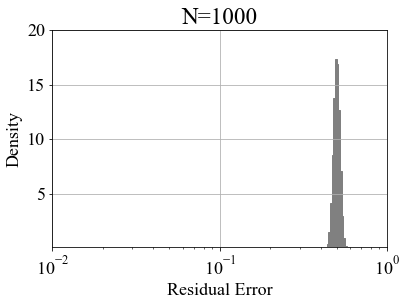}
 \caption{}
 \label{fig:image3}
 \end{subfigure}

 \vspace{5mm} 

 \begin{subfigure}{0.32\textwidth}
 \includegraphics[width=\linewidth]{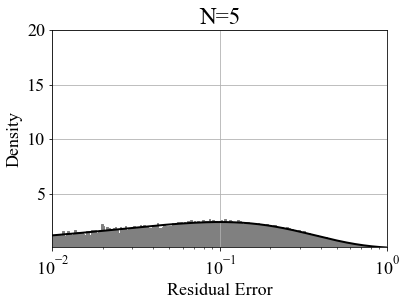}
 \caption{}
 \label{fig:image4}
 \end{subfigure}
 \hfill
 \begin{subfigure}{0.32\textwidth}
 \includegraphics[width=\linewidth]{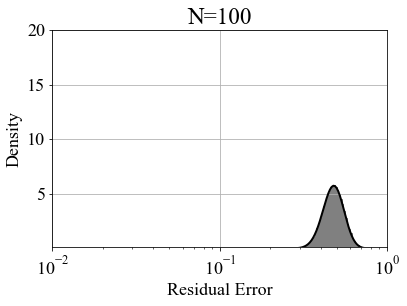}
 \caption{}
 \label{fig:image5}
 \end{subfigure}
 \hfill
 \begin{subfigure}{0.32\textwidth}
 \includegraphics[width=\linewidth]{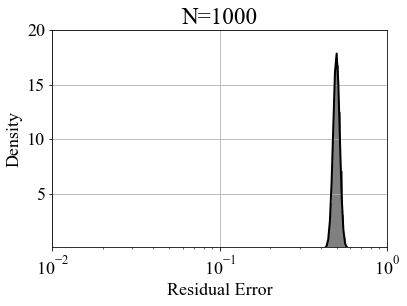}
 \caption{}
 \label{fig:image1-6}
 \end{subfigure}

\caption{Probability distribution of residual errors. (a)--(c): Probability distribution of values of residual errors calculated from the microscopic expression (\ref{residualError}) for 100,000 artificially generated data points with model parameters \(a_0 = 1.0, b_0 = 0.0, \sigma_0^2 = 1.0\). (d)--(f): Probability distribution obtained from 100,000 samples of the probability distribution of residual errors using the mesoscopic expression (\ref{residual_meso}). Solid black lines represent the theoretical lines calculated from the chi-squared distribution (Eq. (\ref{residual_meso})).}
\label{fig:six-images}
\end{figure}

Next, Figure 2 presents the probability distributions of free energy calculated from the microscopic expression (\ref{FreeEstrict}) and the mesoscopic expression (\ref{Freemeso}) for stochastically generated data. Panels (a)--(c) of Figure 2 show the probability distribution of normalized values of free energy calculated using the microscopic expression (\ref{FreeEstrict}) for 100,000 artificially generated data points with model parameters \(a_0 = 1.0, b_0 = 0.0, \sigma_0^2 = 1.0\). Meanwhile, panels (d)--(f) display the probability distribution obtained from 100,000 samples of the probability distribution of free energy using the mesoscopic expression (\ref{Freemeso}). A comparison between the top and bottom rows of Figure 2 confirms that the distributions of free energy from both the microscopic and mesoscopic expressions match.

\begin{figure}[ht]
\centering
\begin{subfigure}{0.32\textwidth}
\includegraphics[width=\linewidth]{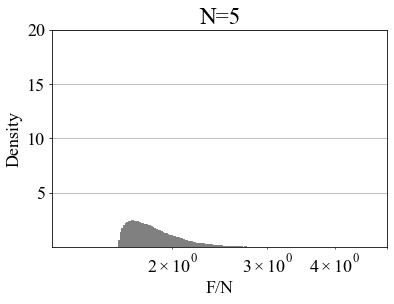}
\caption{}
\label{fig:image7}
\end{subfigure}
\hfill
 \begin{subfigure}{0.32\textwidth}
\includegraphics[width=\linewidth]{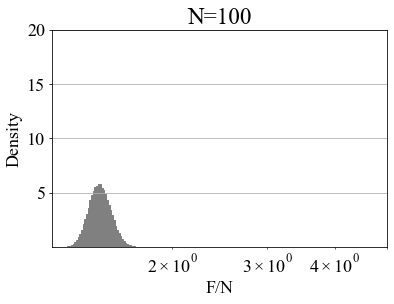}
\caption{}
\label{fig:image8}
 \end{subfigure}
\hfill
\begin{subfigure}{0.32\textwidth}
\includegraphics[width=\linewidth]{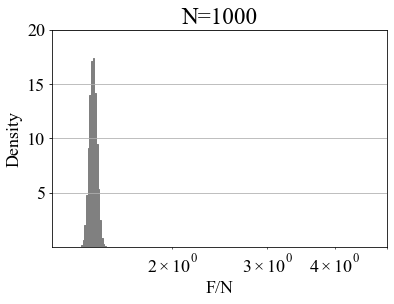}
\caption{}
\label{fig:image9}
\end{subfigure}

\vspace{5mm} 

\begin{subfigure}{0.32\textwidth}
\includegraphics[width=\linewidth]{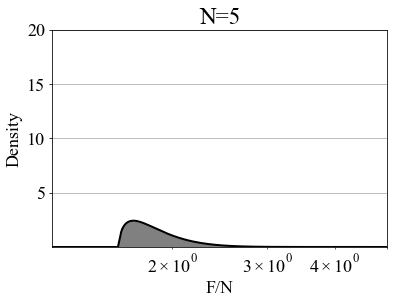}
\caption{}
\label{fig:image10}
\end{subfigure}
\hfill
\begin{subfigure}{0.32\textwidth}
\includegraphics[width=\linewidth]{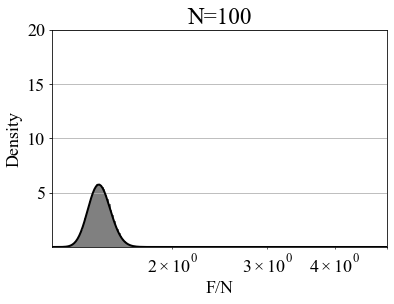}
\caption{}
\label{fig:image11}
\end{subfigure}
\hfill
\begin{subfigure}{0.32\textwidth}
\includegraphics[width=\linewidth]{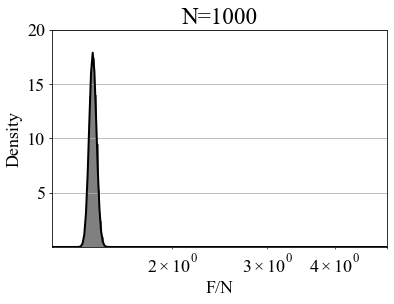}
\caption{}
\label{fig:image12}
\end{subfigure}

\caption{Probability distribution of free energy density. (a)--(c): Probability distribution of normalized values of free energy, where the normalization is performed by dividing the free energy by the number of data points \(N\). These values are calculated from the microscopic expression (\ref{FreeEstrict}) using 100,000 artificially generated data points with model parameters \(a_0 = 1.0, b_0 = 0.0, \sigma_0^2 = 1.0\). (d)--(f): Probability distribution obtained from 100,000 samples of the probability distribution of free energy using the mesoscopic expression (\ref{Freemeso}). Solid black lines represent the theoretical lines calculated from the chi-squared distribution, where the terms in the second and third lines of the mesoscopic expression (\ref{Freemeso}) were each approximated as \(-\log 2\).}
\label{fig:six-images2}
\end{figure}

\clearpage
\section{Model Selection}
\label{sec:modelselection}
This section explores model selection between a two- and one-variable linear regression model using the Bayesian free energy, as discussed in previous sections. That is, we deal with the problem of which model best fits a given dataset \(D = \{(x_i,y_i)\}_{i=1}^N\). Here, both models are defined as follows.
\begin{eqnarray}
 y_i & = & a x_i \\
 y_i & = & a x_i + b
\end{eqnarray}
Since the theoretical analysis of the two-variable model was covered in the previous section, this section first discusses the theoretical analysis of the one-variable model. Then, by considering the relationship between the two models via meso variables, we discuss the difference in free energy and the nature of model selection. In this section, the noise level is assumed to be predefined. The case where the noise level is also estimated is discussed in Appendix B.

\subsection{Representation of the One-Variable Linear Regression Model Using Microscopic Variables}
In this section, we assume that the data are generated from the one-variable model. That is, the following equation is assumed to be generated.
\begin{equation}
y_i = a_0x_i + n_i
\end{equation}
where \(\{n_i\}_{i=1}^N\) are normally distributed with mean zero and variance \(\sigma^2_0\).

\subsubsection{Microscopic Notation of Mean Squared Error for One-Variable Linear Model}
The MSE, similar to the discussions in the previous sections, can be written as
\begin{eqnarray}
 E(a) &=& \frac{1}{2}\left(\bar{y^2}-2a\bar{xy}+a^2\bar{x^2}\right),\\
 &=& \frac{1}{2} \left[\bar{x^2}\left(a-\frac{\bar{xy}}{\bar{x^2}}\right)^2-\frac{\bar{xy}^2}{\bar{x^2}}+\bar{y^2}\right],\\
 &=& \mathcal{E}_a(a) + E(\hat{a}).
\end{eqnarray}

Given that \(\bar{x}=0\), the empirical means of input and output can be described as
\begin{eqnarray}
 \bar{x}&=&\frac{1}{N}\sum_{i=1}^N x_i = 0,\\
 \bar{xy}&=&\frac{1}{N}\sum_{i=1}^N x_iy_i,\\
 &=&a_0\bar{x^2}+\bar{xn},\\
 \bar{y}&=&\frac{1}{N}\sum_{i=1}^N y_i, \\
 &=& \bar{n},\\
 \bar{y^2}&=&\frac{1}{N}\sum_{i=1}^N y_i^2,\\
 &=&a_0^2\bar{x^2}+\bar{n^2}+2a_0\bar{xn}.
\end{eqnarray}
Here, the residual error \(E(\hat{a})\) can be expressed as
\begin{eqnarray}
 E(\hat{a}) &=& \frac{1}{2}\left[-\frac{\bar{xn}^2}{\bar{x^2}}+\bar{n^2}\right].
\end{eqnarray}

\subsubsection{Bayesian Inference for One-Variable Linear Model}
Assuming that each noise \(n_i\) added to the data \(D = \{(x_i,y_i)\}_{i=1}^N\) independently follows a normal distribution with mean zero and variance \(\sigma^2_0\), the conditional probability of the output given the input variables and model parameters can be written as
\begin{eqnarray}
 p(Y|a) &=& \prod_{i=1}^N p(y_i|a),\\
 &=& \left(\frac{1}{\sqrt{2\pi\sigma^2_0}}\right)^N \exp\left(-\frac{N}{\sigma^2_0}E(a)\right).
\end{eqnarray}
Therefore, the joint conditional probability of all output data \(Y = \{y_i\}_{i=1}^N\) can be expressed as
\begin{eqnarray}
 p(Y) = \int \mathrm{d}a \ p(Y|a)p(a). \label{posterior2}
\end{eqnarray}
According to Bayes' theorem, the posterior distribution is
\begin{eqnarray}
 p(a|Y) &=& \left(\frac{1}{\sqrt{2\pi\sigma^2_0}}\right)^N \exp\left(-\frac{N}{\sigma^2_0}E(a)\right) \nonumber \\
 &\times& \frac{1}{2\xi_a}\left\{\Theta(a+\xi_a)-\Theta(a-\xi_a)\right\} \frac{1}{p(Y)}\\
 &=& \sqrt{\frac{2N\bar{x^2}}{\sigma^2_0\pi}}\exp\left\{-\frac{N}{\sigma^2_0}\mathcal{E}_a(a)\right\} \left\{\Theta(a+\xi_a)-\Theta(a-\xi_a)\right\}\nonumber\\
 &\times& \left[\mathrm{erfc} \left(\sqrt{\frac{N\bar{x^2}}{2\sigma^2_0}}\left(-\xi_a-\hat{a}\right)\right)-\mathrm{erfc} \left(\sqrt{\frac{N\bar{x^2}}{2\sigma^2_0}}\left(\xi_a-\hat{a}\right)\right) \right]^{-1}.\label{posteriorstrict2}
\end{eqnarray}

Here, we derive the Bayesian free energy for a one-variable linear regression model. The Bayesian free energy is obtained by taking the negative logarithm of the marginal likelihood.
\begin{eqnarray}
 &&F(Y) = \frac{N}{2}\ln(2\pi\sigma^2_0) - \frac{1}{2} \ln\left(\frac{\sigma^2_0\pi}{2N\bar{x^2}}\right) + \ln(2\xi_a) +\frac{N}{\sigma^2_0}E(\hat{a})\nonumber\\
 &&-\ln \left[\mathrm{erfc} \left(\sqrt{\frac{N\bar{x^2}}{2\sigma^2_0}}\left(-\xi_a-\hat{a}\right)\right)-\mathrm{erfc} \left(\sqrt{\frac{N\bar{x^2}}{2\sigma^2_0}}\left(\xi_a-\hat{a}\right)\right) \right]. \label{FreeEstrict2}
\end{eqnarray}

\subsection{Representation of the One-Variable Linear Regression Model Using Mesoscopic Variables}
Up to this point, each statistical quantity has been considered as an empirical mean. This section, following the approach of the previous one, introduces mesoscopic variables to provide a theoretical framework for handling these quantities.

\subsubsection{Residual Error in One-Variable Linear Regression Model Through Mesoscopic Variables}
In the previous sections, the residual error was obtained as a probabilistic variable dependent on the stochastic variables \(\{n_i\}_{i=1}^N\). Here, we discuss the probability distribution of the value \(E(\hat{a})\times \frac{2N}{\sigma^2_0}\) and demonstrate that it follows a chi-squared distribution. The residual error was given by
\begin{eqnarray}
 E(\hat{a}) = \frac{1}{2}\left(-\frac{\bar{xn}^2}{\bar{x^2}}+\bar{n^2}\right) \label{residualError2}.
\end{eqnarray}
The terms on the right side of Equation (\ref{residualError2}) are independently distributed. Therefore, \(E(\hat{a}) \times \frac{2N}{\sigma^2_0}\) follows a chi-squared distribution with \(N-1\) degrees of freedom. Introducing a probability variable \(\upsilon_2\) that follows a chi-squared distribution with \(N-1\) degrees of freedom, we can write
\begin{eqnarray}
 p(\upsilon_2) = \frac{1}{2^{\frac{N-1}{2}}\Gamma(\frac{N-1}{2})}\upsilon^{\frac{N-3}{2}}\exp\left(-\frac{\upsilon}{2}\right).
\end{eqnarray}
Thus, the left side of Equation (\ref{residualError2}), which is the residual error, can be expressed as
\begin{eqnarray}
 E(\hat{a}) = \frac{\sigma^2_0}{2N}\upsilon_2.
\end{eqnarray}
Furthermore, the first term on the right side of Equation (\ref{residualError2}) can be expressed using an independent stochastic variable \(\tau_1\), following a normal distribution \(\mathcal{N}(0,1)\), as
\begin{eqnarray}
 \frac{\bar{xn}^2}{\bar{x^2}} &=& \frac{\sigma^2_0}{N}\tau_1^2.
\end{eqnarray}

\subsubsection{Posterior Distribution in One-Variable Linear Regression Model Through Mesoscopic Variables}
Using the mesoscopic variables introduced in the previous section, we can reformulate the posterior distribution. From Equation (\ref{posteriorstrict2}), the posterior distribution \(p(a|Y)\) can be rewritten as
\begin{align}
 p(a|Y) &= \sqrt{\frac{2N\bar{x^2}}{\sigma^2_0\pi}} \exp\left\{-\frac{N\bar{x^2}}{2\sigma^2_0}\left(a-\hat{a}(\tau_1)\right)^2\right\} \nonumber\\
 &\quad \times \left\{\Theta(a+\xi_a) - \Theta(a-\xi_a)\right\} \nonumber\\
 &\quad \times \left[\mathrm{erfc} \left(\sqrt{\frac{N\bar{x^2}}{2\sigma^2_0}}\left(-\xi_a-\hat{a}(\tau_1)\right)\right) \right. \nonumber\\
 &\quad \quad \left. - \mathrm{erfc} \left(\sqrt{\frac{N\bar{x^2}}{2\sigma^2_0}}\left(\xi_a-\hat{a}(\tau_1)\right)\right) \right]^{-1} \label{posteriormeso2}
\end{align}
Thus, the posterior distribution is determined solely by the stochastic variable \(\tau_1\). Moreover, the distribution of the model parameter \(a\), given the model, denoted as \(p_\mathrm{m}(a)\), can be expressed as
\begin{eqnarray}
 p_\mathrm{m}(a) &=& \int \mathrm{d} \tau_1 \delta(a-\hat{a}(\tau_1))p(\tau_1)\\
 &=& \sqrt{\frac{N\bar{x^2}}{2\pi\sigma^2_0}}\exp\left(-\frac{N\bar{x^2}}{2\sigma^2_0}(a-a_0)^2\right).
\end{eqnarray}

Here, we reformulate the Bayesian free energy using mesoscopic variables. From Equation (\ref{FreeEstrict2}), the Bayesian free energy can be rewritten as
\begin{eqnarray}
 \hspace{-2cm} F(Y) &=& \frac{N}{2}\ln(2\pi\sigma^2_0) - \frac{1}{2}\ln\left(\frac{\sigma^2_0\pi}{2N\bar{x^2}}\right) + \ln(2\xi_a) +\frac{\upsilon_2}{2}\nonumber\\
 &-&\ln \left[\mathrm{erfc} \left(\sqrt{\frac{N\bar{x^2}}{2\sigma^2_0}}\left(-\xi_a-\hat{a}(\tau_1)\right)\right)-\mathrm{erfc} \left(\sqrt{\frac{N\bar{x^2}}{2\sigma^2_0}}\left(\xi_a-\hat{a}(\tau_1)\right)\right) \right]. \label{Freemeso2}
\end{eqnarray}

Therefore, the Bayesian free energy is determined by two stochastic variables, \(\upsilon_2\) and \(\tau_1\), and can be expressed as \(F(Y) = F(\upsilon_2, \tau_1)\). The probability distribution of the Bayesian free energy can be described as
\begin{eqnarray}
 p(F) = \int \mathrm{d}\upsilon_2\mathrm{d}\tau_1 \delta(F - F(\upsilon_2,\tau_1))p(\upsilon_2)p(\tau_1).
\end{eqnarray}

\subsubsection{Model Selection Through Bayesian Free Energy}
This section compares the Bayesian free energy of a two- and one-variable linear regression model to perform model selection. First, we will discuss the relationship between mesoscopic variables \(\upsilon_1, \upsilon_2\). The residual error for the two models can be expressed as
\begin{eqnarray}
 E(\hat{a},\hat{b}) &=& \frac{1}{2N}\sum_{i=1}^N\{y_i-(\hat{a}x_i+\hat{b})\}^2\\
 &=& E(\hat{a}) - \frac{1}{2} \left(b_0+\sqrt{\frac{{\sigma_0}^2}{N}}\tau_2\right)^2 \label{relation_residual}
\end{eqnarray}
leading to the relationship between \(\upsilon_1\) and \(\upsilon_2\) as
\begin{eqnarray}
 \upsilon_1 = \upsilon_2 - \frac{N}{\sigma_0^2}\left(b_0+\sqrt{\frac{{\sigma_0}^2}{N}}\tau_2\right)^2.
\end{eqnarray}
The Bayesian free energy for each model, from Equations (\ref{Freemeso}) and (\ref{Freemeso2}), is given by:
\begin{eqnarray}
 \hspace{-5cm}F_{y=ax+b}(\upsilon_1,\tau_1,\tau_2) &=& \frac{N}{2}\ln(2\pi\sigma^2_0) - \ln\left(\frac{\sigma^2_0\pi}{2N}\right) + \frac{1}{2}\ln\left(\bar{x^2}\right) + \ln(2\xi_a) + \ln(2\xi_b)+\frac{\upsilon_1}{2}\nonumber\\
 &-&\ln \left[\mathrm{erfc} \left(\sqrt{\frac{N\bar{x^2}}{2\sigma^2_0}}\left(-\xi_a-\hat{a}(\tau_1)\right)\right)-\mathrm{erfc} \left(\sqrt{\frac{N\bar{x^2}}{2\sigma^2_0}}\left(\xi_a-\hat{a}(\tau_1)\right)\right) \right]\nonumber\\
 &-&\ln \left[\mathrm{erfc} \left(\sqrt{\frac{N}{2\sigma^2_0}}\left(-\xi_b-\hat{b}(\tau_2)\right)\right)-\mathrm{erfc} \left(\sqrt{\frac{N}{2\sigma^2_0}}\left(\xi_b-\hat{b}(\tau_2)\right)\right) \right]
\end{eqnarray}
\begin{align}
 F_{y=ax}(\upsilon_2,\tau_1) &= \frac{N}{2}\ln(2\pi\sigma^2_0) - \frac{1}{2}\ln\left(\frac{\sigma^2_0\pi}{2N\bar{x^2}}\right) + \ln(2\xi_a) +\frac{\upsilon_2}{2}\nonumber\\
 &-\ln \left[\mathrm{erfc} \left(\sqrt{\frac{N\bar{x^2}}{2\sigma^2_0}}\left(-\xi_a-\hat{a}(\tau_1)\right)\right)-\mathrm{erfc} \left(\sqrt{\frac{N\bar{x^2}}{2\sigma^2_0}}\left(\xi_a-\hat{a}(\tau_1)\right)\right) \right]
\end{align}
Hence, the difference in the Bayesian free energy (\(\Delta F\)) depends only on the stochastic variable \(\tau_2\), and can be expressed as
\begin{eqnarray}
 \hspace{-2cm}\Delta F(\tau_2) &=& F_{y=ax}(\upsilon_2,\tau_1) - F_{y=ax+b}(\upsilon_1,\tau_1,\tau_2)\\
 &=& \frac{1}{2}\ln\left(\frac{\sigma^2_0\pi}{2N}\right) - \ln(2\xi_b) + \frac{N}{2\sigma_0^2}\hat{b}(\tau_2)^2\nonumber\\
 &+& \ln \left[\mathrm{erfc} \left(\sqrt{\frac{N}{2\sigma^2_0}}\left(-\xi_b-\hat{b}(\tau_2)\right)\right)-\mathrm{erfc} \left(\sqrt{\frac{N}{2\sigma^2_0}}\left(\xi_b-\hat{b}(\tau_2)\right)\right) \right]. \label{delta_F_model_selection}
\end{eqnarray}
Note that the logarithmic term in the second line converges to \(\log 2\) in the limit of large \(N\), thereby indicating that the stochastic fluctuations are primarily affected by the third term \(\frac{N}{2\sigma^2_0}\hat{b}(\tau_2)^2\). Since \(\hat{b}(\tau_2)\) follows a normal distribution with mean \(b_0\) and variance \(\sigma^2/N\), \(\hat{b}(\tau_2)^2\) follows a non-central chi-squared distribution. This enables analytical treatment of the distribution of the free energy difference. The probability distribution of the difference in the Bayesian free energy is
\begin{eqnarray}
 p(\Delta F) = \int \mathrm{d}\tau_2 \delta(\Delta F - \Delta F(\tau_2)) p(\tau_2). \label{prob_modelselection}
\end{eqnarray}
We can effectively assess the fluctuations of model selection by mesoscopic variables as described in Section \ref{sec:bayesianinference}.

\subsection{Numerical Experiments: Model Selection}
\begin{figure}[ht]
 \centering
 \begin{subfigure}{0.32\textwidth}
 \includegraphics[width=\linewidth]{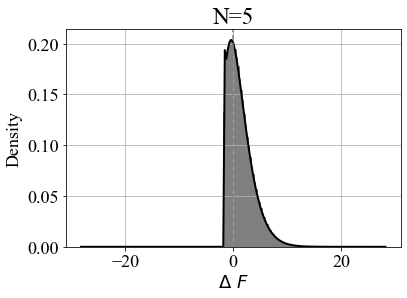}
 \caption{}
 \label{fig:image13}
 \end{subfigure}
 \hfill 
 \begin{subfigure}{0.32\textwidth}
 \includegraphics[width=\linewidth]{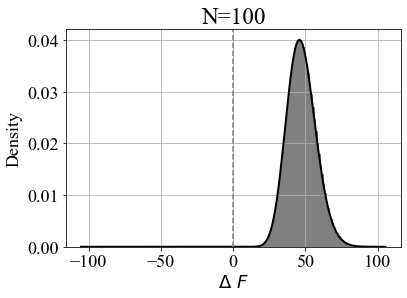}
 \caption{}
 \label{fig:image14}
 \end{subfigure}
 \hfill 
 \begin{subfigure}{0.32\textwidth}
 \includegraphics[width=\linewidth]{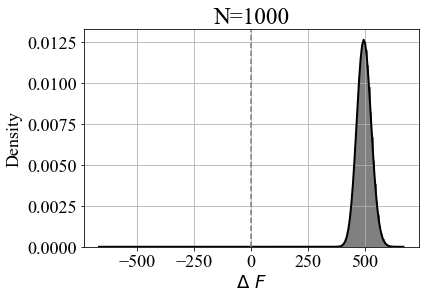}
 \caption{}
 \label{fig:image15}
 \end{subfigure}

\caption{Probability distribution of differences in the Bayesian free energy for model selection. Probability distribution from 100,000 samples of the Bayesian free energy difference (Equation (\ref{prob_modelselection})) with model parameters \(a_0 = 1.0, b_0 = 1.0, \sigma_0^2 = 1.0\) for data sizes \(N=5, 100, 1000\). Solid black lines represent the theoretical lines calculated from the non-central chi-squared distribution, where the second line of Equation (\ref{prob_modelselection}) was approximated as \(\ln 2\).}
 \label{fig:delta_F_two_variable}
\end{figure}

Here, we examine the impact of data quantity and noise intensity inherent in the data on the outcomes of model selection using the mesoscopic representation. Figure \ref{fig:delta_F_two_variable} shows the two-dimensional frequency distribution from 100,000 samples of the Bayesian free energy difference (Equation (\ref{prob_modelselection})) and \(\tau_2\). The model parameters are \(a_0 = 1.0\), \(b_0 = 1.0\), and \(\sigma_0^2 = 1.0\), for data sizes \(N = 5, 100, 1000\). The vertical and horizontal axes represent the frequency distributions of the free energy difference and of \(\tau_2\), respectively. Figure \ref{fig:delta_F_two_variable}(a) shows that with a small number of data points, the frequency of \(\Delta F < 0\) is high, indicating frequent failures in model selection. Conversely, Figures \ref{fig:delta_F_two_variable}(b) and (c) show that with a larger number of data points, failures in model selection become negligible.

\begin{figure}[ht]
 \centering
 \begin{subfigure}{0.32\textwidth}
 \includegraphics[width=\linewidth]{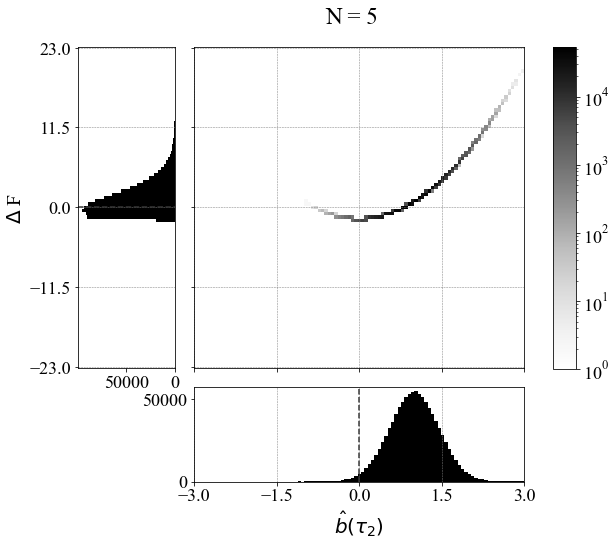}
 \caption{}
 \label{fig:image13}
 \end{subfigure}
 \hfill 
 \begin{subfigure}{0.32\textwidth}
 \includegraphics[width=\linewidth]{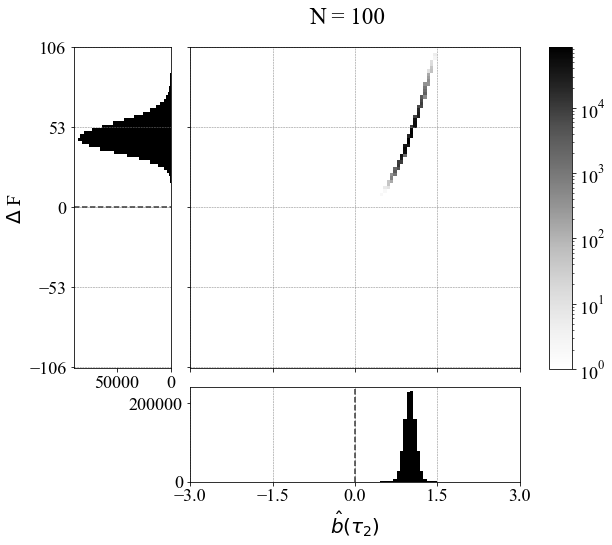}
 \caption{}
 \label{fig:image14}
 \end{subfigure}
 \hfill 
 \begin{subfigure}{0.32\textwidth}
 \includegraphics[width=\linewidth]{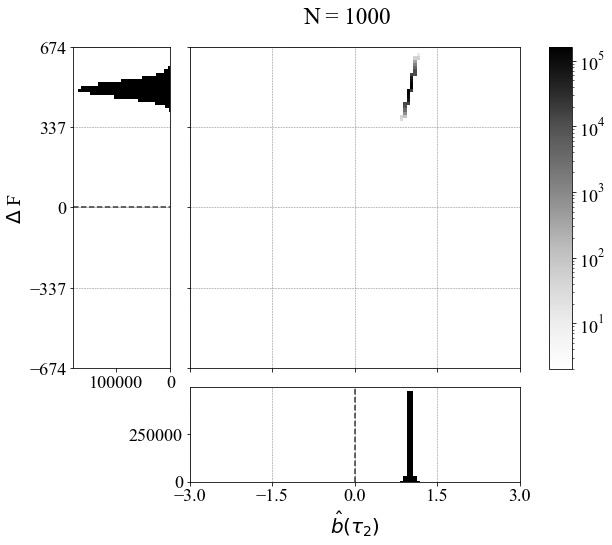}
 \caption{}
 \label{fig:image15}
 \end{subfigure}
\caption{Two-dimensional frequency distribution from 100,000 samples of the Bayesian free energy difference (Equation (\ref{prob_modelselection})) and \(\hat{b}(\tau_2)\). The model parameters are \(a_0 = 1.0, b_0 = 1.0\), and \(\sigma_0^2 = 1.0\), for data sizes \(N = 5, 100, 1000\). The vertical and horizontal axes represent the frequency distributions of the free energy difference and of \(\tau_2\), respectively.}

 \label{fig:delta_F_two_variable2D}
\end{figure}

\begin{figure}[ht]
 \centering
 \begin{subfigure}{0.32\textwidth}
 \includegraphics[width=\linewidth]{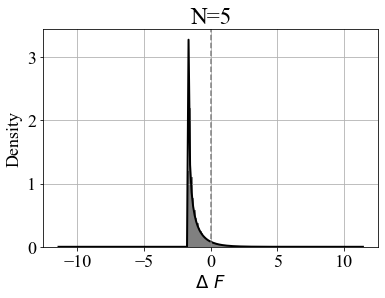}
 \caption{}
 \label{fig:image16}
 \end{subfigure}
 \hfill 
 \begin{subfigure}{0.32\textwidth}
 \includegraphics[width=\linewidth]{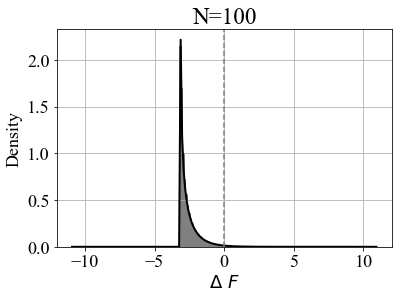}
 \caption{}
 \label{fig:image17}
 \end{subfigure}
 \hfill 
 \begin{subfigure}{0.32\textwidth}
 \includegraphics[width=\linewidth]{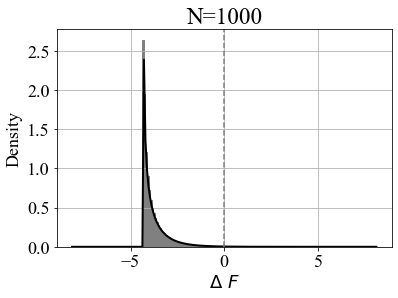}
 \caption{}
 \label{fig:image18}
 \end{subfigure}

\caption{Probability distribution of differences in the Bayesian free energy for model selection. Probability distribution from 100,000 samples of the Bayesian free energy difference (Equation (\ref{prob_modelselection})) with model parameters \(a_0 = 1.0, b_0 = 0.0, \sigma_0^2 = 1.0\) for data sizes \(N=5, 100, 1000\). Solid black lines represent the theoretical lines calculated from the non-central chi-squared distribution, where the second line of Equation (\ref{prob_modelselection}) was approximated as \(\ln 2\).}
\label{fig:delta_F_one_variable}
\end{figure}

\begin{figure}[ht]
 \centering
 \begin{subfigure}{0.32\textwidth}
 \includegraphics[width=\linewidth]{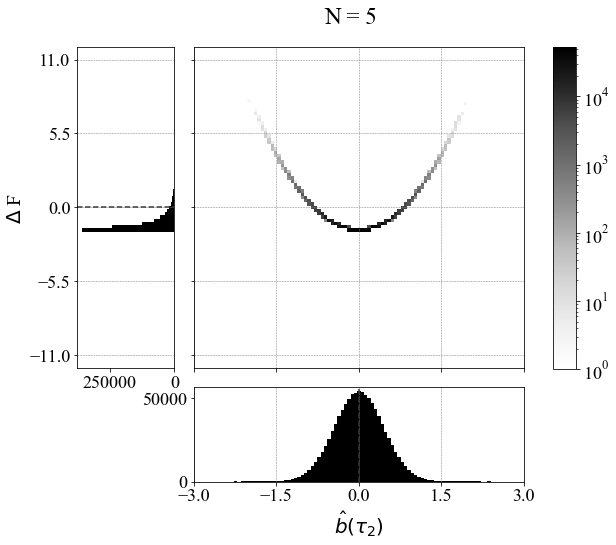}
 \caption{}
 \label{fig:image16}
 \end{subfigure}
 \hfill 
 \begin{subfigure}{0.32\textwidth}
 \includegraphics[width=\linewidth]{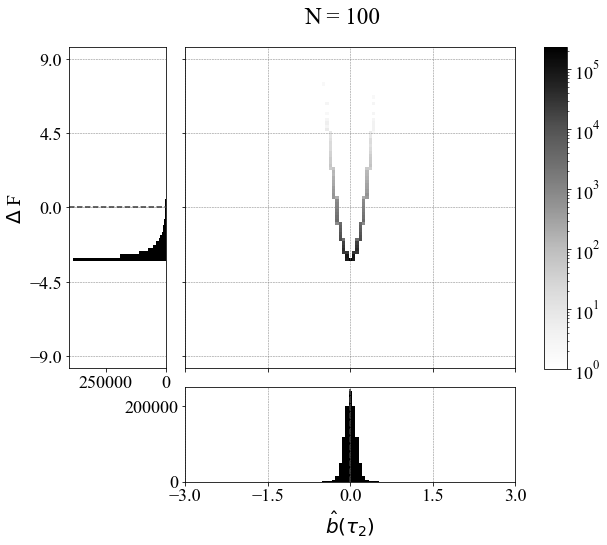}
 \caption{}
 \label{fig:image17}
 \end{subfigure}
 \hfill 
 \begin{subfigure}{0.32\textwidth}
 \includegraphics[width=\linewidth]{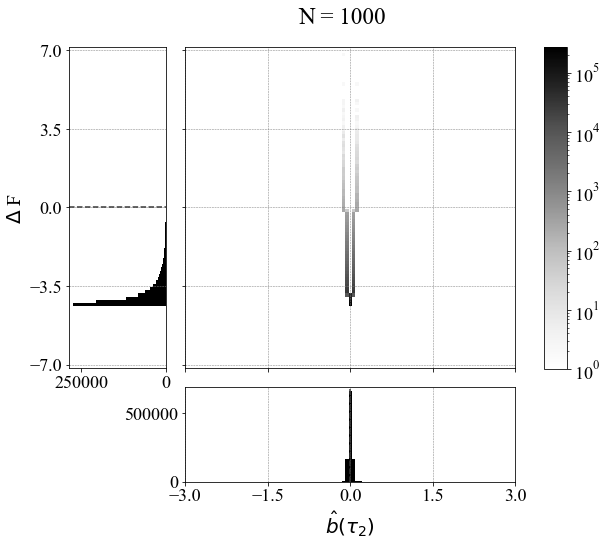}
 \caption{}
 \label{fig:image18}
 \end{subfigure}
\caption{Two-dimensional frequency distribution from 100,000 samples of the Bayesian free energy difference (Equation (\ref{prob_modelselection})) and \(\hat{b}(\tau_2)\). The model parameters are \(a_0 = 1.0, b_0 = 0.0\), and \(\sigma_0^2 = 1.0\), for data sizes \(N = 5, 100, 1000\). The vertical and horizontal axes represent the frequency distributions of the free energy difference and of \(\tau_2\), respectively.}

 \label{fig:delta_F_two_variable1D}
\end{figure}

Figure \ref{fig:delta_F_two_variable2D} shows the relationship between the differential free energy \(\Delta F (\tau_2)\) and the estimated parameter \(\hat{b}(\tau_2)\) for the case \(b_0=1.0\). Formula (\ref{delta_F_model_selection}) shows that \(\hat{b}(\tau_2)\) is distributed with the true parameter \(b_0\) in the center. Therefore, should Figures \ref{fig:delta_F_two_variable2D}(\(b_0=1.0\)), \(\hat{b}(\tau_2)^2\) is distributed with positive values, so that \(\Delta F(\tau_2)\) is concentrated at positive values. This enables us to understand why the two-variable model is mainly selected should Figures \ref{fig:delta_F_two_variable2D} in this analysis.

Next, Figure \ref{fig:delta_F_one_variable} shows the two-dimensional frequency distribution from 100,000 samples of the Bayesian free energy difference (Equation (\ref{prob_modelselection})) and \(\tau_2\). The model parameters are \(a_0 = 1.0\), \(b_0 = 0.0\), and \(\sigma_0^2 = 1.0\), for data sizes \(N = 5, 100, 1000\). The vertical and horizontal axes represent the frequency distributions of the free energy difference and of \(\tau_2\), respectively. Figures \ref{fig:delta_F_one_variable}(a)--(c) show that as the number of data points increases, the frequency of \(\Delta F > 0\) gradually decreases, but even at \(N=1000\), the occurrence of \(\Delta F > 0\) remains, indicating failures in model selection are still present. The minimum value of the distribution, as evident from Equation (\ref{prob_modelselection}), shifts negatively on a \(\mathrm{log}(N)\) scale. Therefore, when \(y=ax\) is the true model, the pace of improvement in model selection by increasing the number of data points is slower compared with when \(y=ax+b\) is the true model.

In the case shown in Figures \ref{fig:delta_F_two_variable1D}(\(b_0=0.0\)), the value of \(\hat{b}(\tau_2)^2\) is distributed around zero, so that the value of \(\Delta F(\tau_2)\) becomes negatively distributed due to the effect of other terms. This suggests that the one-variable model is selected should Figures \ref{fig:delta_F_two_variable1D}.

\begin{figure}[ht]
 \centering
 \begin{subfigure}{0.32\textwidth}
 \includegraphics[width=\linewidth]{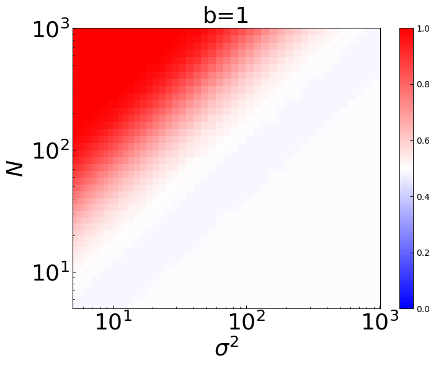}
 \caption{}
 \label{fig:image19}
 \end{subfigure}
 \hfill 
 \begin{subfigure}{0.32\textwidth}
 \includegraphics[width=\linewidth]{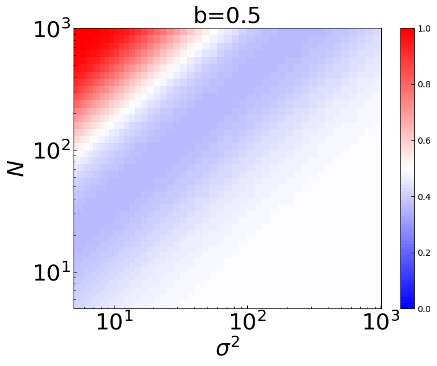}
 \caption{}
 \label{fig:image20}
 \end{subfigure}
 \hfill 
 \begin{subfigure}{0.32\textwidth}
 \includegraphics[width=\linewidth]{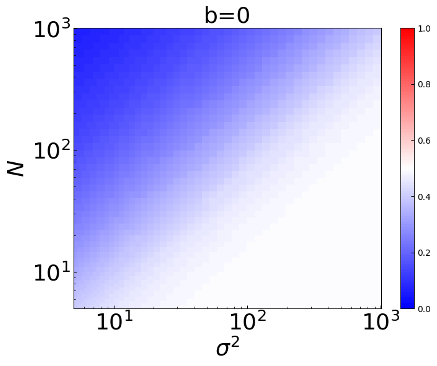}
 \caption{}
 \label{fig:image21}
 \end{subfigure}

\caption{Average selection probability of the two-variable model \(y = ax + b\) derived from the probability distribution of the difference in the Bayesian free energy with model parameters \(b_0 = 1.0, 0.5, 0.0\) (Equation (\ref{prob_modelselection})). The parameters are set as \(a_0 = 1.0, \sigma_0^2 = 1.0\), and the frequency distribution is shown as a two-dimensional histogram of the number of data points \(N\) and the data noise intensity \(\sigma^2\).}

 \label{fig:heatmap}
\end{figure}

Figure \ref{fig:heatmap} shows the probability of selecting the two-variable model \(y=ax+b\) on the basis of the Bayesian free energy difference (Equation (\ref{prob_modelselection})) for model parameters \(b_0 = 1.0, 0.5, 0.0\). Here, we set \(a_0 = 1.0\) and display the frequency distribution as a two-dimensional histogram from 100,000 samples across the dimensions of data number \(N\) and data noise intensity \(\sigma^2\). Figure \ref{fig:heatmap}(a) shows that model selection tends to fail along the diagonal line where \(N\) and \(\sigma^2\) have similar values. As the data number increases from this line, appropriate model selection gradually becomes possible. Conversely, as the data number decreases away from this diagonal line, discerning the correct model selection becomes challenging. This diagonal line, as shown in Figures \ref{fig:heatmap}(b) and (c), broadens as the value of \(b\) decreases, and at \(b=0.0\), the probability of selecting the model \(y=ax+b\) disappears.

\clearpage

\section{Bayesian Integration}
\label{sec:bayesianintegration}
In this section, we explore a framework for Bayesian inference of model parameters by integrating multiple sets of measurement data under varying conditions. The primary focus here is to demonstrate that the use of Bayesian free energy can determine whether integrated or independent analysis of multiple measurement datasets can be performed using a small number of variables, similar to the previous sections.

We specifically address the regression problem involving two sets of one-dimensional data: \(D_1 = \{(x_{i}^{(1)}, y_{i}^{(1)})\}_{i=1}^{N_1}\) with sample size \(N_1\), and \(D_2 = \{(x_{i}^{(2)}, y_{i}^{(2)})\}_{i=1}^{N_2}\) with sample size \(N_2\). The regression is formulated with a two-variable linear model as follows:

\begin{align}
 y_{i}^{(1)} &= a_0^{(1)} x_{i}^{(1)} + b_0^{(1)} + n_{i}^{(1)}, \label{bayes_int_1}\\
 y_{i}^{(2)} &= a_0^{(2)} x_{i}^{(2)} + b_0^{(2)} + n_{i}^{(2)}, \label{bayes_int_2}
\end{align}

where the noise terms \(n_{i}^{(1)}\) and \(n_{i}^{(2)}\) are assumed to follow normal distributions with mean zero and variances \((\sigma_{0}^{(1)})^2\) and \((\sigma_{0}^{(2)})^2\), respectively. This setup enables us to perform integrated analysis that considers different noise levels and relationships in the data from two distinct experimental conditions.

As in the previous section, the noise variances \((\sigma_{0}^{(1)})^2\) and \((\sigma_{0}^{(2)})^2\) are assumed to be known in this section. However, we will discuss the estimation of these noise variances for each model in Appendix B, and how these estimates affect the Bayesian inference process.

\subsection{Representation through Microscopic Variables in Bayesian Integration}
\subsubsection{Microscopic Notation of Mean Squared Error for Bayesian Integration}
In this case, we define the MSE as
\begin{align}
 E_m(a,b) &= \frac{1}{2N_m}\sum_{i=1}^{N_m}(y_{i}^{(m)}-ax_{i}^{(m)}-b)^2 \nonumber \\
 &= \frac{1}{2}\left(\bar{{x^{(m)}}^2}\left(a^{(m)}-\frac{\bar{x^{(m)}y^{(m)}}}{\bar{{x^{(m)}}^2}}\right)^2 + (b^{(m)}-\bar{y^{(m)}})^2 - \frac{\bar{x^{(m)}y^{(m)}}^2}{\bar{{x^{(m)}}^2}} - \bar{y^{(m)}}^2 + \bar{{y^{(m)}}^2}\right).
\end{align}
When each dataset has independent parameters \((a^{(1)}, b^{(1)}), (a^{(2)}, b^{(2)})\), the combined MSE after integration can be written as
\begin{eqnarray}
 E(a^{(1)}, b^{(1)}, a^{(2)}, b^{(2)}) = \sum_{m=1}^2 \frac{N_m}{{\sigma^{(m)}}_0^2} E_m(a^{(m)},b^{(m)}).
\end{eqnarray}
Since there are two two-variable linear regression models, we can complete the square independently for each model. Therefore, the expression for the total MSE is:
\begin{align}
 E(a^{(1)}, b^{(1)}, a^{(2)}, b^{(2)}) &= \frac{N_1}{2{\sigma^{(1)}}_0^2}\left(\bar{{x^{(1)}}^2}\left(a^{(1)}-\frac{\bar{x^{(1)}y^{(1)}}}{\bar{{x^{(1)}}^2}}\right)^2 + (b^{(1)}-\bar{y^{(1)}})^2 \right. \nonumber \\
 &\quad \left. - \frac{\bar{x^{(1)}y^{(1)}}^2}{\bar{{x^{(1)}}^2}} - \bar{y^{(1)}}^2 + \bar{{y^{(1)}}^2}\right) \nonumber \\
 &+ \frac{N_2}{2{\sigma^{(2)}}_0^2}\left(\bar{{x^{(2)}}^2}\left(a^{(2)}-\frac{\bar{x^{(2)}y^{(2)}}}{\bar{{x^{(2)}}^2}}\right)^2 + (b^{(2)}-\bar{y^{(2)}})^2 \right. \nonumber \\
 &\quad \left. - \frac{\bar{x^{(2)}y^{(2)}}^2}{\bar{{x^{(2)}}^2}} - \bar{y^{(2)}}^2 + \bar{{y^{(2)}}^2}\right).
\end{align}
This matches the results that would be obtained by treating the datasets \(D_1\) and \(D_2\) independently with linear regression models. If we infer a common model parameter \(a, b\) from each dataset, then the expression for the MSE becomes:
\begin{align}
 E(a, b) &= \sum_{m=1}^2 \frac{N_m}{{\sigma^{(m)}}_0^2} E_m(a, b) \nonumber \\
 &= \frac{N_1}{2{\sigma^{(1)}}_0^2}\left(\bar{{x^{(1)}}^2}\left(a-\frac{\bar{x^{(1)}y^{(1)}}}{\bar{{x^{(1)}}^2}}\right)^2 + (b-\bar{y^{(1)}})^2 - \frac{\bar{x^{(1)}y^{(1)}}^2}{\bar{{x^{(1)}}^2}} - \bar{y^{(1)}}^2 + \bar{{y^{(1)}}^2}\right) \nonumber \\
 &+ \frac{N_2}{2{\sigma^{(2)}}_0^2}\left(\bar{{x^{(2)}}^2}\left(a-\frac{\bar{x^{(2)}y^{(2)}}}{\bar{{x^{(2)}}^2}}\right)^2 + (b-\bar{y^{(2)}})^2 - \frac{\bar{x^{(2)}y^{(2)}}^2}{\bar{{x^{(2)}}^2}} - \bar{y^{(2)}}^2 + \bar{{y^{(2)}}^2}\right)
\end{align}
Further transformations will be applied to \(a\) and \(b\). Let \(\beta^{(1)} = \frac{N_1}{{\sigma^{(1)}}^2}\), \(\beta^{(2)} = \frac{N_2}{{\sigma^{(2)}}^2}\), \(\beta^{(1)}_0 = \frac{N_1}{{\sigma^{(1)}}_0^2}\), \(\beta^{(2)}_0 = \frac{N_2}{{\sigma^{(2)}}_0^2}\), \(\hat{a}^{(m)}=\frac{\bar{x^{(m)}y^{(m)}}}{\bar{{x^{(m)}}^2}}\), and \(b^{(m)}=\bar{y^{(m)}}\). Then, the error function can be written as:
\begin{align}
 E(a, b) &= \frac{\left(\beta^{(1)}\bar{{x^{(1)}}^2}+\beta^{(2)}\bar{{x^{(2)}}^2}\right)}{2} \left(a - \frac{\left(\beta^{(1)}\bar{{x^{(1)}}^2}\hat{a}^{(1)}+\beta^{(2)}\bar{{x^{(2)}}^2}\hat{a}^{(2)}\right)}{\left(\beta^{(1)}\bar{{x^{(1)}}^2}+\beta^{(2)}\bar{{x^{(2)}}^2}\right)}\right)^2 \nonumber \\
 &\quad + \frac{\beta^{(1)}+\beta^{(2)}}{2}\left(b - \frac{\left(\beta^{(1)}\hat{b}^{(1)}+\beta^{(2)}\hat{b}^{(2)}\right)}{\beta^{(1)}+\beta^{(2)}}\right)^2 \nonumber \\
 &\quad + \frac{1}{2} \left(\frac{(\beta^{(1)}\bar{{x^{(1)}}^2})(\beta^{(2)}\bar{{x^{(2)}}^2})}{\beta^{(1)}\bar{{x^{(1)}}^2}+\beta^{(2)}\bar{{x^{(2)}}^2}}(\hat{a}^{(1)}-\hat{a}^{(2)})^2 + \frac{\beta^{(1)}\beta^{(2)}}{(\beta^{(1)}+\beta^{(2)})}(\hat{b}^{(1)}-\hat{b}^{(2)})^2 \right. \nonumber \\
 &\quad \left. - \frac{\beta_0^{(1)}}{\beta^{(1)}}{\tau_1^{(1)}}^2 - \frac{\beta_0^{(1)}}{\beta^{(1)}}{\tau_2^{(1)}}^2 + \frac{\bar{{n^{(1)}}^2}}{\beta_1} -\frac{\beta_0^{(2)}}{\beta^{(2)}}{\tau_1^{(2)}}^2 - \frac{\beta_0^{(2)}}{\beta^{(2)}}{\tau_2^{(2)}}^2 + \frac{\bar{{n^{(2)}}^2}}{\beta^{(2)}}\right).
\end{align}
Let us define the integrated errors for parameters \(a\) and \(b\) as follows:
\begin{align}
 \mathcal{E}^{\mathrm{int}}_a(a) &= \frac{\left(\beta^{(1)}\bar{{x^{(1)}}^2}+\beta^{(2)}\bar{{x^{(2)}}^2}\right)}{2} \left(a - \frac{\left(\beta^{(1)}\bar{{x^{(1)}}^2}\hat{a}^{(1)}+\beta^{(2)}\bar{{x^{(2)}}^2}\hat{a}^{(2)}\right)}{\left(\beta^{(1)}\bar{{x^{(1)}}^2}+\beta^{(2)}\bar{{x^{(2)}}^2}\right)}\right)^2 \\
 \mathcal{E}^{\mathrm{int}}_b(b) &= \frac{\beta^{(1)}+\beta^{(2)}}{2}\left(b - \frac{\left(\beta^{(1)}\hat{b}^{(1)}+\beta^{(2)}\hat{b}^{(2)}\right)}{\beta^{(1)}+\beta^{(2)}}\right)^2
\end{align}
The optimal parameters \(\hat{a}\) and \(\hat{b}\) are given by:
\begin{eqnarray}
 \hat{a} &=& \frac{\left(\beta^{(1)}\bar{{x^{(1)}}^2}\hat{a}^{(1)}+\beta^{(2)}\bar{{x^{(2)}}^2}\hat{a}^{(2)}\right)}{\left(\beta^{(1)}\bar{{x^{(1)}}^2}+\beta^{(2)}\bar{{x^{(2)}}^2}\right)}\\
 \hat{b} &=& \frac{\left(\beta^{(1)}\hat{b}^{(1)}+\beta^{(2)}\hat{b}^{(2)}\right)}{\beta^{(1)}+\beta^{(2)}}
\end{eqnarray}
The residual error can be expressed as:
\begin{align}
 E(\hat{a},\hat{b}) &= \frac{1}{2} \left(\frac{(\beta^{(1)}\bar{{x^{(1)}}^2})(\beta^{(2)}\bar{{x^{(2)}}^2})}{\beta^{(1)}\bar{{x^{(1)}}^2}+\beta^{(2)}\bar{{x^{(2)}}^2}}(\hat{a}^{(1)}-\hat{a}^{(2)})^2 \right. \nonumber \\
 &\quad + \frac{\beta^{(1)}\beta^{(2)}}{(\beta^{(1)}+\beta^{(2)})}(\hat{b}^{(1)}-\hat{b}^{(2)})^2 \nonumber \\
 &\quad - \frac{\beta_0^{(1)}}{\beta^{(1)}}{\tau_1^{(1)}}^2 - \frac{\beta_0^{(1)}}{\beta^{(1)}}{\tau_2^{(1)}}^2 + \frac{\bar{{n^{(1)}}^2}}{\beta_1} \nonumber \\
 &\quad \left. -\frac{\beta_0^{(2)}}{\beta^{(2)}}{\tau_1^{(2)}}^2 - \frac{\beta_0^{(2)}}{\beta^{(2)}}{\tau_2^{(2)}}^2 + \frac{\bar{{n^{(2)}}^2}}{\beta^{(2)}}\right)
\end{align}

\subsubsection{Bayesian Inference in Bayesian Integration}
Given the input variables and model parameters, the conditional probability of the output can be expressed as:
\begin{equation}
 p(y_{i}^{(m)}|a^{(m)},b^{(m)}) = \frac{1}{\sqrt{2\pi(\sigma_{0}^{(m)})^2}} \exp\left[-\frac{(y_{i}^{(m)}-a^{(m)}x_{i}^{(m)}-b^{(m)})^2}{2(\sigma^{(m)}_{0})^2}\right]
\end{equation}
Therefore, when we have independent parameters \(a^{(1)}, a^{(2)}, b^{(1)}, b^{(2)}\), the joint conditional probability of all output data \(Y = \{D_1, D_2\}\) can be expressed as:
\begin{align}
 \hspace{-1cm}p(Y|a^{(1)},a^{(2)},b^{(1)},b^{(2)}) &= \prod_{m=1}^2\prod_{i=1}^{N_m} p(y_{i}^{(m)}|a^{(m)}, b^{(m)}) \nonumber \\
 &= \left(\frac{1}{\sqrt{2\pi(\sigma^{(1)}_{0})^2}}\right)^{N_1}
 \left(\frac{1}{\sqrt{2\pi(\sigma^{(2)}_{0})^2}}\right)^{N_2}
 \exp\left(-\frac{N_1}{(\sigma^{(1)}_{0})^2}E_1(a^{(1)},b^{(1)})
 -\frac{N_2}{(\sigma^{(2)}_{0})^2}E_2(a^{(2)},b^{(2)})\right)
\end{align}
The posterior distribution can be independently analyzed for each model parameter \(a^{(1)}, a^{(2)}, b^{(1)}, b^{(2)}\), and can be computed as:
\begin{align}
p(a^{(1)},a^{(2)},b^{(1)},b^{(2)}|Y) &= \prod_{m=1}^2 \frac{2N_m\sqrt{\bar{{x^{(m)}}^2}}}{{\sigma^{(m)}}^2_0\pi}\exp\left\{-\frac{N_m}{{\sigma^{(m)}}^2_0}\left[\mathcal{E}_a(a^{(m)}) + \mathcal{E}_b(b^{(m)})\right]\right\} \nonumber\\ 
 &\quad \times \left\{\Theta(a^{(m)}+\xi_{a^{(m)}})-\Theta(a^{(m)}-\xi_{a^{(m)}})\right\} \nonumber\\
 &\quad \times \left\{\Theta(b^{(m)}+\xi_{b^{(m)}})-\Theta(b^{(m)}-\xi_{b^{(m)}})\right\} \nonumber\\
 &\quad \times \left[\mathrm{erfc} \left(\sqrt{\frac{N_m\bar{{x^{(m)}}^2}}{2{\sigma^{(m)}}^2_0}}\left(-\xi_{a^{(m)}}-\frac{\bar{x^{(m)}y^{(m)}}}{\bar{{x^{(m)}}^2}}\right)\right) \right. \nonumber\\
 &\quad \quad \left. -\mathrm{erfc} \left(\sqrt{\frac{N_m\bar{{x^{(m)}}^2}}{2{\sigma^{(m)}}^2_0}}\left(\xi_{a^{(m)}}-\frac{\bar{x^{(m)}y^{(m)}}}{\bar{{x^{(m)}}^2}}\right)\right) \right]^{-1} \nonumber\\
 &\quad \times \left[\mathrm{erfc} \left(\sqrt{\frac{N_m}{2{\sigma^{(m)}}^2_0}}\left(-\xi_{b^{(m)}}-\bar{y^{(m)}}\right)\right) \right. \nonumber\\
 &\quad \quad \left. -\mathrm{erfc} \left(\sqrt{\frac{N_m}{2{\sigma^{(m)}}^2_0}}\left(\xi_{b^{(m)}}-\bar{y^{(m)}}\right)\right) \right]^{-1}
\end{align}

When the range of the prior distribution is sufficiently large, the posterior distributions of each model parameter can be described as Gaussian distributions centered around \(\hat{a}^{(1)}, \hat{a}^{(2)}, \hat{b}^{(1)}, \hat{b}^{(2)}\). On the other hand, when the estimated parameters from both datasets share common model parameters \(a, b\), the joint conditional probability is given by:
\begin{align}
 p(Y|a,b) &= \prod_{m=1}^2\left(\frac{1}{\sqrt{2\pi(\sigma^{(m)}_{0})^2}}\right)^{N_m} \exp\left(-\frac{N_m}{(\sigma^{(m)}_{0})^2}E_m(a,b)\right) \nonumber \\
 &= \left(\frac{1}{\sqrt{2\pi(\sigma^{(1)}_{0})^2}}\right)^{N_1} \left(\frac{1}{\sqrt{2\pi(\sigma^{(2)}_{0})^2}}\right)^{N_2} \exp\left(- \sum_{m=1}^2\frac{N_m}{(\sigma^{(m)}_{0})^2}E_m(a,b)\right)
\end{align}
The posterior distribution can then be expressed as:
\begin{eqnarray}
 p(a,b|Y) &=& \frac{2}{\pi}\sqrt{(\beta^{(1)}\bar{{x^{(1)}}^2}+\beta^{(2)}\bar{{x^{(2)}}^2})(\beta^{(1)}+\beta^{(2)})}\exp(-\mathcal{E}^\mathrm{int}_a(a)-\mathcal{E}^\mathrm{int}_b(b))\nonumber\\
 &\times& \left\{\Theta(a+\xi_a)-\Theta(a-\xi_a)\right\}
 \left\{\Theta(b+\xi_b)-\Theta(b-\xi_b)\right\}\nonumber\\
 &\times& \left[\mathrm{erfc}\left(\sqrt{\frac{\beta^{(1)}\bar{{x^{(1)}}^2}+\beta^{(2)}\bar{{x^{(2)}}^2}}{2}}(-\xi_a-\hat{a})\right)\right.\nonumber\\
 &-&\left.\mathrm{erfc}\left(\sqrt{\frac{\beta^{(1)}\bar{{x^{(1)}}^2}+\beta^{(2)}\bar{{x^{(2)}}^2}}{2}}(\xi_a-\hat{a})\right)\right]^{-1}\nonumber\\
 &\times& \left[\mathrm{erfc}\left(\sqrt{\frac{\beta^{(1)}+\beta^{(2)}}{2}}(-\xi_b-\hat{b})\right)\right.\nonumber\\
 &-&\left.\mathrm{erfc}\left(\sqrt{\frac{\beta^{(1)}+\beta^{(2)}}{2}}(\xi_b-\hat{b})\right)\right]^{-1}. \label{intposterior}
\end{eqnarray}

Here, we derive the Bayesian free energy from the results of the previous section, which is used as a criterion for model selection. The Bayesian free energy is the negative logarithm of the marginal likelihood. Assuming a uniform prior distribution and that model parameters are independent for each dataset, the Bayesian free energy can be expressed as:
\begin{eqnarray}
 \hspace{-1cm}&&F^{(m)}(Y) = \frac{N_m}{2}\ln(2\pi{\sigma^{(m)}}^2_0) - \ln\left(\frac{{\sigma^{(m)}}^2_0\pi}{2N_m}\right) + \frac{1}{2}\ln\left(\bar{{x^{(m)}}^2}\right) + \ln(2\xi^{(m)}_a) + \ln(2\xi^{(m)}_b)+\frac{N_m}{{\sigma^{(m)}}^2_0}E_m(\hat{a}^{(m)},\hat{b}^{(m)})\nonumber\\
 \hspace{-1cm}&&\hspace{3cm}-\ln \left[\mathrm{erfc} \left(\sqrt{\frac{N_m\bar{{x^{(m)}}^2}}{2{\sigma^{(m)}}^2_0}}\left(-\xi^{(m)}_a-\hat{a}^{(m)}\right)\right)-\mathrm{erfc} \left(\sqrt{\frac{N_m\bar{{x^{(m)}}^2}}{2{\sigma^{(m)}}^2_0}}\left(\xi^{(m)}_a-\hat{a}^{(m)}\right)\right) \right]\nonumber\\
 \hspace{-1cm}&&\hspace{3cm}-\ln \left[\mathrm{erfc} \left(\sqrt{\frac{N_m}{2{\sigma^{(m)}}^2_0}}\left(-\xi^{(m)}_b-\hat{b}^{(m)}\right)\right)-\mathrm{erfc} \left(\sqrt{\frac{N_m}{2{\sigma^{(m)}}^2_0}}\left(\xi^{(m)}_b-\hat{b}^{(m)}\right)\right) \right]\\
 \hspace{-2cm}&&F(Y) = \sum_{m=1}^2 F^{(m)}(Y).
\end{eqnarray}
On the other hand, if this model has common model parameters, the Bayesian free energy is obtained by taking the negative logarithm of the marginal likelihood:
\begin{eqnarray}
 F(Y) &=& \frac{N_1}{2}\ln 2\pi (\sigma_0^{(1)})^2 + \frac{N_2}{2}\ln 2\pi (\sigma_0^{(2)})^2 + \ln 2\xi_a + \ln 2\xi_b + E(\hat{a},\hat{b})\nonumber\\
 &+&\frac{1}{2}\ln \frac{2(\beta^{(1)}\bar{{x^{(1)}}^2}+\beta^{(2)}\bar{{x^{(2)}}^2})}{\pi}+\frac{1}{2}\ln \frac{2(\beta^{(1)}+\beta^{(2)})}{\pi}\nonumber\\
 &-&\ln \left[\mathrm{erfc}\left(\sqrt{\frac{\beta^{(1)}\bar{{x^{(1)}}^2}+\beta^{(2)}\bar{{x^{(2)}}^2}}{2}}(-\xi_a-\hat{a})\right)-\mathrm{erfc}\left(\sqrt{\frac{\beta^{(1)}\bar{{x^{(1)}}^2}+\beta^{(2)}\bar{{x^{(2)}}^2}}{2}}(\xi_a-\hat{a})\right)\right]\nonumber\\
 &-& \ln \left[\mathrm{erfc}\left(\sqrt{\frac{\beta^{(1)}+\beta^{(2)}}{2}}(-\xi_b-\hat{b})\right)-\mathrm{erfc}\left(\sqrt{\frac{\beta^{(1)}+\beta^{(2)}}{2}}(\xi_b-\hat{b})\right)\right]\label{intFreeEstrict}.
\end{eqnarray}

\subsection{Representation through Mesoscopic Variables in Bayesian Integration}
Up to this point, each statistical measure has been handled as an empirical average. In this section, similar to the previous section, we introduce mesoscopic variables to theoretically manage these statistical measures.

\subsubsection{Residual Error with Mesoscopic Variables in Bayesian Integration}
In the previous sections, the residual error was derived as a probability variable dependent on the set of random variables \(\{n^{(m)}_i\}_{i=1}^{N_m}\). In this section, we discuss the probability distribution of the residual error \(E(\hat{a},\hat{b})\) and demonstrate that it follows a chi-squared distribution. The expression for the residual error is given by:
\begin{align}
 E(\hat{a},\hat{b}) &= \frac{1}{2} \left(\frac{(\beta^{(1)}\bar{{x^{(1)}}^2})(\beta^{(2)}\bar{{x^{(2)}}^2})}{\beta^{(1)}\bar{{x^{(1)}}^2}+\beta^{(2)}\bar{{x^{(2)}}^2}}(\hat{a}^{(1)}-\hat{a}^{(2)})^2 + \frac{\beta^{(1)}\beta^{(2)}}{(\beta^{(1)}+\beta^{(2)})}(\hat{b}^{(1)}-\hat{b}^{(2)})^2 \right.\nonumber\\
 &- \left. \frac{\beta^{(1)}}{\beta_0^{(1)}}{\tau_1^{(1)}}^2 - \frac{\beta^{(1)}}{\beta_0^{(1)}}{\tau_2^{(1)}}^2 + \beta^{(1)}\bar{{n^{(1)}}^2} -\frac{\beta^{(2)}}{\beta_0^{(2)}}{\tau_1^{(2)}}^2 - \frac{\beta^{(2)}}{\beta_0^{(2)}}{\tau_2^{(2)}}^2 + \beta^{(2)}\bar{{n^{(2)}}^2}\right).
\end{align}
In this case,
\begin{eqnarray}
 E_m(\hat{a}^{(m)},\hat{b}^{(m)}) &=& \frac{1}{2}\left(-\frac{\bar{x^{(m)}n^{(m)}}^2}{\bar{{x^{(m)}}^2}}-\bar{n^{(m)}}^2+\bar{{n^{(m)}}^2}\right)
\end{eqnarray}
is established. From the content of the previous sections,
\begin{eqnarray}
 E_m(\hat{a}^{(m)},\hat{b}^{(m)}) &=& \frac{{\sigma_0^{(m)}}^2}{2N}\upsilon^{(m)} = \frac{1}{2\beta_0^{(m)}}\upsilon^{(m)}\\
 \frac{\bar{x^{(m)}n^{(m)}}^2}{\bar{{x^{(m)}}^2}} &=& \frac{{\sigma_0^{(m)}}^2}{N}{\tau_1^{(m)}}^2 = \frac{{\tau_1^{(m)}}^2}{\beta_0^{(m)}} \\
 \bar{n^{(m)}}^2 &=& \frac{{\sigma_0^{(m)}}^2}{N}{\tau_2^{(m)}}^2 = \frac{{\tau_2^{(m)}}^2}{\beta_0^{(m)}}
\end{eqnarray}
can be expressed by introducing mesoscopic variables. Thus, \(\hat{a}^{(m)}\) and \(\hat{b}^{(m)}\) can be expressed similarly to the previous section as:
\begin{eqnarray}
 \hat{a}^{(m)}(\tau_1^{(m)}) &=& a_0 + \sqrt{\frac{1}{{\bar{{x^{(m)}}^2}}\beta_0^{(m)}}} \tau_1^{(m)}\\
 \hat{b}^{(m)}(\tau_2^{(m)}) &=& b_0 + \sqrt{\frac{1}{\beta_0^{(m)}}} \tau_2^{(m)}
\end{eqnarray}
Hence, the residual error in Bayesian integration can be expressed as:
\begin{align}
 E(\hat{a},\hat{b}) &= \frac{1}{2} \left(\frac{(\beta^{(1)}\bar{{x^{(1)}}^2})(\beta^{(2)}\bar{{x^{(2)}}^2})}{\beta^{(1)}\bar{{x^{(1)}}^2}+\beta^{(2)}\bar{{x^{(2)}}^2}}(\hat{a}^{(1)}(\tau_1^{(1)})-\hat{a}^{(2)}(\tau_1^{(2)}))^2 + \frac{\beta^{(1)}\beta^{(2)}}{(\beta^{(1)}+\beta^{(2)})}(\hat{b}^{(1)}(\tau_2^{(1)})-\hat{b}^{(2)}(\tau_2^{(2)}))^2 \right)\nonumber\\
 &+ \beta^{(1)}E_1(\hat{a}^{(1)},\hat{b}^{(1)})+\beta^{(2)}E_2(\hat{a}^{(2)},\hat{b}^{(2)})\\
 &= \frac{1}{2} \left(\frac{(\beta^{(1)}\bar{{x^{(1)}}^2})(\beta^{(2)}\bar{{x^{(2)}}^2})}{\beta^{(1)}\bar{{x^{(1)}}^2}+\beta^{(2)}\bar{{x^{(2)}}^2}}(\hat{a}^{(1)}(\tau_1^{(1)})-\hat{a}^{(2)}(\tau_1^{(2)}))^2 + \frac{\beta^{(1)}\beta^{(2)}}{(\beta^{(1)}+\beta^{(2)})}(\hat{b}^{(1)}(\tau_2^{(1)})-\hat{b}^{(2)}(\tau_2^{(2)}))^2 \right)\nonumber\\
 &+ \frac{\beta^{(1)}}{2\beta_0^{(1)}}\upsilon^{(1)}+\frac{\beta^{(2)}}{2\beta_0^{(2)}}\upsilon^{(2)}
\end{align}
and can be described by six mesoscopic variables \(\tau_1^{(1)}, \tau_1^{(2)}, \tau_2^{(1)}, \tau_2^{(2)}, \upsilon^{(1)}, \upsilon^{(2)}\).

\subsubsection{Posterior Distribution with Mesoscopic Variables in Bayesian Integration}
In this section, we use the mesoscopic variables introduced in the previous section to reformulate the posterior distribution. The error functions can be expressed using mesoscopic variables as:
\begin{eqnarray}
 \mathcal{E}^{\mathrm{int}}_a(a,\tau_1^{(1)},\tau_1^{(2)}) &=& \frac{\left(\beta^{(1)}\bar{{x^{(1)}}^2}+\beta^{(2)}\bar{{x^{(2)}}^2}\right)}{2} \left(a - \hat{a}(\tau_1^{(1)},\tau_1^{(2)}) \right)^2\\
 \mathcal{E}^{\mathrm{int}}_b(b,\tau_2^{(1)},\tau_2^{(2)}) &=& \frac{\beta^{(1)}+\beta^{(2)}}{2}\left(b - \hat{b}(\tau_2^{(1)},\tau_2^{(2)})\right)^2\\
 \hat{a}(\tau_1^{(1)},\tau_1^{(2)}) &=& \frac{\left(\beta^{(1)}\bar{{x^{(1)}}^2}\hat{a}^{(1)}(\tau_1^{(1)})+\beta^{(2)}\bar{{x^{(2)}}^2}\hat{a}^{(2)}(\tau_1^{(2)})\right)}{\left(\beta^{(1)}\bar{{x^{(1)}}^2}+\beta^{(2)}\bar{{x^{(2)}}^2}\right)}\\
 \hat{b}(\tau_2^{(1)},\tau_2^{(2)}) &=& \frac{\left(\beta^{(1)}\hat{b}^{(1)}(\tau_2^{(1)})+\beta^{(2)}\hat{b}^{(2)}(\tau_2^{(2)})\right)}{\beta^{(1)}+\beta^{(2)}}
\end{eqnarray}
Therefore, the posterior distribution as per Equation (\ref{intposterior}) can be described as:
\begin{eqnarray}
 p(a,b|Y) &=& \frac{2}{\pi}\sqrt{(\beta^{(1)}\bar{{x^{(1)}}^2}+\beta^{(2)}\bar{{x^{(2)}}^2})(\beta^{(1)}+\beta^{(2)})}\exp(-\mathcal{E}^\mathrm{int}_a(a,\tau_1^{(1)},\tau_1^{(2)})-\mathcal{E}^\mathrm{int}_b(b,\tau_2^{(1)},\tau_2^{(2)}))\nonumber\\
 &\times& \left\{\Theta(a+\xi_a)-\Theta(a-\xi_a)\right\}
 \left\{\Theta(b+\xi_b)-\Theta(b-\xi_b)\right\}\nonumber\\
 &\times& \left[\mathrm{erfc}\left(\sqrt{\frac{\beta^{(1)}\bar{{x^{(1)}}^2}+\beta^{(2)}\bar{{x^{(2)}}^2}}{2}}(-\xi_a-\hat{a}(\tau_1^{(1)},\tau_1^{(2)}))\right)\right.\nonumber\\
 &-&\left.\mathrm{erfc}\left(\sqrt{\frac{\beta^{(1)}\bar{{x^{(1)}}^2}+\beta^{(2)}\bar{{x^{(2)}}^2}}{2}}(\xi_a-\hat{a}(\tau_1^{(1)},\tau_1^{(2)}))\right)\right]^{-1}\nonumber\\
 &\times& \left[\mathrm{erfc}\left(\sqrt{\frac{\beta^{(1)}+\beta^{(2)}}{2}}(-\xi_b-\hat{b}(\tau_2^{(1)},\tau_2^{(2)}))\right)\right.\nonumber\\
 &-&\left.\mathrm{erfc}\left(\sqrt{\frac{\beta^{(1)}+\beta^{(2)}}{2}}(\xi_b-\hat{b}(\tau_2^{(1)},\tau_2^{(2)}))\right)\right]^{-1}
\end{eqnarray}

Thus, the posterior distribution is determined solely by the four stochastic variables \(\tau_1^{(1)},\tau_1^{(2)},\tau_2^{(1)},\tau_2^{(2)}\). Also, given the model, the distributions of the model parameters \(a, b\), \(p_{\mathrm{m}}(a), p_{\mathrm{m}}(b)\) can be described as:
\begin{align}
 p_{\mathrm{m}}(a) &= \int \mathrm{d}\tau_1^{(1)}\mathrm{d}\tau_1^{(2)} \delta(a-\hat{a}(\tau_1^{(1)},\tau_1^{(2)}))p(\tau_1^{(1)})p(\tau_1^{(2)})\\
 &\propto \exp\left(-\frac{1}{2(\beta^{(1)}\bar{{x^{(1)}}^2}+\beta^{(2)}\bar{{x^{(2)}}^2})}\left(\sqrt{\frac{\beta^{(1)}}{\bar{{x^{(1)}}^2}}}+\sqrt{\frac{\beta^{(2)}}{\bar{{x^{(2)}}^2}}}\right)^2 (a-a_0)^2\right)\\
 p_{\mathrm{m}}(b) &= \int \mathrm{d}\tau_2^{(1)}\mathrm{d}\tau_2^{(2)} \delta(b-\hat{b}(\tau_2^{(1)},\tau_2^{(2)}))p(\tau_2^{(1)})p(\tau_2^{(2)})\\
 &\propto \exp\left(-\frac{1}{2(\beta^{(1)}+\beta^{(2)})}\left(\sqrt{\beta^{(1)}}+\sqrt{\beta^{(2)}}\right)^2 (b-b_0)^2\right)
\end{align}

Here, we reformulate the Bayesian free energy using mesoscopic variables. From Equation (\ref{intFreeEstrict}),
\begin{align}
 F(Y) &= \frac{N_1}{2}\ln (2\pi (\sigma^{(1)})^2) + \frac{N_2}{2}\ln (2\pi (\sigma^{(2)})^2) + \ln (2\xi_a) + \ln (2\xi_b) \nonumber\\
 &\quad + E(\hat{a}(\tau_1^{(1)},\tau_1^{(2)}), \hat{b}(\tau_2^{(1)},\tau_2^{(2)}), \upsilon^{(1)}, \upsilon^{(2)})\nonumber\\
 &\quad + \frac{1}{2}\ln \left(\frac{2(\beta^{(1)}\bar{{x^{(1)}}^2}+\beta^{(2)}\bar{{x^{(2)}}^2})}{\pi}\right) + \frac{1}{2}\ln \left(\frac{2(\beta^{(1)}+\beta^{(2)})}{\pi}\right)\nonumber\\
 &\quad - \ln \left[\mathrm{erfc}\left(\sqrt{\frac{\beta^{(1)}\bar{{x^{(1)}}^2}+\beta^{(2)}\bar{{x^{(2)}}^2}}{2}}(-\xi_a - \hat{a}(\tau_1^{(1)}, \tau_1^{(2)}))\right)\right. \nonumber\\
 &\quad \quad \left. - \mathrm{erfc}\left(\sqrt{\frac{\beta^{(1)}\bar{{x^{(1)}}^2}+\beta^{(2)}\bar{{x^{(2)}}^2}}{2}}(\xi_a - \hat{a}(\tau_1^{(1)}, \tau_1^{(2)}))\right)\right]\nonumber\\
 &\quad - \ln \left[\mathrm{erfc}\left(\sqrt{\frac{\beta^{(1)}+\beta^{(2)}}{2}}(-\xi_b - \hat{b}(\tau_2^{(1)}, \tau_2^{(2)}))\right)\right. \nonumber\\
 &\quad \quad \left. - \mathrm{erfc}\left(\sqrt{\frac{\beta^{(1)}+\beta^{(2)}}{2}}(\xi_b - \hat{b}(\tau_2^{(1)}, \tau_2^{(2)}))\right)\right]
\end{align}
is rewritten. Therefore, the Bayesian free energy is determined by six stochastic variables \(\tau_1^{(1)},\tau_1^{(2)},\tau_2^{(1)},\tau_2^{(2)},\upsilon^{(1)},\upsilon^{(2)}\), and can be expressed as \(F(Y)=F(\tau_1^{(1)},\tau_1^{(2)},\tau_2^{(1)},\tau_2^{(2)},\upsilon^{(1)},\upsilon^{(2)})\). Consequently, the probability distribution of the Bayesian free energy is
\begin{align}
 p(F) = \int & \mathrm{d}\tau_1^{(1)} \mathrm{d}\tau_1^{(2)} \mathrm{d}\tau_2^{(1)} \mathrm{d}\tau_2^{(2)} \mathrm{d}\upsilon^{(1)} \mathrm{d}\upsilon^{(2)} \nonumber \\
 &\delta\left(F - F(\tau_1^{(1)}, \tau_1^{(2)}, \tau_2^{(1)}, \tau_2^{(2)}, \upsilon^{(1)}, \upsilon^{(2)})\right) \nonumber \\
 &p(\tau_1^{(1)}) p(\tau_1^{(2)}) p(\tau_2^{(1)}) p(\tau_2^{(2)}) p(\upsilon^{(1)}) p(\upsilon^{(2)})
\end{align}

\subsubsection{Model Selection in Bayesian Integration}
In this section, we compare the Bayesian free energy of the linear regression model by Bayesian integration with that of the independent analysis linear regression model, to perform model selection. The Bayesian free energy for independent analysis is given by:
\begin{eqnarray}
 &&\hspace{-2cm}F^{\mathrm{iso}}(\tau_1^{(1)}, \tau_1^{(2)}, \tau_2^{(1)}, \tau_2^{(2)}, \upsilon^{(1)}, \upsilon^{(2)}) \\
 &=& \sum_{m=1}^2 \frac{N_m}{2}\ln(2\pi{\sigma^{(m)}}^2) - \ln\left(\frac{{\sigma^{(m)}}^2\pi}{2N_m}\right) + \frac{1}{2}\ln\left(\bar{{x^{(m)}}^2}\right) + \ln(2\xi^{(m)}_a) + \ln(2\xi^{(m)}_b)+\frac{N_m}{{\sigma^{(m)}}^2}E_m(\hat{a}^{(m)},\hat{b}^{(m)})\nonumber\\
 \hspace{-4cm}&&\hspace{1cm}-\ln \left[\mathrm{erfc} \left(\sqrt{\frac{N_m\bar{{x^{(m)}}^2}}{2{\sigma^{(m)}}^2}}\left(-\xi^{(m)}_a-\hat{a}^{(m)}\right)\right)-\mathrm{erfc} \left(\sqrt{\frac{N_m\bar{{x^{(m)}}^2}}{2{\sigma^{(m)}}^2}}\left(\xi^{(m)}_a-\hat{a}^{(m)}\right)\right) \right]\nonumber\\
 \hspace{-4cm}&&\hspace{1cm}-\ln \left[\mathrm{erfc} \left(\sqrt{\frac{N_m}{2{\sigma^{(m)}}^2}}\left(-\xi^{(m)}_b-\hat{b}^{(m)}\right)\right)-\mathrm{erfc} \left(\sqrt{\frac{N_m}{2{\sigma^{(m)}}^2}}\left(\xi^{(m)}_b-\hat{b}^{(m)}\right)\right) \right]
\end{eqnarray}
and the Bayesian free energy through integration is:
\begin{align}
 \hspace{-2cm}&F^{\mathrm{int}}(\tau_1^{(1)}, \tau_1^{(2)}, \tau_2^{(1)}, \tau_2^{(2)}, \upsilon^{(1)}, \upsilon^{(2)}) = \frac{N_1}{2}\ln 2\pi (\sigma^{(1)})^2 + \frac{N_2}{2}\ln 2\pi (\sigma^{(2)})^2 + \ln 2\xi_a + \ln 2\xi_b + E(\hat{a},\hat{b})\nonumber\\
 \hspace{-2cm}&+\frac{1}{2}\ln \frac{2(\beta^{(1)}\bar{{x^{(1)}}^2}+\beta^{(2)}\bar{{x^{(2)}}^2})}{\pi}+\frac{1}{2}\ln \frac{2(\beta^{(1)}+\beta^{(2)})}{\pi}\nonumber\\
 \hspace{-2cm}&-\ln \left[\mathrm{erfc}\left(\sqrt{\frac{\beta^{(1)}\bar{{x^{(1)}}^2}+\beta^{(2)}\bar{{x^{(2)}}^2}}{2}}(-\xi_a-\hat{a}(\tau_1^{(1)},\tau_1^{(2)}))\right)-\mathrm{erfc}\left(\sqrt{\frac{\beta^{(1)}\bar{{x^{(1)}}^2}+\beta^{(2)}\bar{{x^{(2)}}^2}}{2}}(\xi_a-\hat{a}(\tau_1^{(1)},\tau_1^{(2)}))\right)\right]\nonumber\\
 \hspace{-2cm}&- \ln \left[\mathrm{erfc}\left(\sqrt{\frac{\beta^{(1)}+\beta^{(2)}}{2}}(-\xi_b-\hat{b}(\tau_2^{(1)},\tau_2^{(2)}))\right)-\mathrm{erfc}\left(\sqrt{\frac{\beta^{(1)}+\beta^{(2)}}{2}}(\xi_b-\hat{b}(\tau_2^{(1)},\tau_2^{(2)}))\right)\right]
\end{align}
If the noise intensity is known, \(\beta^{(m)}\) is equal to \(\beta^{(m)}_0\), leading to a difference in the residual error contributions given by:
\begin{align}
 &E(\hat{a}(\tau_1^{(1)}, \tau_1^{(2)}), \hat{b}(\tau_2^{(1)}, \tau_2^{(2)}), \upsilon^{(1)}, \upsilon^{(2)}) - \sum_{m=1}^2 \frac{1}{\beta_0^{(m)}}E_m(\hat{a}^{(m)}, \hat{b}^{(m)}) \nonumber\\
 &= \frac{1}{2} \left(\frac{(\beta^{(1)}\bar{{x^{(1)}}^2})(\beta^{(2)}\bar{{x^{(2)}}^2})}{\beta^{(1)}\bar{{x^{(1)}}^2}+\beta^{(2)}\bar{{x^{(2)}}^2}}(\hat{a}^{(1)}(\tau_1^{(1)})-\hat{a}^{(2)}(\tau_1^{(2)}))^2 + \frac{\beta^{(1)}\beta^{(2)}}{(\beta^{(1)}+\beta^{(2)})}(\hat{b}^{(1)}(\tau_2^{(1)})-\hat{b}^{(2)}(\tau_2^{(2)}))^2 \right),\label{difresume}
\end{align}
which is found that mesoscopic variables \(\upsilon^{(1)}\) and \(\upsilon^{(2)}\) disappear from this equation. Therefore, the difference in the Bayesian free energy is determined by four stochastic variables \(\tau_1^{(1)}, \tau_1^{(2)}, \tau_2^{(1)}, \tau_2^{(2)}\), and can be expressed as
\begin{align}
 \Delta F(\tau_1^{(1)}, \tau_1^{(2)}, \tau_2^{(1)}, \tau_2^{(2)}) = F^{\mathrm{int}}(\tau_1^{(1)}, \tau_1^{(2)}, \tau_2^{(1)}, \tau_2^{(2)}, \upsilon^{(1)}, \upsilon^{(2)}) - F^{\mathrm{iso}}(\tau_1^{(1)}, \tau_1^{(2)}, \tau_2^{(1)}, \tau_2^{(2)}, \upsilon^{(1)}, \upsilon^{(2)}). 
\end{align}
Consequently, the probability distribution of the difference in the Bayesian free energy can be expressed as:
\begin{align}
 p(\Delta F) = \int & \mathrm{d}\tau_1^{(1)}\mathrm{d}\tau_1^{(2)}\mathrm{d}\tau_2^{(1)}\mathrm{d}\tau_2^{(2)} \nonumber \\
 & \delta(\Delta F - \Delta F(\tau_1^{(1)}, \tau_1^{(2)}, \tau_2^{(1)}, \tau_2^{(2)})) \nonumber \\
 & p(\tau_1^{(1)})p(\tau_1^{(2)})p(\tau_2^{(1)})p(\tau_2^{(2)}) \label{prob_modelselection_int}
\end{align}

We can effectively assess the fluctuations of Bayesian integration by mesoscopic variables as described in Section \ref{sec:bayesianinference}.

\subsection{Numerical Experiment: Bayesian Integration}
\begin{figure}[ht]
 \centering
 \begin{subfigure}{0.32\textwidth}
 \includegraphics[width=\linewidth]{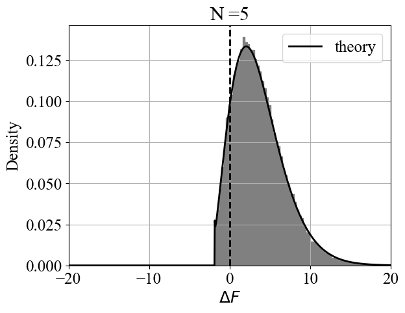}
 \caption{}
 \label{fig:image25}
 \end{subfigure}
 \hfill 
 \begin{subfigure}{0.32\textwidth}
 \includegraphics[width=\linewidth]{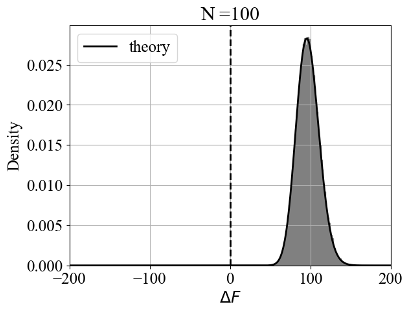}
 \caption{}
 \label{fig:image26}
 \end{subfigure}
 \hfill 
 \begin{subfigure}{0.32\textwidth}
 \includegraphics[width=\linewidth]{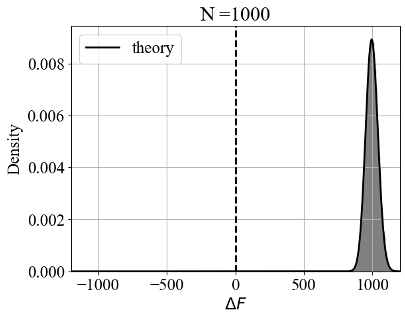}
 \caption{}
 \label{fig:image27}
 \end{subfigure}

\caption{Probability distribution of differences in the Bayesian free energy for Bayesian integration. Probability distribution from 100,000 samples of the Bayesian free energy difference (Equation (\ref{num_prob_modelselection_int})) with model parameters \(a_0^{(1)} = 2.0, a_0^{(2)} = 4.0, \sigma_0^2 = 1.0 \) for data sizes \(N = 5, 100, 1000\). Solid black lines represent the theoretical lines calculated from the non-central chi-squared distribution.}

\label{fig:delta_F_iso}
\end{figure}

Here, we examine the effects of the number of data points and the noise intensity of the data on estimation from the results of Bayesian integration via meso-expression. Figure \ref{fig:delta_F_iso} shows the frequency distribution from 100,000 samples of the probability distribution of the difference of the Bayesian free energy with model parameters \(a_0^{(1)} = 2.0, a_0^{(2)} = 4.0, \sigma_0^2 = 1.0\). Here, for simplicity, we omit the terms \(b_0^{(1)}\) and \(b_0^{(2)}\) from Equations (\ref{bayes_int_1}) and (\ref{bayes_int_2}). Consequently, we proceed to calculate the difference in the residual error
\begin{align}
 E(\hat{a}(\tau_1^{(1)}, \tau_1^{(2)}), \upsilon^{(1)}, \upsilon^{(2)}) - \sum_{m=1}^2 \frac{1}{\beta_0^{(m)}}E_m(\hat{a}^{(m)}) = \frac{1}{2} \left(\frac{(\beta^{(1)}\bar{{x^{(1)}}^2})(\beta^{(2)}\bar{{x^{(2)}}^2})}{\beta^{(1)}\bar{{x^{(1)}}^2}+\beta^{(2)}\bar{{x^{(2)}}^2}}(\hat{a}^{(1)}(\tau_1^{(1)})-\hat{a}^{(2)}(\tau_1^{(2)}))^2 \right),\label{num_difresume}
\end{align}
and the difference of the Bayesian free energy
\begin{equation}
 \Delta F(\tau_1^{(1)}, \tau_1^{(2)}) = F^{\mathrm{int}}(\tau_1^{(1)}, \tau_1^{(2)}, \upsilon^{(1)}, \upsilon^{(2)}) - F^{\mathrm{iso}}(\tau_1^{(1)}, \tau_1^{(2)}, \upsilon^{(1)}, \upsilon^{(2)}). \label{num_delta_f_int}
\end{equation}
\begin{align}
 p(\Delta F) = \int & \mathrm{d}\tau_1^{(1)}\mathrm{d}\tau_1^{(2)} \delta(\Delta F - \Delta F(\tau_1^{(1)}, \tau_1^{(2)})) p(\tau_1^{(1)})p(\tau_1^{(2)}) \label{num_prob_modelselection_int}
\end{align}
which is derived from the Bayesian integration of these simplified equations and thus does not account for the effect of mesoscopic variables \(\tau_2^{(1)}\) and \(\tau_2^{(2)}\) associated with \(b_0^{(1)}\) and \(b_0^{(2)}\).

Figure \ref{fig:delta_F_iso}(a) shows that with a small number of data points, the frequency of \(\Delta F < 0\) is high, and failures in model selection occur frequently. Conversely, Figures \ref{fig:delta_F_iso}(b) and \ref{fig:delta_F_iso}(c) show that with a larger number of data points, failures in model selection do not occur.

\begin{figure}[ht]
 \centering
 \begin{subfigure}{0.32\textwidth}
 \includegraphics[width=\linewidth]{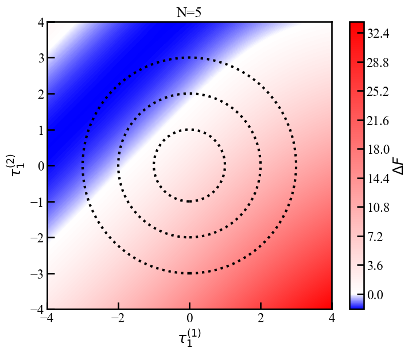}
 \caption{}
\end{subfigure}
 \hfill 
 \begin{subfigure}{0.32\textwidth}
 \includegraphics[width=\linewidth]{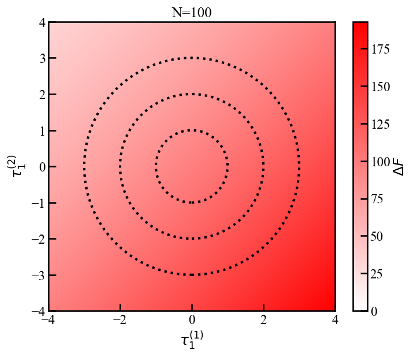}
 \caption{}
\end{subfigure}
 \hfill 
 \begin{subfigure}{0.32\textwidth}
 \includegraphics[width=\linewidth]{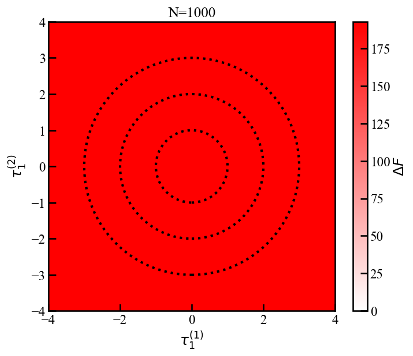}
 \caption{}
\end{subfigure}

\caption{Heat map of \(\Delta F\) as a function of \(\tau_1^{(1)}\) and \(\tau_1^{(2)}\) with model parameters \(a_0^{(1)} = 2.0, a_0^{(2)} = 4.0, \sigma_0^2 = 1.0 \) for data sizes \(N = 5, 100, 1000\). as the small, medium, and large circles represent those with radii of 1, 2, and 3, respectively. According to the properties of the chi-square distribution with 2 degrees of freedom, the probabilities for \(\tau_1^{(1)}\) and \(\tau_1^{(2)}\) to lie within these circles are approximately 39, 86, and 98\%, respectively.
}

\label{fig:delta_F_2d_iso}
\end{figure}

Figure \ref{fig:delta_F_2d_iso} shows the relationship between the differential free energy \(\Delta F (\tau_1^{(1)},\tau_2^{(2)})\) and \((\tau_1^{(1)},\tau_1^{(2)})\). Here, model parameters are set as \(a_0^{(1)} = 2.0, a_0^{(2)} = 4.0, \sigma_0^2 = 1.0 \) for data sizes \(N = 5, 100, 1000\). The region being depicted corresponds to the main area where \(\tau_1^{(1)}\) and \(\tau_1^{(2)}\) are generated. In this case, Equation (\ref{num_difresume}) shows that the difference in free energy is primarily due to the difference in the estimated parameter \(a\). In this case, since there is a significant difference with \(a_0^{(1)} - a_0^{(2)} = 2.0\), the result selected by the free energy suggests that they should be treated as almost independent. However, in situations with a small amount of data, the effect of \(\tau_1^{(1)}\) and \(\tau_1^{(2)}\) due to the fluctuation in the estimation of the parameter \(a\) becomes relatively large, which also indicates that the posterior probability of treating them as independent slightly decreases.

\begin{figure}[ht]
 \centering
 \begin{subfigure}{0.32\textwidth}
 \includegraphics[width=\linewidth]{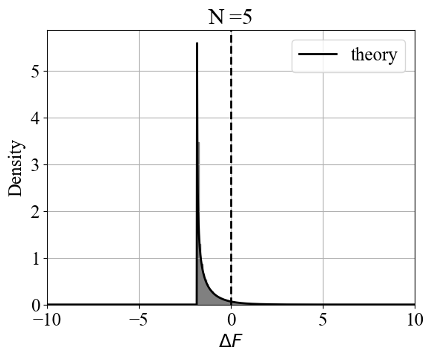}
 \caption{}
 \label{fig:image22}
 \end{subfigure}
 \hfill 
 \begin{subfigure}{0.32\textwidth}
 \includegraphics[width=\linewidth]{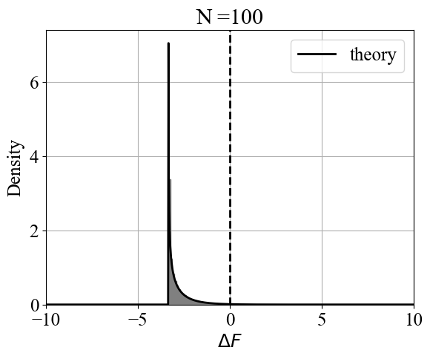}
 \caption{}
 \label{fig:image23}
 \end{subfigure}
 \hfill 
 \begin{subfigure}{0.32\textwidth}
 \includegraphics[width=\linewidth]{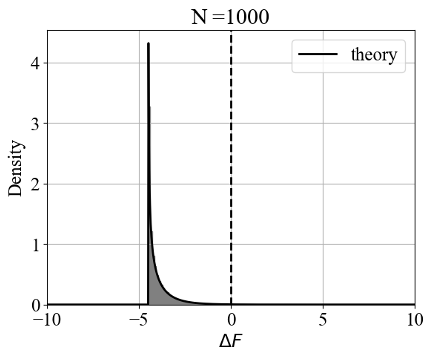}
 \caption{}
 \label{fig:image24}
 \end{subfigure}

\caption{Probability distribution of differences in the Bayesian free energy for Bayesian integration. Frequency distribution from 100,000 samples of the Bayesian free energy difference (Equation (\ref{num_prob_modelselection_int})) with model parameters \(a_0^{(1)} = 2.0, a_0^{(2)} = 2.0, \sigma_0^2 = 1.0 \) for data sizes \(N = 5, 100, 1000\). Solid black lines represent the theoretical lines calculated from the non-central chi-squared distribution.}

\label{fig:delta_F_int}
\end{figure}

Next, Figure \ref{fig:delta_F_int} shows the frequency distribution from 100,000 samples of the probability distribution of the difference in the Bayesian free energy with model parameters \(a_0^{(1)} = 2.0, a_0^{(2)} = 2.0, \sigma_0^2 = 1.0\) (Equation (\ref{num_prob_modelselection_int})). Figures \ref{fig:delta_F_int}(a)--(c) show that as the number of data points increases, the frequency of \(\Delta F > 0\) gradually decreases. However, even at \(N=1000\), the frequency of \(\Delta F > 0\) remains, indicating that failures in model selection are occurring. The minimum value of the distribution transitions negatively on a \(\mathrm{log}(N)\) scale, as evident from Equation (\ref{prob_modelselection_int}).

\begin{figure}[ht]
 \centering
 \begin{subfigure}{0.32\textwidth}
 \includegraphics[width=\linewidth]{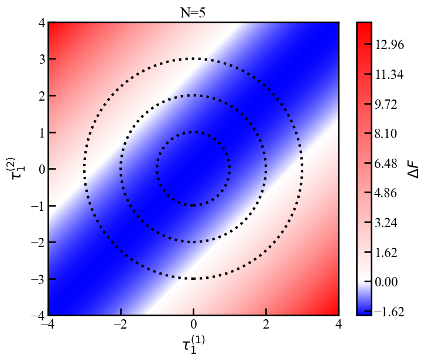}
 \caption{}
\end{subfigure}
 \hfill 
 \begin{subfigure}{0.32\textwidth}
 \includegraphics[width=\linewidth]{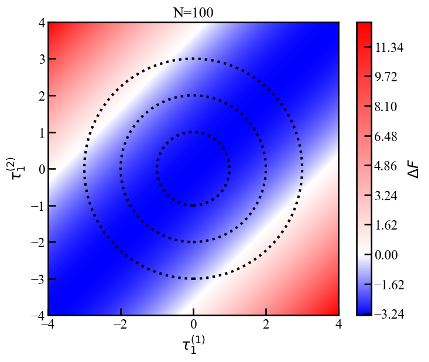}
 \caption{}
\end{subfigure}
 \hfill 
 \begin{subfigure}{0.32\textwidth}
 \includegraphics[width=\linewidth]{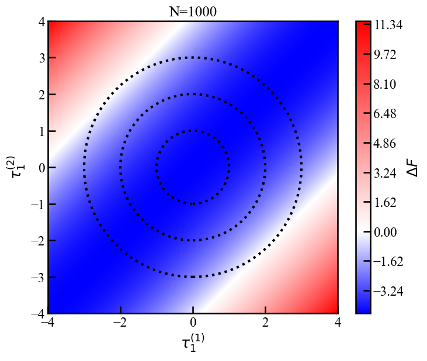}
 \caption{}
\end{subfigure}

\caption{Heat map of \(\Delta F\) as a function of \(\tau_1^{(1)}\) and \(\tau_1^{(2)}\) with model parameter \(a_0^{(1)} = 2.0, a_0^{(2)} = 2.0,\sigma_0^2 = 1.0 \) for data sizes \(N = 5, 100, 1000\).
The small, medium, and large circles represent those with radii of 1, 2, and 3, respectively. According to the properties of the chi-square distribution with 2 degrees of freedom, the probabilities for \(\tau_1^{(1)}\) and \(\tau_1^{(2)}\) to lie within these circles are approximately 39, 86, and 98\%, respectively.
}

\label{fig:delta_F_2d_int}
\end{figure}

Figure \ref{fig:delta_F_2d_int} shows the relationship between the differential free energy \(\Delta F (\tau_1^{(1)}, \tau_2^{(2)})\) and \((\tau_1^{(1)},\tau_1^{(2)})\). Here, model parameters are set as \(a_0^{(1)} = 2.0, a_0^{(2)} = 2.0, \sigma_0^2 = 1.0 \) for data sizes \(N = 5, 100, 1000\). In this case, since there is no significant difference between \(a_0^{(1)} - a_0^{(2)} = 0.0\), the result indicates that they should be integrated, as shown in this figure. Furthermore, as the amount of data increases, the difference in free energy decreases in the negative direction, suggesting that the posterior probability of integration increases.

\begin{figure}[ht]
 \centering
 \begin{subfigure}{0.32\textwidth}
 \includegraphics[width=\linewidth]{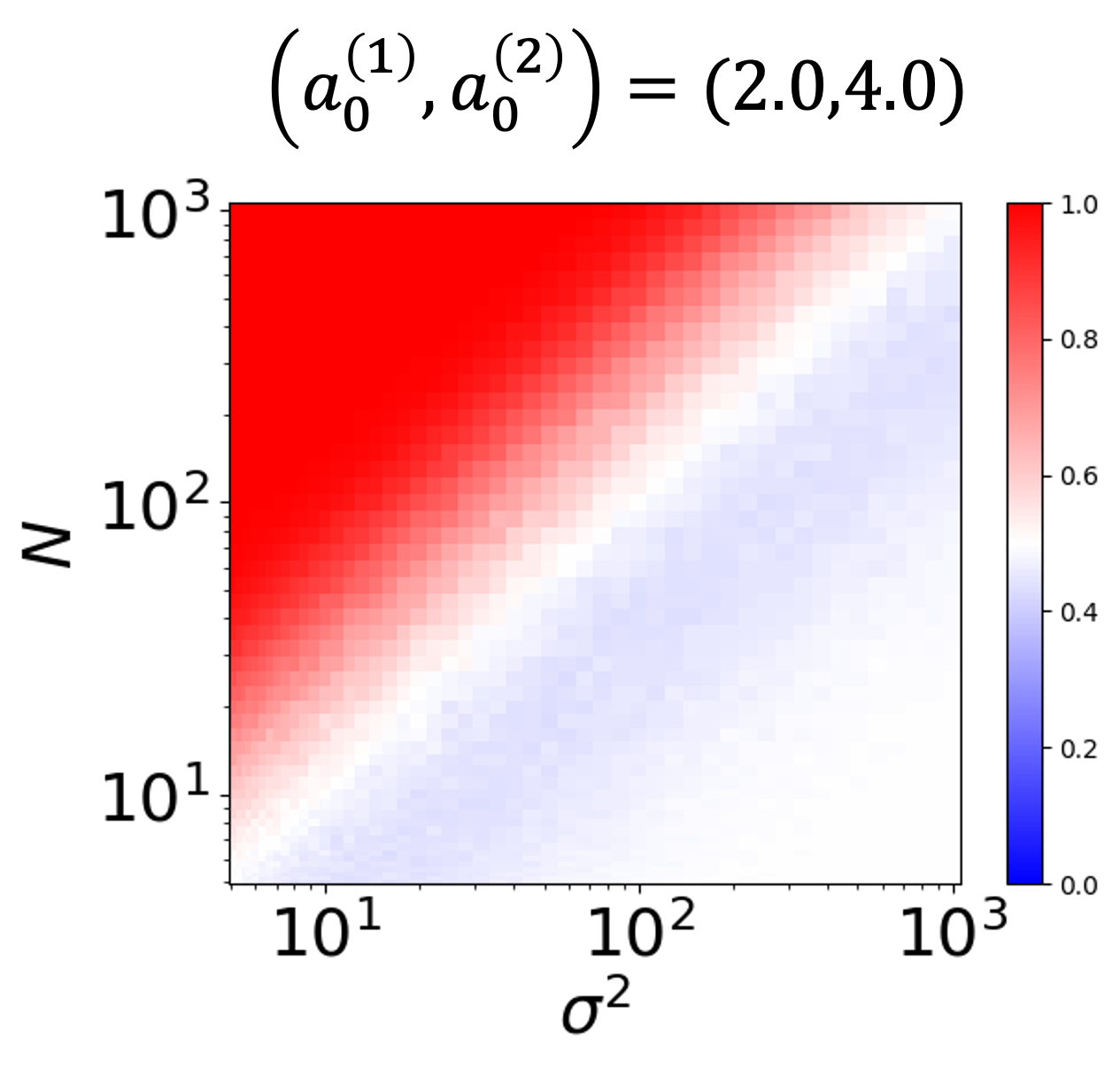}
 \caption{}
 \label{fig:image28}
 \end{subfigure}
 \hfill 
 \begin{subfigure}{0.32\textwidth}
 \includegraphics[width=\linewidth]{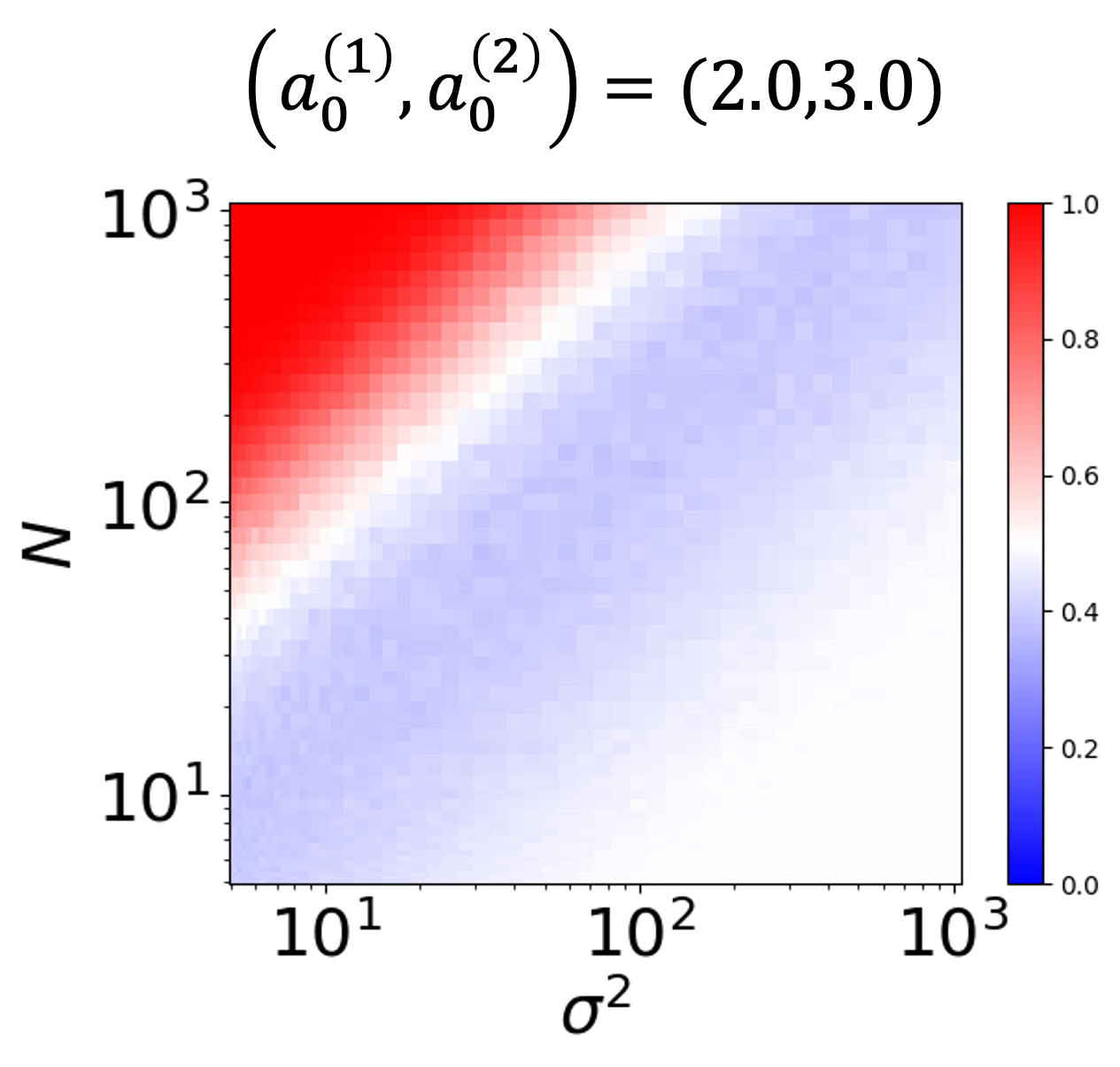}
 \caption{}
 \label{fig:image29}
 \end{subfigure}
 \hfill 
 \begin{subfigure}{0.32\textwidth}
 \includegraphics[width=\linewidth]{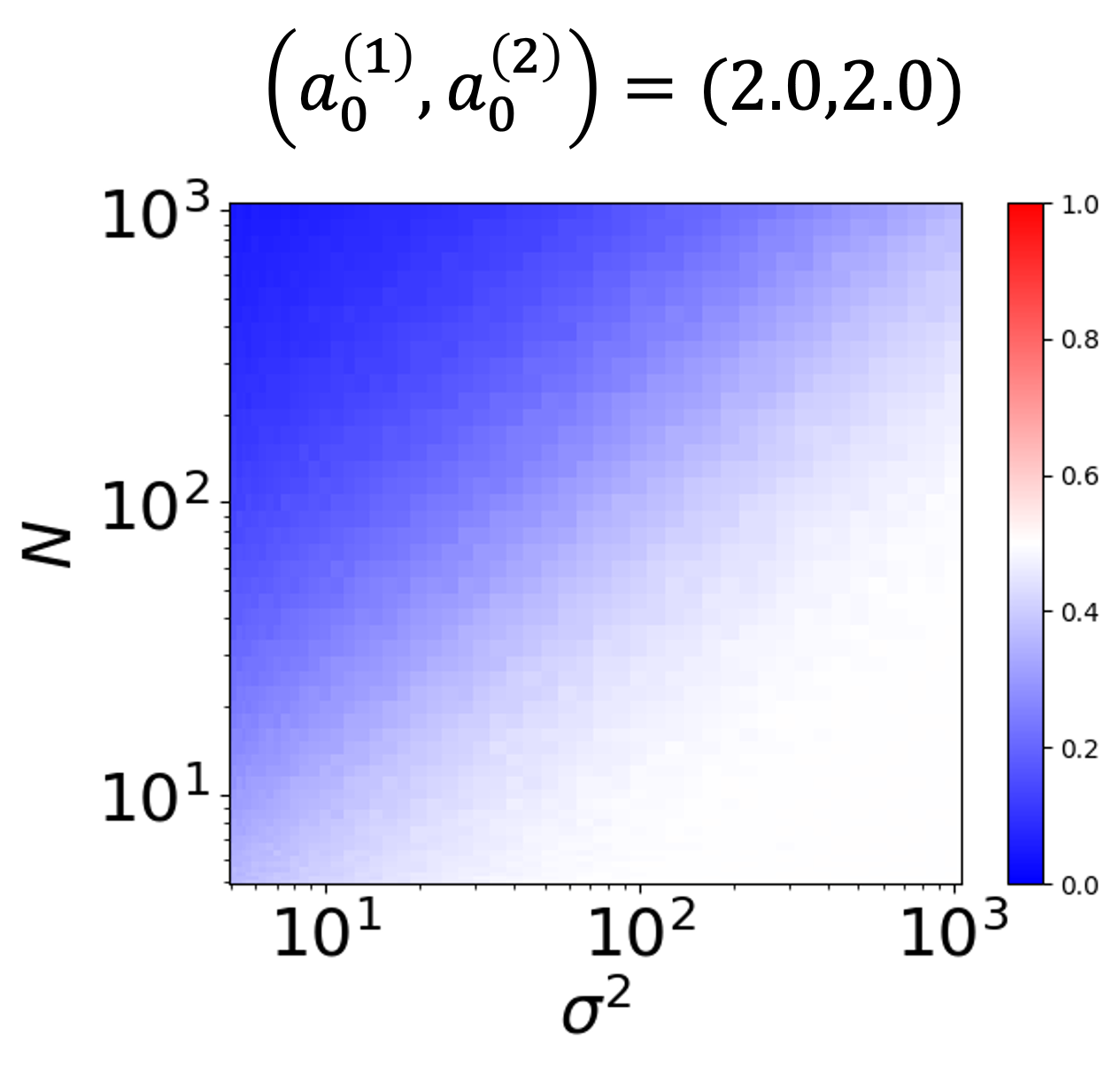}
 \caption{}
 \label{fig:image30}
 \end{subfigure}

\caption{Selection probabilities of the separate model derived from the probability distribution of the difference in the Bayesian free energy with model parameters \(a_0^{(1)}=2.0, a_0^{(2)} = 4.0, 3.0, 2.0\) (Equation (\ref{num_prob_modelselection_int})). Here, we set \(b_0^{(1)} = 0.0, b_0^{(2)} = 0.0\) and display the frequency distribution as a two-dimensional histogram from 100,000 samples in a two-dimensional space of the number of data points \(N\) and data noise intensity \(\sigma^2\).}
 \label{fig:heatmap_int}
\end{figure}

Finally, Figure \ref{fig:heatmap_int} shows the selection probabilities of the separate model derived from the probability distribution of the difference in the Bayesian free energy with model parameters \(a_0^{(1)}=2.0, a_0^{(2)} = 4.0, 3.0, 2.0\) (Equation (\ref{num_prob_modelselection_int})). Here, we display the frequency distribution as a two-dimensional histogram from 100,000 samples in a two-dimensional space of the number of data points \(N\) and data noise intensity \(\sigma^2\). Figure \ref{fig:heatmap_int}(a) shows that model selection tends to fail along the diagonal line where \(N\) and \(\sigma^2\) have similar values. As the number of data points increases from this line, model selection gradually becomes more effective. Conversely, as the number of data points decreases from the diagonal line, discrimination in model selection is eliminated. This diagonal line, as shown in Figures \ref{fig:heatmap_int}(b) and (c), widens as the value of \(a_0^{(2)}\) approaches that of \(a_0^{(1)}\), and when \(a_0^{(2)}=2.0\), only the selection probabilities of the integrated model remain.

\clearpage
\section{Conclusion}
\label{sec:conclusion}
In this study, we proposed an innovative theory that can address finite measurement datasets \(N\) in a linear regression model, a previously unaddressed challenge. This is the first step towards a new theoretical framework that goes beyond the conventional Bayesian measurement framework. Through our results, we confirmed that the outcomes of Bayesian estimation using theoretically derived microscopic and mesoscopic representations are consistent.

We summarize the insights gained from our results in the following. In conventional theoretical frameworks, only asymptotic characteristics such as the number of observation data \(N\) approaches infinity are discussed, making it difficult to consider fluctuations due to finite data \cite{schwarz1978estimating}. By introducing \(O(1)\) mesoscopic variables defined from \(N\) Gaussian noises, we succeeded in theoretically determining how important statistics such as the free energy difference \(\Delta F\) converge to a limit as \(N\) increases to infinity. We have established a theoretical foundation that can handle fluctuations due to finite data in the estimation of the posterior probability distribution of parameters, model selection, and Bayesian integration—the three core principles of Bayesian measurement. This is a groundbreaking achievement in the history of Bayesian inference.

As a result, the estimation of the posterior probability distribution of parameters in a linear regression model could be analytically expressed in terms of mesoscopic variables consisting of a sum of \(N\) Gaussian variables when using microscopic and mesoscopic representations. The residual error can be described using Gaussian and chi-squared distributions of mesoscopic variables, enabling the posterior probability distribution to be analytically derived even for a finite number of observed data \(N\). This is particularly important in real measurements where data is limited, demonstrating the potential for practical applications.

Regarding model selection, the proposed theory proved particularly useful. In conventional Bayesian measurements based on numerical calculations, fluctuations due to the finite number \(N\) are often overlooked, leading to erroneous model selection results. The proposed theory addressed this by analytically evaluating the Bayesian free energy difference \(\Delta F\), which depends on the number of observed data \(N\). This enables us to quantitatively evaluate the variation in the free energy difference distribution obtained from the microscopic and mesoscopic representations, demonstrating how the number of observational data \(N\) and the observation noise variance \(\sigma^2\) affect model selection. Theoretically, we demonstrate that the model selection results are stable when the number of observational data \(N\) is about 100, suggesting that the proposed theory can provide guidelines for actual measurements.

The proposed theory also proved useful in Bayesian integration, enabling analytical evaluation even when the number of observational data \(N\) is finite, and showing how the number of observational data \(N\) and the observation noise variance \(\sigma^2\) affect the results of Bayesian integration. Theoretically, we demonstrated that the Bayesian integration results are stable when the number of observational data \(N\) is about 100, again suggesting that the proposed theory can provide guidelines for data analysis and design of actual measurements.

The proposed theory establishes a new paradigm in Bayesian measurement, leading to more accurate and reliable scientific and technological results. The linear regression model \(y = ax + b\), while seemingly simple, is not merely for theoretical analysis. It is widely used in real measurement settings, such as linear system responses. Furthermore, insights gained from this model can be extended to general nonlinear models. We hope that this research will contribute to the development of measurement and data analysis and design in various fields of natural science.

\begin{acknowledgments}
This work was supported by JSPS KAKENHI Grant Numbers JP23H00486, 23KJ0723, and 23K16959, and CREST Grant Numbers JPMJCR1761 and JPMJCR1861 from the Japan Science and Technology Agency (JST).
\end{acknowledgments}

\bibliographystyle{unsrt}

\newpage

\appendix

\section{Proof that Residual Errors Follow a Chi-squared Distribution}
\label{sec:appA}
\label{append1}
Consider an orthogonal matrix $Q \in \mathbb{R^{N \times N}}$ whose first and second rows are defined as follows:
\begin{eqnarray}
 \bm{q}_1^T = \left(
 \frac{1}{\sqrt{N}},
 \frac{1}{\sqrt{N}},
 \cdots,
 \frac{1}{\sqrt{N}}
 \right)
\end{eqnarray}

\begin{eqnarray}
 \bm{q}_2^T = \left(
 \frac{x_1}{\sqrt{N\bar{x^2}}},
 \frac{x_2}{\sqrt{N\bar{x^2}}},
 \cdots,
 \frac{x_N}{\sqrt{N\bar{x^2}}}
 \right)
\end{eqnarray}
The existence of such an orthogonal matrix is guaranteed by Basis Extension Theorem in linear algebra\cite{lang1987linear,seber2012linear}. From the properties of orthogonal matrices, we have:
\begin{eqnarray}
 Q^TQ=QQ^T = I 
\end{eqnarray}
The random variables \(\{n_i\}_{i=1}^N\) independently follow a Gaussian distribution \(\mathcal{N}(0,\sigma^2)\), so let \(\bm{n}^T = (n_1, n_2, \cdots, n_N)\). The probability density function of \(\bm{n}\) is given by:
\begin{eqnarray}
 f(\bm{n}) = \left(\frac{1}{\sqrt{2\pi\sigma^2}}\right)^N \exp\left(-\frac{1}{2\sigma^2}\bm{n}^T\bm{n}\right)
\end{eqnarray}
Applying the orthogonal transformation \(\tilde{\bm{n}} = Q\bm{n}\), we have:
\begin{eqnarray}
 f(\tilde{\bm{n}}) = \left(\frac{1}{\sqrt{2\pi\sigma^2}}\right)^N \exp\left(-\frac{1}{2\sigma^2}\tilde{\bm{n}}^T\tilde{\bm{n}}\right)
\end{eqnarray}

At this time, the elements \(\tilde{n}_i\) of \(\tilde{\bm{n}}\) obtained by the orthogonal transformation are independent. In addition, \(\tilde{n}_1\) and \(\tilde{n}_2\) are given by:
\begin{eqnarray}
 \tilde{n}_1 &=& \bm{q}_1^T \bm{n} \\
 &=& \frac{1}{\sqrt{N}}\sum_{i=1}^N n_i \\
 \tilde{n}_2 &=& \bm{q}_2^T \bm{n} \\
 &=& \frac{1}{\sqrt{N\bar{x^2}}}\sum_{i=1}^N x_i n_i
\end{eqnarray}
where each corresponds to the second and first terms of the right-hand side of Equation (\ref{residualError}), respectively. From this, the residual error is:
\begin{eqnarray}
 \hspace{-2 cm}E(\hat{a},\hat{b}) \times 2N &=& 
 - \left(\frac{1}{\sqrt{N\bar{x^2}}}\sum_{i=1}^N x_i n_i \right)^2
 - \left(\frac{1}{\sqrt{N}}\sum_{i=1}^N n_i \right)^2
 + \sum_{i=1}^N n_i^2 \\
 &=& - \tilde{n}_2^2 - \tilde{n}_1^2 + \sum_{i=1}^N n_i^2 \\
 &=& - \tilde{n}_2^2 - \tilde{n}_1^2 + \sum_{i=1}^N \tilde{n}_i^2 \\
 &=& \sum_{i=3}^N \tilde{n}_i^2 
\end{eqnarray}
Thus, \(E(\hat{a},\hat{b}) \times 2N/\sigma^2\) follows a chi-squared distribution with \(N-2\) degrees of freedom, independent of the first and second terms on the right-hand side of Equation (\ref{residualError}).

\section{Noise Estimation}
\label{sec:appB}
In this section, we examine how the inclusion of noise estimation affects the results of model selection and Bayesian integration, using mesoscopic variables for Bayesian representation.

\subsection{Bayesian Inference with Noise Estimation}
In this subsection, we explore the impact of noise estimation on Bayesian inference. We extend the previous models to include noise variance as a probabilistic variable, making the framework applicable to realistic situations where the noise intensity is unknown beforehand.

\subsubsection{Noise Variance Estimation}
Up to this point, the noise variance \(\sigma^2_0\) has been treated as a constant. By considering the noise variance as a probabilistic variable \(\sigma^2\), this section demonstrates how to estimate the noise variance from the data by maximizing posterior distribution \(p(\sigma^2|Y)\) on the basis of Bayesian inference.

The posterior probability of the noise variance \(p(\sigma^2|Y)\) can be determined from the joint probability \(p(\sigma^2,a,b,Y)\) as
\begin{eqnarray}
 p(\sigma^2,a,b,Y) &=& p(Y|\sigma^2,a,b)p(a)p(b)p(\sigma^2).
\end{eqnarray}

The dependency part of the posterior probability \(p(\sigma^2|Y)\) on \(\sigma^2\) is
\begin{eqnarray}
 p(\sigma^2|Y) &\propto& p(\sigma^2) p(Y|\sigma^2), \label{sigmaposterior}\\
 &=& p(\sigma^2) \int \mathrm{d}a\mathrm{d}b\ p(Y|\sigma^2,a,b)p(a)p(b), \\
 &=& p(\sigma^2) \left(\frac{1}{\sqrt{2\pi\sigma^2}}\right)^N \exp\left(-\frac{N}{\sigma^2}E(\hat{a},\hat{b})\right)\nonumber\\
 &\times&\frac{1}{2\xi_a} \sqrt{\frac{\sigma^2\pi}{2N\bar{x^2}}}\left[\mathrm{erfc} \left(\sqrt{\frac{N\bar{x^2}}{2\sigma^2}}\left(-\xi_a-\frac{\bar{xy}}{\bar{x^2}}\right)\right)-\mathrm{erfc} \left(\sqrt{\frac{N\bar{x^2}}{2\sigma^2}}\left(\xi_a-\frac{\bar{xy}}{\bar{x^2}}\right)\right) \right]\nonumber\\
 &\times&\frac{1}{2\xi_b} \sqrt{\frac{\sigma^2\pi}{2N}}\left[\mathrm{erfc} \left(\sqrt{\frac{N}{2\sigma^2}}\left(-\xi_b-\bar{y}\right)\right)-\mathrm{erfc} \left(\sqrt{\frac{N}{2\sigma^2}}\left(\xi_b-\bar{y}\right)\right) \right].
\end{eqnarray}

When the prior distribution \(p(\sigma^2)\) is considered uniform, the posterior probability of the noise variance can be equated to the marginal likelihood for model parameters \(a,b\), thus enabling us to treat Equation (\ref{sigmaposterior}) similarly to the calculation of Equation (\ref{FreeEstrict}). The free energy \(F(\sigma^2)\), obtained by taking the negative log of \(p(\sigma^2|Y)\), is
\begin{eqnarray}
 F(\sigma^2) &=& -\ln p(\sigma^2|Y), \\
 &\sim& \frac{N}{2}\ln(2\pi\sigma^2) - \ln\left(\frac{\sigma^2\pi}{2N}\right) + \frac{1}{2}\ln\left(\bar{x^2}\right) + \ln(2\xi_a) + \ln(2\xi_b)\nonumber + \frac{N}{\sigma^2}E(\hat{a},\hat{b})\nonumber\\
 &-&\ln \left[\mathrm{erfc} \left(\sqrt{\frac{N\bar{x^2}}{2\sigma^2}}\left(-\xi_a-\hat{a}\right)\right)-\mathrm{erfc} \left(\sqrt{\frac{N\bar{x^2}}{2\sigma^2}}\left(\xi_a-\hat{a}\right)\right) \right]\nonumber\\
 &-&\ln \left[\mathrm{erfc} \left(\sqrt{\frac{N}{2\sigma^2}}\left(-\xi_b-\hat{b}\right)\right)-\mathrm{erfc} \left(\sqrt{\frac{N}{2\sigma^2}}\left(\xi_b-\hat{b}\right)\right) \right]. \label{freeenergysigma}
\end{eqnarray}
The optimal noise variance \(\sigma^2\) can be obtained by
\begin{eqnarray}
 \hat{\sigma}^2(\upsilon, \tau_1, \tau_2) &=& \arg\max_{\sigma^2} p(\sigma^2|Y),\\
 &=& \arg\min_{\sigma^2} F(\sigma^2).
\end{eqnarray}

\subsubsection{Noise Variance Through Mesoscopic Variables}
Here, we describe the noise variance using mesoscopic variables. Since \(\hat{\sigma}^2(\upsilon, \tau_1, \tau_2)\) cannot be analytically determined, we assume it has been numerically estimated. The probability distribution of the noise variance can then be described as
\begin{eqnarray}
 p(\sigma^2) = \int \mathrm{d}\upsilon\mathrm{d}\tau_1\mathrm{d}\tau_2 \delta(\sigma^2 - \hat{\sigma}^2(\upsilon,\tau_1,\tau_2))p(\upsilon)p(\tau_1)p(\tau_2).\label{deltasigma}
\end{eqnarray}

\subsubsection{Numerical Experiment Including Noise Estimation: Bayesian Inference}
Here, we numerically verify the results of Bayesian estimation including noise estimation. Initially, noise estimation is performed using Equation (\ref{freeenergysigma}), and the estimated noise is used to calculate the free energy from Equation (\ref{Freemeso}).

\begin{figure}[ht]
 \centering
 \begin{subfigure}{0.32\textwidth}
 \includegraphics[width=\linewidth]{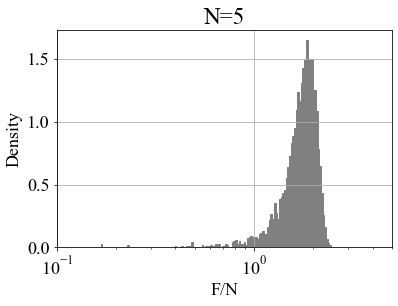}
 \caption{}
 \label{fig:b-1}
 \end{subfigure}
 \hfill
 \begin{subfigure}{0.32\textwidth}
 \includegraphics[width=\linewidth]{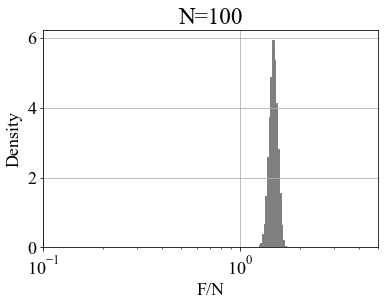}
 \caption{}
 \label{fig:b-2}
 \end{subfigure}
 \hfill
 \begin{subfigure}{0.32\textwidth}
 \includegraphics[width=\linewidth]{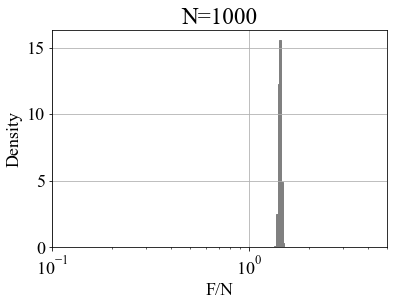}
 \caption{}
 \label{fig:b-3}
 \end{subfigure}

 \vspace{5mm} 

 \begin{subfigure}{0.32\textwidth}
 \includegraphics[width=\linewidth]{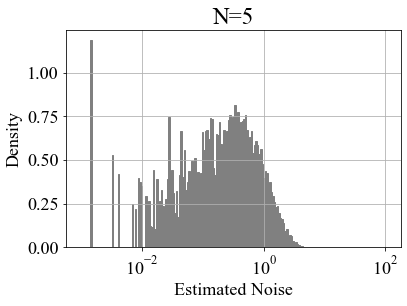}
 \caption{}
 \label{fig:b-4}
 \end{subfigure}
 \hfill
 \begin{subfigure}{0.32\textwidth}
 \includegraphics[width=\linewidth]{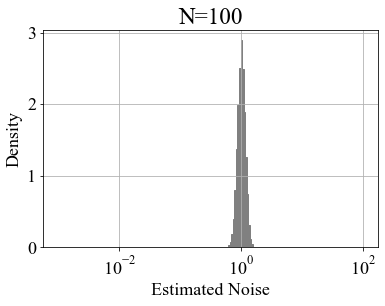}
 \caption{}
 \label{fig:b-5}
 \end{subfigure}
 \hfill
 \begin{subfigure}{0.32\textwidth}
 \includegraphics[width=\linewidth]{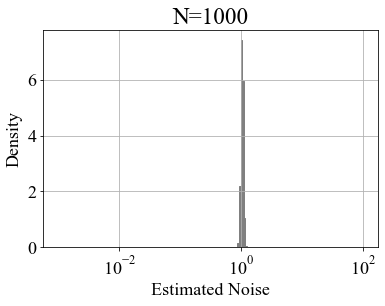}
 \caption{}
 \label{fig:b-6}
 \end{subfigure}
\caption{Probability distribution of free energy density and estimated noise. (a)--(c): Probability distribution of the normalized values of free energy calculated from Equation (\ref{Freemeso}) for 100,000 artificially generated data patterns with model parameters \(a_0 = 1.0, b_0 = 0.0, \sigma_0^2 = 1.0\), where noise estimation was performed. (d)--(f): Probability distribution of noise estimated for each of the 100,000 patterns of artificial data.}

 \label{fig:B1}
\end{figure}

Figure \ref{fig:B1} shows the probability distribution of the free energy density during noise estimation and that of the estimated noise. Figures \ref{fig:B1}(a)--(c) show the probability distributions of the normalized free energy values, calculated from Equation (\ref{Freemeso}) for 100,000 artificially generated data patterns with model parameters \(a_0 = 1.0, b_0 = 0.0, \sigma_0^2 = 1.0\), where noise estimation was performed. Figures \ref{fig:B1}(d)--(f) show the frequency distribution of the noise estimated for each dataset.

\begin{figure}[ht]
 \centering
 \includegraphics[width=0.9\linewidth]{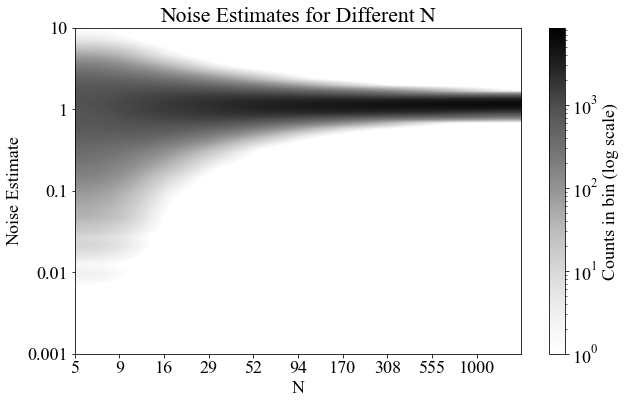}
\caption{Frequency distribution of estimated noise for 10,000 artificially generated data patterns with model parameters \(a_0 = 1.0, b_0 = 0.0, \sigma_0^2 = 1.0\) and \(N \in [5,1000]\).}

 \label{fig:B2}
\end{figure}

Figure \ref{fig:B2} shows the frequency distribution of the estimated noise for 10,000 artificially generated data patterns with model parameters \(a_0 = 1.0, b_0 = 0.0, \sigma_0^2 = 1.0\) and \(N \in [5,1000]\). Figure \ref{fig:B2}shows that as \(N\) increases, the frequency distribution of the estimated noise converges towards the true value.

\begin{figure}[ht]
 \centering
 \begin{subfigure}{0.32\textwidth}
 \includegraphics[width=\linewidth]{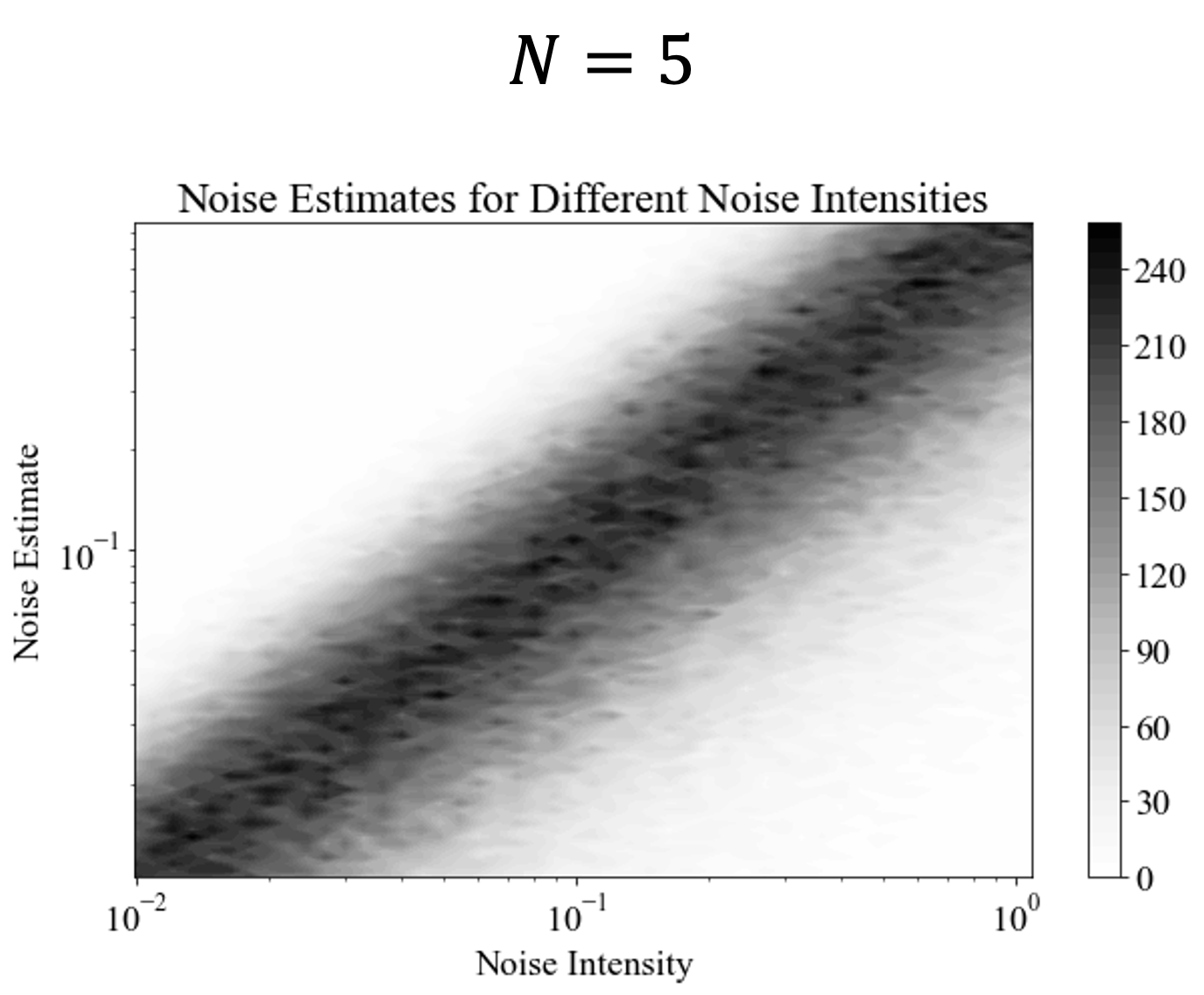}
 \caption{}
 \label{fig:b-8}
 \end{subfigure}
 \hfill
 \begin{subfigure}{0.32\textwidth}
 \includegraphics[width=\linewidth]{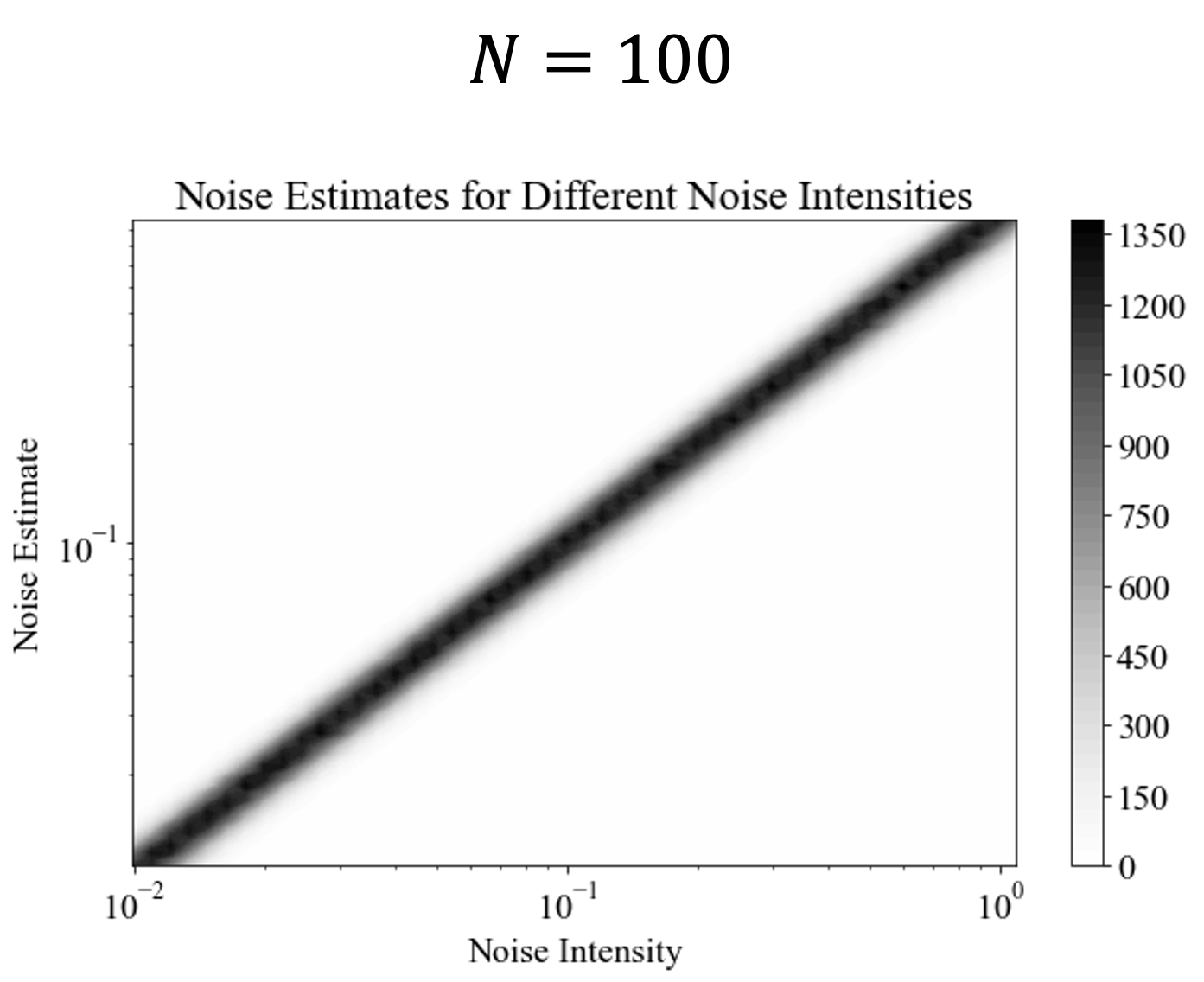}
 \caption{}
 \label{fig:b-9}
 \end{subfigure}
 \hfill
 \begin{subfigure}{0.32\textwidth}
 \includegraphics[width=\linewidth]{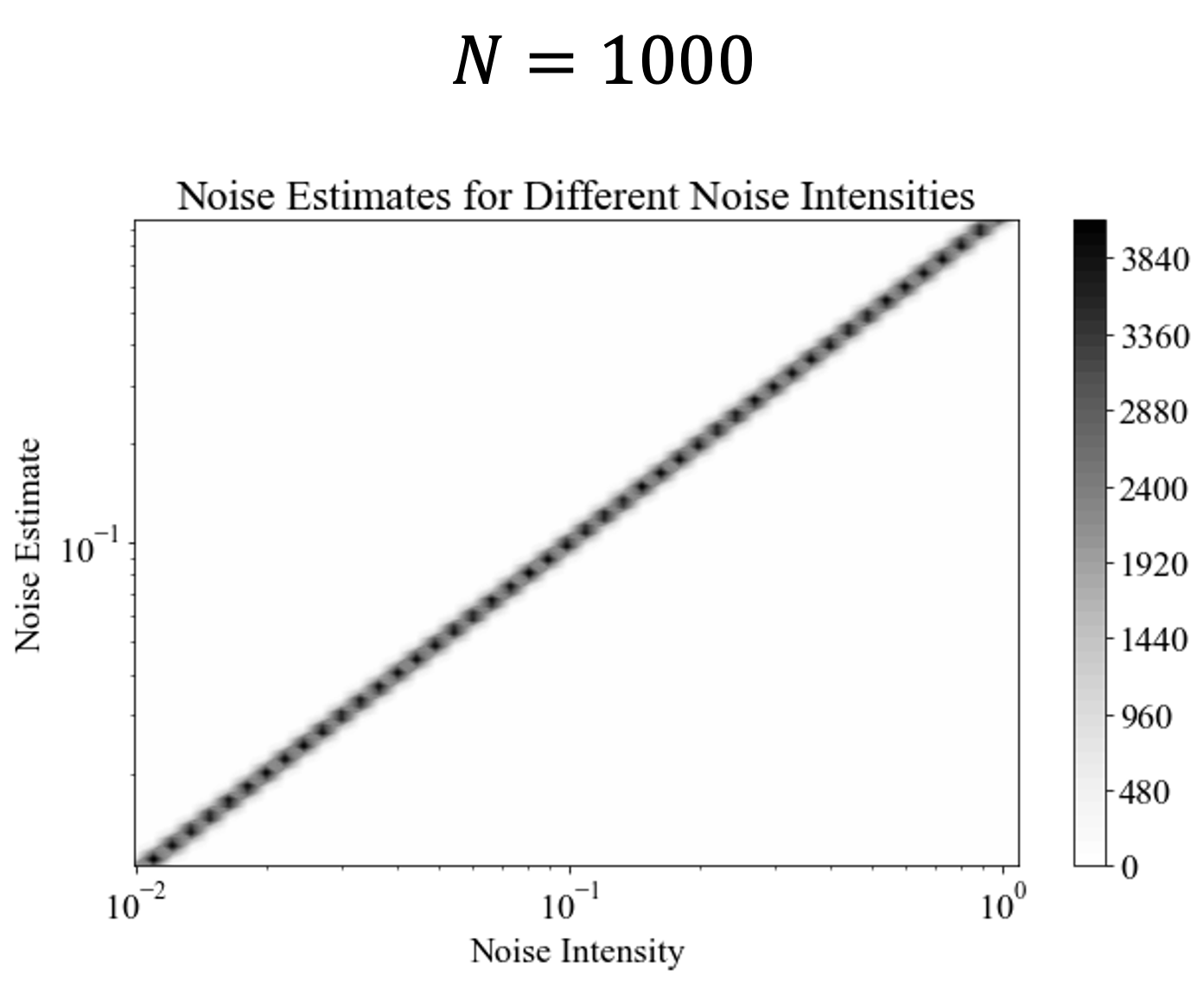}
 \caption{}
 \label{fig:b-10}
 \end{subfigure}
\caption{Frequency distribution of estimated noise for 10,000 artificially generated data patterns with model parameters \(a_0 = 1.0, b_0 = 0.0\) and \(N = 5, 100, 1000\), \(\sigma^2_0 = [0.01,\ldots, 1]\).}
 \label{fig:B3}
\end{figure}

Figure \ref{fig:B3} shows the frequency distribution of the estimated noise for 10,000 artificially generated data patterns with \(N = 5, 100, 1000\), \(\sigma^2_0 \in [0.01, 1]\), and model parameters \(a_0 = 1.0, b_0 = 0.0\). Figure \ref{fig:B3} shows that the estimation accuracy of the frequency distribution of the estimated noise depends only on \(N\) and not on \(\sigma^2_0\).

\clearpage

\subsection{Model Selection with Noise Estimation}
In this subsection, we delve into how noise estimation affects the process of model selection. By incorporating the noise variance as a probabilistic variable, we refine the Bayesian framework to better handle realistic scenarios where the noise intensity is unknown.

We will specifically explore how the estimation of noise variance affects the comparison between different models. This includes demonstrating the application of Bayesian free energy to select between one- and two-variable linear regression models with the additional complexity of noise estimation.

\subsubsection{Noise Variance Estimation for One-Variable Linear Regression Model}
Up to this point, the noise variance \(\sigma_0^2\) was treated as known. By considering the noise variance as a probabilistic variable \(\sigma^2\) and building upon the discussions in previous sections, this section demonstrates a method for estimating the noise variance from data by maximizing posterior distribution \(p(\sigma^2|Y)\) on the basis of Bayesian inference.

Given the joint probability \(p(\sigma^2,a,Y)\), the posterior probability of the noise variance \(p(\sigma^2|Y)\) is derived as
\begin{eqnarray}
 p(\sigma^2,a,Y) &=& p(Y|\sigma^2,a)p(a)p(\sigma^2).
\end{eqnarray}
The portion of the posterior probability \(p(\sigma^2|Y)\) dependent on \(\sigma^2\) is
\begin{eqnarray}
p(\sigma^2|Y) &=& \frac{1}{p(Y)} \int \mathrm{d}a\mathrm{d}b\ p(\sigma^2,a,Y),\\
&=& \frac{p(\sigma^2)}{p(Y)} \int \mathrm{d}a\ p(Y|\sigma^2,a)p(a) \label{sigmaposterior2},\\
p(Y) &=& \int \mathrm{d}a\mathrm{d}\sigma^2 \ p(Y|\sigma^2,a)p(a)p(\sigma^2).
\end{eqnarray}

Assuming a uniform prior distribution \(p(\sigma^2)\), the right side of Equation (\ref{sigmaposterior2}) can be executed similarly to the calculation of Equation (\ref{FreeEstrict2}). Taking the negative logarithm of Equation (\ref{sigmaposterior2}), the free energy \(F(\sigma^2)\) is expressed as
\begin{eqnarray}
 F(\sigma^2) &=& - \ln p(\sigma^2|Y)\\ 
 &\sim& \frac{N}{2}\ln(2\pi\sigma^2) - \frac{1}{2} \ln\left(\frac{\sigma^2\pi}{2N\bar{x^2}}\right) + \ln(2\xi_a) +\frac{N}{\sigma^2}E(\hat{a})\nonumber\\
 &&-\ln \left[\mathrm{erfc} \left(\sqrt{\frac{N\bar{x^2}}{2\sigma^2}}\left(-\xi_a-\hat{a}\right)\right)-\mathrm{erfc} \left(\sqrt{\frac{N\bar{x^2}}{2\sigma^2}}\left(\xi_a-\hat{a}\right)\right) \right].\label{freeenergysigma2}
\end{eqnarray}
Here, note that \(p(Y)\) is constant with respect to \(\sigma^2\).

The optimal noise variance \(\sigma^2\) can be obtained by
\begin{eqnarray}
 \hat{\sigma}^2(\upsilon_2,\tau_1) &=& \arg\max_{\sigma^2} p(\sigma^2|Y),\\
 &=& \arg\min_{\sigma^2} F(\sigma^2).
\end{eqnarray}

\subsubsection{Noise Variance in One-Variable Linear Regression Model Through Mesoscopic Variables}
Here, we describe the noise variance using mesoscopic variables. Since \(\hat{\sigma}^2(\upsilon_2, \tau_1)\) cannot be analytically determined, assuming it has been numerically determined, the probability distribution of the noise variance can be described as:
\begin{equation}
 p(\sigma^2) = \int \mathrm{d}\upsilon_2\mathrm{d}\tau_1 \delta(\sigma^2 - \hat{\sigma}^2(\upsilon_2,\tau_1))p(\upsilon_2)p(\tau_1).
\end{equation}

\subsubsection{Model Selection Through Bayesian Free Energy with Noise Estimation}
Up to this point, we have considered model selection on the basis of known true noise variance. Now, let us consider model selection when also estimating noise variance within each model, denoted as \(\sigma_1^2\) and \(\sigma_2^2\) for the two models, respectively. The Bayesian free energy for each model, after estimating noise variance, is
\begin{align}
 F_{y=ax+b}(\upsilon_1, \tau_1, \tau_2, \sigma^2_1) &= \frac{N}{2}\ln(2\pi\sigma^2_1) - \ln\left(\frac{\sigma^2_1\pi}{2N}\right) + \frac{1}{2}\ln\left(\bar{x^2}\right) + \ln(2\xi_a) + \ln(2\xi_b) + \frac{\sigma^2_0 \upsilon_1}{2\sigma^2_1} \nonumber \\
 &\quad - \ln \left[\mathrm{erfc} \left(\sqrt{\frac{N \bar{x^2}}{2\sigma^2_1}}\left(-\xi_a - \hat{a}(\tau_1)\right)\right) \right. \nonumber \\
 &\quad \left. - \mathrm{erfc} \left(\sqrt{\frac{N \bar{x^2}}{2\sigma^2_1}}\left(\xi_a - \hat{a}(\tau_1)\right)\right) \right] \nonumber \\
 &\quad - \ln \left[\mathrm{erfc} \left(\sqrt{\frac{N}{2\sigma^2_1}}\left(-\xi_b - \hat{b}(\tau_2)\right)\right) \right. \nonumber \\
 &\quad \left. - \mathrm{erfc} \left(\sqrt{\frac{N}{2\sigma^2_1}}\left(\xi_b - \hat{b}(\tau_2)\right)\right) \right]
\end{align}
for the two-variable model and
\begin{align}
 F_{y=ax}(\upsilon_2, \tau_1, \sigma^2_2) &= \frac{N}{2}\ln(2\pi\sigma^2_2) - \frac{1}{2}\ln\left(\frac{\sigma^2_2\pi}{2N\bar{x^2}}\right) + \ln(2\xi_a) + \frac{\sigma^2_0 \upsilon_2}{2\sigma^2_2} \nonumber \\
 &\quad - \ln \left[\mathrm{erfc}\left(\sqrt{\frac{N\bar{x^2}}{2\sigma^2_2}}\left(-\xi_a-\hat{a}(\tau_1)\right)\right) \right. \nonumber \\
 &\quad \left. - \mathrm{erfc}\left(\sqrt{\frac{N\bar{x^2}}{2\sigma^2_2}}\left(\xi_a-\hat{a}(\tau_1)\right)\right)\right]
\end{align}
for the one-variable model.

From these expressions, the difference in the Bayesian free energy, taking into account noise variance estimation, can be described as
\begin{align}
 \Delta F(\upsilon_1, \tau_1, \tau_2, \sigma^2_1, \sigma^2_2) &= F_{y=ax}(\upsilon_2, \tau_1, \sigma^2_2) - F_{y=ax+b}(\upsilon_1, \tau_1, \tau_2, \sigma^2_1) 
\end{align}
This difference is determined on the basis of mesoscopic variables and their relationships as noted in Equation (\ref{relation_residual}), enabling us to depict the probability distribution of the difference in the Bayesian free energy as a function of mesoscopic variables:
\begin{equation}
 p(\Delta F) = \int \mathrm{d}\upsilon_1 \mathrm{d}\tau_1 \mathrm{d}\tau_2 \delta(\Delta F - \Delta F(\upsilon_1, \tau_1, \tau_2, \sigma^2_1(\upsilon_1, \tau_1, \tau_2), \sigma^2_2(\upsilon_1, \tau_1, \tau_2))) p(\upsilon_1) p(\tau_1) p(\tau_2).
\end{equation}

\subsubsection{Numerical Experiment Including Noise Estimation: Model Selection}
Here, we examine the impact of the number of data points and the noise strength of the data on estimation from the results of model selection performed with noise estimation using mesoscopic representation.

\begin{figure}[ht]
 \centering
 \begin{subfigure}{0.32\textwidth}
 \includegraphics[width=\linewidth]{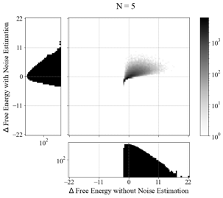}
 \caption{}
 \label{fig:b-11}
 \end{subfigure}
 \hfill
 \begin{subfigure}{0.32\textwidth}
 \includegraphics[width=\linewidth]{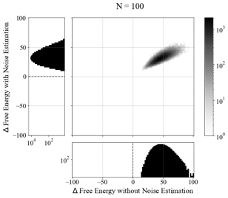}
 \caption{}
 \label{fig:b-12}
 \end{subfigure}
 \hfill
 \begin{subfigure}{0.32\textwidth}
 \includegraphics[width=\linewidth]{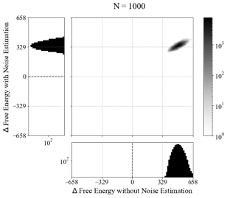}
 \caption{}
 \label{fig:b-13}
 \end{subfigure}
\caption{Frequency distribution from 100,000 samples of the probability distribution of the difference in the Bayesian free energy when noise estimation is performed and when noise is assumed known, with model parameters \(a_0 = 1.0, b_0 = 1.0, \sigma_0^2 = 1.0\) (Equation (\ref{prob_modelselection})). The horizontal and vertical axes represent the frequency distribution of the free energy when noise is known and estimated, respectively.}

 \label{fig:B4}
\end{figure}

Figure \ref{fig:B4} shows the frequency distribution from 100,000 samples of the probability distribution of the difference in the Bayesian free energy when noise estimation is performed and when noise is assumed known, with model parameters \(a_0 = 1.0, b_0 = 1.0, \sigma_0^2 = 1.0\) (Equation (\ref{prob_modelselection})). The horizontal and vertical axes represent the frequency distribution of the free energy when noise is known and estimated, respectively. Figure \ref{fig:B4} shows that as \(N\) increases, the difference in the frequency distributions between the cases of noise estimation and known noise diminishes.

\begin{figure}[ht]
 \centering
 \begin{subfigure}{0.32\textwidth}
 \includegraphics[width=\linewidth]{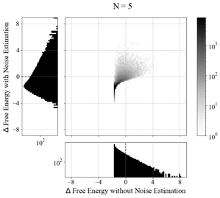}
 \caption{}
 \label{fig:b-14}
 \end{subfigure}
 \hfill
 \begin{subfigure}{0.32\textwidth}
 \includegraphics[width=\linewidth]{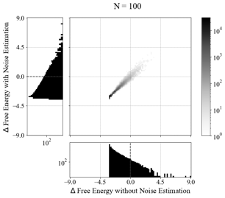}
 \caption{}
 \label{fig:b-15}
 \end{subfigure}
 \hfill
 \begin{subfigure}{0.32\textwidth}
 \includegraphics[width=\linewidth]{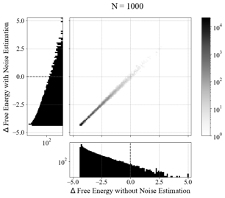}
 \caption{}
 \label{fig:b-16}
 \end{subfigure}
\caption{Frequency distribution from 100,000 samples of the probability distribution of the difference in the Bayesian free energy when noise estimation is performed and when noise is assumed known, with model parameters \(a_0 = 1.0, b_0 = 0.0, \sigma_0^2 = 1.0\) (Equation (\ref{prob_modelselection})). The horizontal and vertical axes represent the frequency distribution of the free energy when noise is known and estimated, respectively.}

 \label{fig:B5}
\end{figure}

Figure \ref{fig:B5} shows the frequency distribution from 100,000 samples of the probability distribution of the difference in the Bayesian free energy when noise estimation is performed and when noise is assumed known, with model parameters \(a_0 = 1.0, b_0 = 0.0, \sigma_0^2 = 1.0\) (Equation (\ref{prob_modelselection})). The horizontal and vertical axes represent the frequency distribution of the free energy when noise is known and estimated, respectively. Figure \ref{fig:B5} shows that as \(N\) increases, the difference in the frequency distributions between the cases of noise estimation and known noise diminishes.

\begin{figure}[ht]
 \centering
 \begin{subfigure}{0.32\textwidth}
 \includegraphics[width=\linewidth]{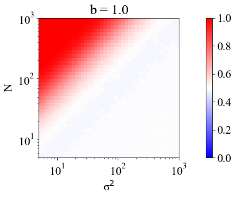}
 \caption{}
 \label{fig:b-17}
 \end{subfigure}
 \hfill 
 \begin{subfigure}{0.32\textwidth}
 \includegraphics[width=\linewidth]{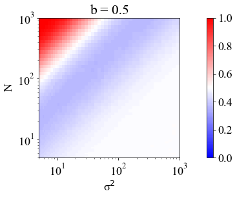}
 \caption{}
 \label{fig:b-18}
 \end{subfigure}
 \hfill 
 \begin{subfigure}{0.32\textwidth}
 \includegraphics[width=\linewidth]{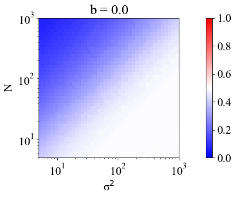}
 \caption{}
 \label{fig:b-19}
 \end{subfigure}

\caption{Average selection probability of the two-variable model \(y = ax + b\) derived from the probability distribution of the difference in the Bayesian free energy with model parameters \(b_0 = 1.0, 0.5, 0.0\) (Equation (\ref{prob_modelselection})). The parameters are set as \(a_0 = 1.0, \sigma_0^2 = 1.0\), and the frequency distribution is shown as a two-dimensional histogram of the number of data points \(N\) and the data noise intensity \(\sigma^2\).}

 \label{fig:B6}
\end{figure}

Finally, Figure \ref{fig:B6} shows the selection probabilities of the two-variable model \(y = ax + b\) derived from the probability distribution of the difference in the Bayesian free energy with model parameters \(b_0 = 1.0, 0.5, 0.0\) (Equation (\ref{prob_modelselection})). Here, we set \(a_0 = 1.0\) and display the frequency distribution as a two-dimensional histogram from 100,000 samples in a two-dimensional space of the number of data points \(N\) and data noise intensity \(\sigma^2\). Figure \ref{fig:B6} (a) shows that along the diagonal line where \(N\) and \(\sigma^2\) have similar values, there is a tendency for model selection to fail. As the number of data points increases from this line, gradually more appropriate model selections become possible. Conversely, as the number of data points decreases from the diagonal line, discrimination in model selection is eliminated. This diagonal line, as shown in Figures \ref{fig:B6} (b) and (c), widens as the value of \(b\) decreases, and at \(b=0.0\), the selection probability of \(y = ax + b\) disappears. The overall behavior of the probability distribution does not change regardless of whether noise estimation is performed.

From the aforementioned results, we found that the difference between simultaneously estimating two noises and estimating each noise independently becomes negligible with large data sizes. The former method involves optimization in a high-dimensional space, while the latter involves that in a one-dimensional space. Estimating multiple noises simultaneously increases the search space exponentially. Optimizing each noise independently ensures sufficient accuracy, which is beneficial for real-world applications.

\clearpage

\subsection{Bayesian Integration with Noise Estimation}
In this subsection, we examine how noise estimation affects the process of Bayesian integration when multiple datasets are involved. By treating the noise variances as probabilistic variables, we enhance the Bayesian framework to accommodate realistic situations where the noise levels are unknown and may vary between datasets.

\subsubsection{Bayesian Integration of Noise Variance Estimation}
In this section, we consider the noise variances \({\sigma^{(1)}}^2, {\sigma^{(2)}}^2\) as random variables and demonstrate a method to estimate these variances by maximizing posterior distribution \(p({\sigma^{(1)}}^2, {\sigma^{(2)}}^2|Y)\) on the basis of Bayesian inference. When the model parameters are inferred independently for each dataset, the noise variances \({\sigma^{(1)}}^2, {\sigma^{(2)}}^2\) should be inferred independently as well. Consider the case where the model parameters are common across datasets.

Let us assume the joint probability of \({\sigma^{(1)}}^2, {\sigma^{(2)}}^2, a, b, Y\) is given by:
\begin{eqnarray}
 p({\sigma^{(1)}}^2, {\sigma^{(2)}}^2, a, b, Y) &=& p(Y|{\sigma^{(1)}}^2, {\sigma^{(2)}}^2, a, b)p(a)p(b)p({\sigma^{(1)}}^2)p({\sigma^{(2)}}^2)
\end{eqnarray}
From this, the posterior probability of the noise variances \(p({\sigma^{(1)}}^2, {\sigma^{(2)}}^2|Y)\) can be expressed as:
\begin{align}
p({\sigma^{(1)}}^2, {\sigma^{(2)}}^2|Y) &= \frac{1}{p(Y)} \int \mathrm{d}a\mathrm{d}b\ p({\sigma^{(1)}}^2, {\sigma^{(2)}}^2, a, b, Y)\\
&= \frac{p({\sigma^{(1)}}^2, {\sigma^{(2)}}^2)}{p(Y)} \int \mathrm{d}a\mathrm{d}b\ p(Y|{\sigma^{(1)}}^2, {\sigma^{(2)}}^2, a, b)p(a)p(b)\\
p(Y) &= \int \mathrm{d}a\mathrm{d}b\mathrm{d}{\sigma^{(1)}}^2\mathrm{d}{\sigma^{(2)}}^2 \ p(Y|{\sigma^{(1)}}^2, {\sigma^{(2)}}^2, a, b)p(a)p(b) p({\sigma^{(1)}}^2)p({\sigma^{(2)}}^2).
\end{align}
Here, note that \(p(Y)\) is constant with respect to \({\sigma^{(1)}}^2\) and \({\sigma^{(2)}}^2\). At this time, the portion of the posterior probability \(p({\sigma^{(1)}}^2, {\sigma^{(2)}}^2 | Y)\) that depends on \({\sigma^{(1)}}^2, {\sigma^{(2)}}^2\) can be expressed as:
\begin{eqnarray}
 p({\sigma^{(1)}}^2, {\sigma^{(2)}}^2 | Y) &\propto& p({\sigma^{(1)}}^2)p({\sigma^{(2)}}^2) \int \mathrm{d}a\mathrm{d}b\ p(Y | {\sigma^{(1)}}^2, {\sigma^{(2)}}^2, a, b)p(a, b) \label{intsigmaposterior}\\
 &=& p({\sigma^{(1)}}^2)p({\sigma^{(2)}}^2) \left(\frac{1}{\sqrt{2\pi(\sigma^{(1)}_{0})^2}}\right)^{N_1}
 \left(\frac{1}{\sqrt{2\pi(\sigma^{(2)}_{0})^2}}\right)^{N_2}\frac{1}{2\xi_a}\frac{1}{2\xi_b}\exp\left(-E(\hat{a},\hat{b})\right)\nonumber\\
 &\times& \sqrt{\frac{\pi}{2(\beta^{(1)}\bar{{x^{(1)}}^2}+\beta^{(2)}\bar{{x^{(2)}}^2})}}\left[\mathrm{erfc}\left(\sqrt{\frac{\beta^{(1)}\bar{{x^{(1)}}^2}+\beta^{(2)}\bar{{x^{(2)}}^2}}{2}}(-\xi_a-\hat{a})\right)\right.\nonumber\\
 &-&\left.\mathrm{erfc}\left(\sqrt{\frac{\beta^{(1)}\bar{{x^{(1)}}^2}+\beta^{(2)}\bar{{x^{(2)}}^2}}{2}}(\xi_a-\hat{a})\right)\right]\nonumber\\
 &\times& \sqrt{\frac{\pi}{2(\beta^{(1)}+\beta^{(2)})}}\left[\mathrm{erfc}\left(\sqrt{\frac{\beta^{(1)}+\beta^{(2)}}{2}}(-\xi_b-\hat{b})\right)\right.\nonumber\\
 &-&\left.\mathrm{erfc}\left(\sqrt{\frac{\beta^{(1)}+\beta^{(2)}}{2}}(\xi_b-\hat{b})\right)\right]
 \label{intsigmaposterior2}
\end{eqnarray}
When the prior distribution \(p({\sigma^{(1)}}^2)p({\sigma^{(2)}}^2)\) is considered uniform, the right-hand side of Equation (\ref{intsigmaposterior}) can be computed similarly to Equation (\ref{intFreeEstrict}), and the free energy derived from taking the negative logarithm of Equation (\ref{intsigmaposterior}) is given by:
\begin{eqnarray}
 F({\sigma^{(1)}}^2, {\sigma^{(2)}}^2) &=& -\ln p({\sigma^{(1)}}^2, {\sigma^{(2)}}^2 | Y)\\
 &\sim& \frac{N_1}{2}\ln 2\pi (\sigma^{(1)})^2 + \frac{N_2}{2}\ln 2\pi (\sigma^{(2)})^2 + \ln 2\xi_a + \ln 2\xi_b + E(\hat{a},\hat{b})\nonumber\\
 &+& \frac{1}{2}\ln \frac{2(\beta^{(1)}\bar{{x^{(1)}}^2} + \beta^{(2)}\bar{{x^{(2)}}^2})}{\pi} + \frac{1}{2}\ln \frac{2(\beta^{(1)} + \beta^{(2)})}{\pi}\nonumber\\
 &-& \ln \left[\mathrm{erfc}\left(\sqrt{\frac{\beta^{(1)}\bar{{x^{(1)}}^2} + \beta^{(2)}\bar{{x^{(2)}}^2}}{2}}(-\xi_a - \hat{a})\right) - \mathrm{erfc}\left(\sqrt{\frac{\beta^{(1)}\bar{{x^{(1)}}^2} + \beta^{(2)}\bar{{x^{(2)}}^2}}{2}}(\xi_a - \hat{a})\right)\right]\nonumber\\
 &-& \ln \left[\mathrm{erfc}\left(\sqrt{\frac{\beta^{(1)} + \beta^{(2)}}{2}}(-\xi_b - \hat{b})\right) - \mathrm{erfc}\left(\sqrt{\frac{\beta^{(1)} + \beta^{(2)}}{2}}(\xi_b - \hat{b})\right)\right]. \label{free_energy_two_data}
\end{eqnarray}
Minimizing the free energy can numerically determine the values of \({\sigma^{(1)}}^2\) and \({\sigma^{(2)}}^2\) that maximize the posterior probability. The optimal noise variance \({\sigma^{(1)}}^2\) and \({\sigma^{(2)}}^2\) can be obtained by
\begin{align}
\hspace{-3cm}({\hat{\sigma}^{(1)}}(\tau_1^{(1)}, \tau_1^{(2)}, \tau_2^{(1)}, \tau_2^{(2)}, \upsilon^{(1)}, \upsilon^{(2)}),\ {\hat{\sigma}^{(2)}}(\tau_1^{(1)}, \tau_1^{(2)}, & \tau_2^{(1)}, \tau_2^{(2)}, \upsilon^{(1)}, \upsilon^{(2)}))\nonumber\\
&= \arg\max_{({\sigma^{(1)}}, {\sigma^{(2)}})} p({\sigma^{(1)}}^2, {\sigma^{(2)}}^2 \mid Y), \\
 &= \arg\min_{({\sigma^{(1)}}, {\sigma^{(2)}})} F({\sigma^{(1)}}^2, {\sigma^{(2)}}^2).
\end{align}

\subsubsection{Noise Variance Through Mesoscopic Variables in Bayesian Integration}
Noise estimation numerically determines \({\sigma^{(m)}}^2\) that minimizes Equation \ref{free_energy_two_data}. Since the estimated noise variance depends on six random variables \(\tau_1^{(1)}, \tau_1^{(2)}, \tau_2^{(1)}, \tau_2^{(2)}, \upsilon^{(1)}, \upsilon^{(2)}\), it can be expressed as \(\hat{\sigma}^{(m)}(\tau_1^{(1)}, \tau_1^{(2)}, \tau_2^{(1)}, \tau_2^{(2)}, \upsilon^{(1)}, \upsilon^{(2)})\). The probability distribution of the noise variance is given by

\begin{align}
 p({\sigma^{(m)}}^2) &= \int \mathrm{d}\tau_1^{(1)}\mathrm{d}\tau_1^{(2)}\mathrm{d}\tau_2^{(1)}\mathrm{d}\tau_2^{(2)}\mathrm{d}\upsilon^{(1)}\mathrm{d}\upsilon^{(2)} \nonumber \\
 &\quad \delta\left({\sigma^{(m)}}^2 - \left(\hat{\sigma}^{(m)}(\tau_1^{(1)}, \tau_1^{(2)}, \tau_2^{(1)}, \tau_2^{(2)}, \upsilon^{(1)}, \upsilon^{(2)})\right)^2 \right)\nonumber \\
 &\quad \times p(\tau_1^{(1)})p(\tau_1^{(2)})p(\tau_2^{(1)})p(\tau_2^{(2)})p(\upsilon^{(1)})p(\upsilon^{(2)})
\end{align}

\subsubsection{Model Selection in Bayesian Integration with Noise Estimation}
The difference in the Bayesian free energy, \(\Delta F\), is determined by six stochastic variables \(\tau_1^{(1)}, \tau_1^{(2)}, \tau_2^{(1)}, \tau_2^{(2)}, \upsilon^{(1)}, \upsilon^{(2)}\), and can be expressed as:
\begin{eqnarray}
 \Delta F = \Delta F(\tau_1^{(1)}, \tau_1^{(2)}, \tau_2^{(1)}, \tau_2^{(2)}, \upsilon^{(1)}, \upsilon^{(2)})
\end{eqnarray}
This indicates that the calculation of the Bayesian free energy differences incorporates these six variables, reflecting the complex dynamics when noise intensity is part of the estimation process.

Consequently, the probability distribution of the difference in the Bayesian free energy can be expressed as:
\begin{align}
 p(\Delta F) = \int & \mathrm{d}\tau_1^{(1)}\mathrm{d}\tau_1^{(2)}\mathrm{d}\tau_2^{(1)}\mathrm{d}\tau_2^{(2)}\mathrm{d}\upsilon^{(1)}\mathrm{d}\upsilon^{(2)} \nonumber \\
 & \delta(\Delta F - \Delta F(\tau_1^{(1)}, \tau_1^{(2)}, \tau_2^{(1)}, \tau_2^{(2)}, \upsilon^{(1)}, \upsilon^{(2)})) \nonumber \\
 & p(\tau_1^{(1)})p(\tau_1^{(2)})p(\tau_2^{(1)})p(\tau_2^{(2)})p(\upsilon^{(1)})p(\upsilon^{(2)})
\end{align}

\subsubsection{Numerical Experiment Including Noise Estimation: Bayesian Integration}
Here, we examine the impact of the number of data and the noise intensity inherent in the data on estimation through the results of Bayesian integration, which includes noise estimation using meso-expressions.

\begin{figure}[ht]
 \centering
 \begin{subfigure}{0.32\textwidth}
 \includegraphics[width=\linewidth]{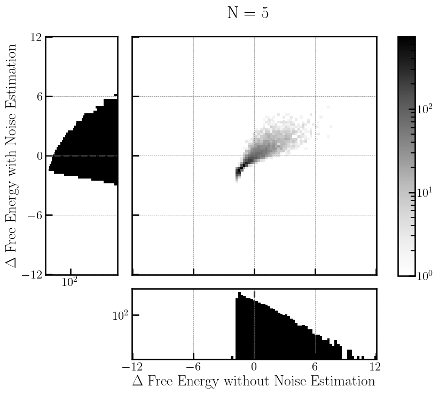}
 \caption{}
 \label{fig:b-20}
 \end{subfigure}
 \hfill
 \begin{subfigure}{0.32\textwidth}
 \includegraphics[width=\linewidth]{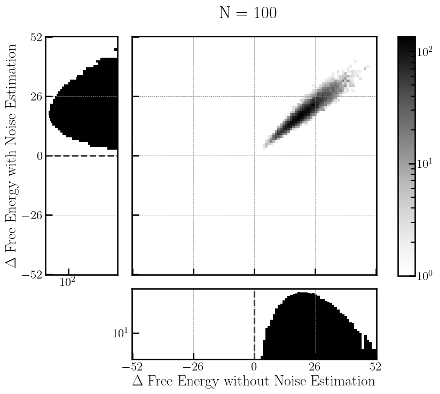}
 \caption{}
 \label{fig:b-21}
 \end{subfigure}
 \hfill
 \begin{subfigure}{0.32\textwidth}
 \includegraphics[width=\linewidth]{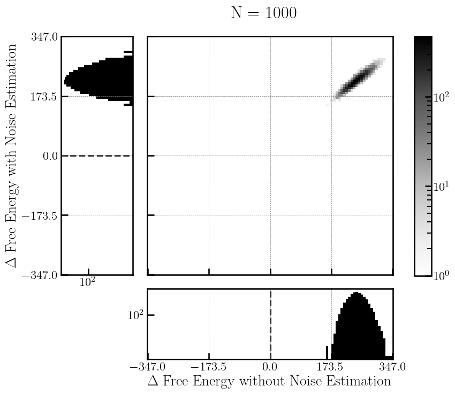}
 \caption{}
 \label{fig:b-22}
 \end{subfigure}
 \begin{subfigure}{0.32\textwidth}
 \includegraphics[width=\linewidth]{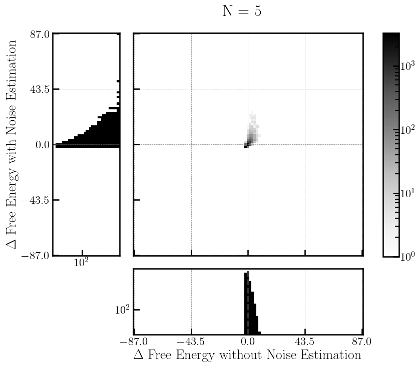}
 \caption{}
 \label{fig:b-26}
 \end{subfigure}
 \hfill
 \begin{subfigure}{0.32\textwidth}
 \includegraphics[width=\linewidth]{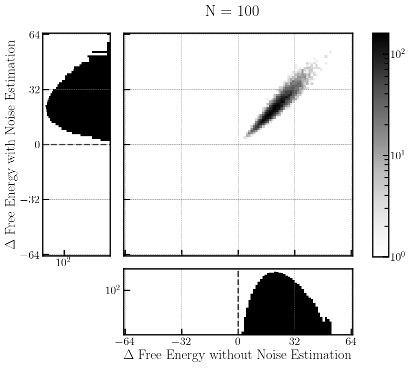}
 \caption{}
 \label{fig:b-27}
 \end{subfigure}
 \hfill
 \begin{subfigure}{0.32\textwidth}
 \includegraphics[width=\linewidth]{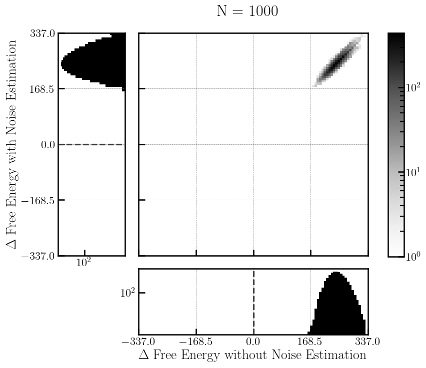}
 \caption{}
 \label{fig:b-28}
 \end{subfigure}
\caption{Frequency distribution sampled from 100,000 instances on the basis of the probability distribution of the difference in the Bayesian free energy with model parameters \(a_0^{(1)} = 2.0, a_0^{(2)} = 4.0, \sigma_0^2 = 1.0\) (see Equation (\ref{prob_modelselection_int})). The horizontal and vertical axes represent the frequency distribution of the free energy when noise is known and estimated, respectively. Noise estimation methods include simultaneous estimation of two noises from two datasets by minimizing free energy using Equation (\ref{free_energy_two_data}), and independent estimation of each noise from each dataset using Equation (\ref{freeenergysigma}). The results of simultaneous estimation are shown in the upper section and those of independent noise estimation in the lower section.}
 \label{fig:B7}
\end{figure}

Figure \ref{fig:B7} shows the frequency distribution of the difference in the Bayesian free energy, sampled from 100,000 instances with model parameters \(a_0^{(1)} = 2.0, a_0^{(2)} = 4.0, \sigma_0^2 = 1.0\) (see Equation (\ref{prob_modelselection_int})). The methods for noise estimation include simultaneous estimation of two noises from two datasets by minimizing free energy using Equation (\ref{free_energy_two_data}), and independent estimation of each noise from each dataset using Equation (\ref{freeenergysigma}). The results of simultaneous estimation are shown in the upper section and those of independent noise estimation in the lower section. The horizontal and vertical axes represent the frequency distribution of the free energy when noise is known and estimated, respectively. Figure \ref{fig:B7}shows that as \(N\) increases, the difference in the frequency distribution between the cases with estimated noise and known noise diminishes.

\begin{figure}[ht]
 \centering
 \begin{subfigure}{0.32\textwidth}
 \includegraphics[width=\linewidth]{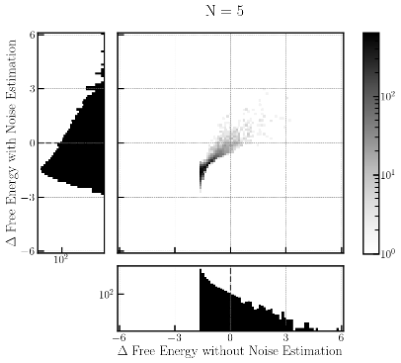}
 \caption{}
 \label{fig:b-23}
 \end{subfigure}
 \hfill
 \begin{subfigure}{0.32\textwidth}
 \includegraphics[width=\linewidth]{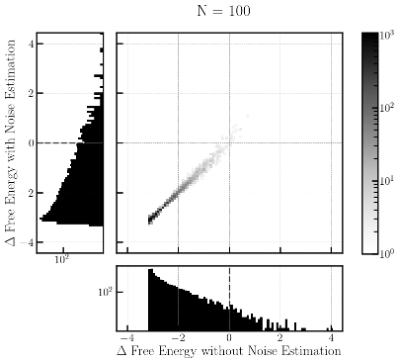}
 \caption{}
 \label{fig:b-24}
 \end{subfigure}
 \hfill
 \begin{subfigure}{0.32\textwidth}
 \includegraphics[width=\linewidth]{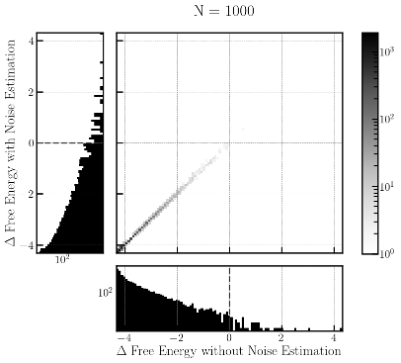}
 \caption{}
 \label{fig:b-25}
 \end{subfigure}
 \begin{subfigure}{0.32\textwidth}
 \includegraphics[width=\linewidth]{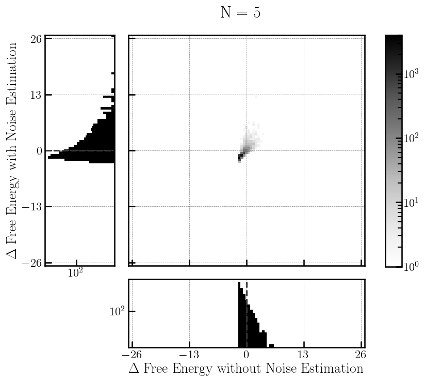}
 \caption{}
 \label{fig:b-29}
 \end{subfigure}
 \hfill
 \begin{subfigure}{0.32\textwidth}
 \includegraphics[width=\linewidth]{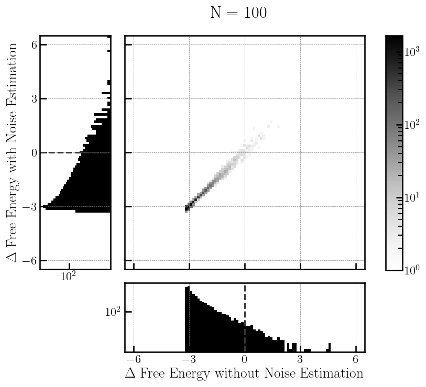}
 \caption{}
 \label{fig:b-30}
 \end{subfigure}
 \hfill
 \begin{subfigure}{0.32\textwidth}
 \includegraphics[width=\linewidth]{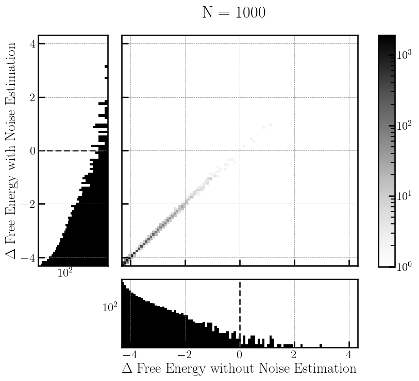}
 \caption{}
 \label{fig:b-31}
 \end{subfigure}
\caption{Frequency distribution sampled from the probability distribution of the difference in the Bayesian free energy, given the model parameters \(a_0^{(1)} = 2.0, a_0^{(2)} = 2.0, \sigma_0^2 = 1.0\) (Equation~(\ref{prob_modelselection_int})), with 100,000 samples. The horizontal and vertical axes represent the frequency distribution of the free energy when noise is known and estimated, respectively. There are two methods for noise estimation: one involves simultaneously estimating two noises by minimizing free energy using Equation~(\ref{free_energy_two_data}) from two data points, and the other involves independently estimating each noise from each data using Equation~(\ref{freeenergysigma}). The upper section shows the results of simultaneous estimation, and the lower section shows the results of independent noise estimation.}

 \label{fig:B8}
\end{figure}

Next, Figure~\ref{fig:B8} shows the frequency distribution obtained from sampling 100,000 times from the probability distribution of the difference in the Bayesian free energy, with model parameters \(a_0^{(1)} = 2.0, a_0^{(2)} = 2.0, \sigma_0^2 = 1.0\) (Equation~(\ref{prob_modelselection_int})). The horizontal and vertical axes represent the frequency distribution of the free energy when noise is known and estimated, respectively. There are two methods for estimating noise: one is to estimate two noises simultaneously by minimizing the free energy using Equation~(\ref{free_energy_two_data}) from two data points, and the other is to independently estimate each noise from each data using Equation~(\ref{freeenergysigma}). The results of simultaneous estimation are shown in the upper section, and those of independent noise estimation are displayed in the lower section. Figure~\ref{fig:B8}shows that as \(N\) increases, the difference in the frequency distributions between the cases of estimated noise and known noise disappears.

\begin{figure}[ht]
 \centering
 \begin{subfigure}{0.32\textwidth}
 \includegraphics[width=\linewidth]{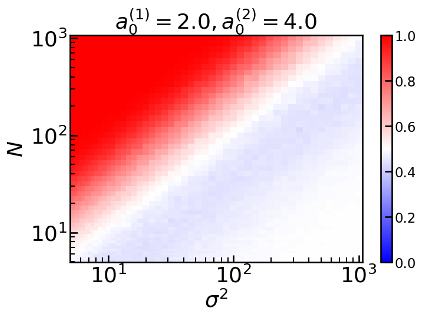}
 \caption{}
 \label{fig:b-32}
 \end{subfigure}
 \hfill 
 \begin{subfigure}{0.32\textwidth}
 \includegraphics[width=\linewidth]{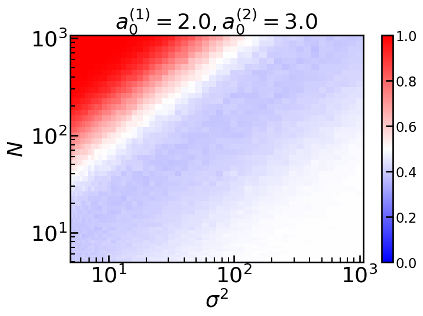}
 \caption{}
 \label{fig:b-33}
 \end{subfigure}
 \hfill 
 \begin{subfigure}{0.32\textwidth}
 \includegraphics[width=\linewidth]{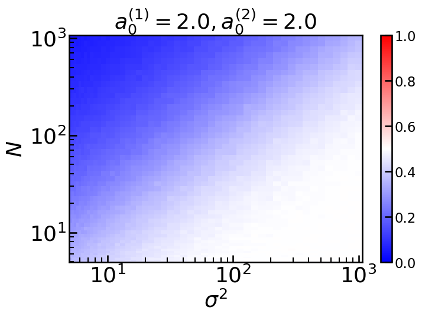}
 \caption{}
 \label{fig:b-34}
 \end{subfigure}
 \begin{subfigure}{0.32\textwidth}
 \includegraphics[width=\linewidth]{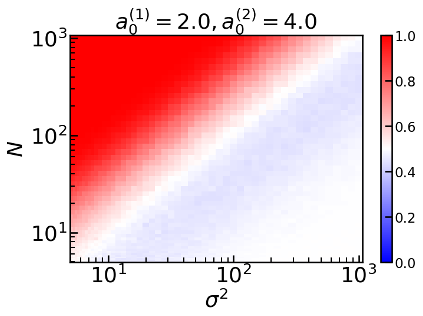}
 \caption{}
 \label{fig:b-35}
 \end{subfigure}
 \hfill 
 \begin{subfigure}{0.32\textwidth}
 \includegraphics[width=\linewidth]{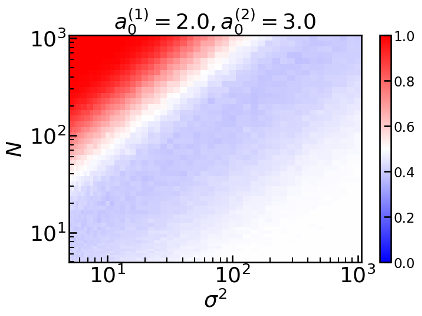}
 \caption{}
 \label{fig:b-36}
 \end{subfigure}
 \hfill 
 \begin{subfigure}{0.32\textwidth}
 \includegraphics[width=\linewidth]{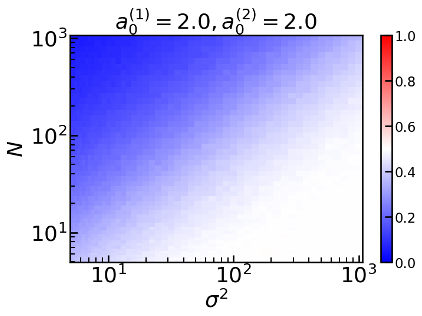}
 \caption{}
 \label{fig:b-37}
 \end{subfigure}

\caption{Probabilities of selecting the separated model derived from the probability distribution of the difference in the Bayesian free energy, given model parameters \(a_0^{(1)}=2.0, a_0^{(2)} = 4.0, 3.0, 2.0\) (Equation~(\ref{prob_modelselection_int})). Setting \(b_0^{(1)} = 0.0, b_0^{(2)} = 0.0\), the frequency distribution is shown as a two-dimensional histogram sampled 100,000 times from the bi-dimensional space of data number \(N\) and data noise strength \(\sigma^2\). Two noise estimation methods are used: one using Equation~(\ref{free_energy_two_data}) for simultaneously estimating two noises by minimizing the free energy from two data points, and another using Equation~(\ref{freeenergysigma}) for independently estimating each noise from each data. The upper section shows results from simultaneous estimation, while the lower section shows results from independent noise estimation.}
 \label{fig:B9}
\end{figure}

Finally, Figure~\ref{fig:B9} displays the probabilities of selecting the integrated model obtained from the probability distribution of the difference in the Bayesian free energy with model parameters \(a_0^{(1)}=2.0, a_0^{(2)} = 4.0, 3.0, 2.0\) (Equation~(\ref{prob_modelselection_int})). Here, we set \(b_0^{(1)} = 0.0, b_0^{(2)} = 0.0\) and show the frequency distribution as a two-dimensional histogram, sampled 100,000 times from the two-dimensional space of the number of data points \(N\) and the noise strength \(\sigma^2\). Two methods for estimating noise are used: one involves simultaneously estimating two noises by minimizing free energy using Equation~(\ref{free_energy_two_data}) from two data points, and the other involves independently estimating each noise from each data using Equation~(\ref{freeenergysigma}). The results of simultaneous estimation are shown in the upper section, and those of independent noise estimation are shown in the lower section.

From the aforementioned results, we found that the difference between simultaneously estimating two noises and estimating each noise independently becomes negligible with large data sizes. The former method involves optimization in a high-dimensional space, while the latter involves optimization in a one-dimensional space. Estimating multiple noises simultaneously increases the search space exponentially. Optimizing each noise independently ensures sufficient accuracy, and it is beneficial for real-world applications.

\end{document}